\pdfoutput=1
\documentclass[twocolumn]{aastex63}

\usepackage{amsmath}
\usepackage{graphicx}
\usepackage{url}
\usepackage{tabularx}
\usepackage{footmisc}

\graphicspath{{figures/}} 

\newcommand{\Kepler}{{Kepler}}

\submitjournal{AAS Journals}

\shorttitle{Robust \Kepler\ Occurrence Rates}
\shortauthors{Bryson et al.}

\begin{document}

\title{Reliability Correction is Key for Robust \Kepler\ Occurrence Rates}

\correspondingauthor{Steve Bryson}
\email{steve.bryson@nasa.gov}

\author[0000-0003-0081-1797]{Steve Bryson}
\affiliation{NASA Ames Research Center, Moffett Field, CA 94901}

\author[0000-0003-1634-9672]{Jeffrey L. Coughlin}
\affiliation{NASA Ames Research Center, Moffett Field, CA 94901}
\affiliation{SETI Institute, Mountain View, CA}

\author[0000-0001-9269-8060]{Michelle Kunimoto}
\affiliation{Department of Physics and Astronomy, University of British Columbia, 6224 Agricultural Road, Vancouver, BC V6T 1Z1, Canada}

\author[0000-0001-7106-4683]{Susan E. Mullally}
\affiliation{Space Telescope Science Institute, 3700 San Martin Drive, Baltimore, MD, 21218, USA}

\begin{abstract}
The \Kepler\ DR25 planet candidate catalog was produced using an automated method of planet candidate identification based on various tests.  These tests were tuned to obtain a reasonable but arbitrary balance between catalog completeness and reliability. We produce new catalogs with differing balances of completeness and reliability by varying these tests, and study the impact of these alternative catalogs on occurrence rates.  We find that if there is no correction for reliability, different catalogs give statistically inconsistent occurrence rates, while if we correct for both completeness and reliability, we get statistically consistent occurrence rates.  This is a strong indication that correction for completeness and reliability is critical for the accurate computation of occurrence rates.  Additionally, we find that this result is the same whether using Bayesian Poisson likelihood MCMC or Approximate Bayesian Computation methods.  We also examine the use of a Robovetter disposition score cut as an alternative to reliability correction, and find that while a score cut does increase the reliability of the catalog, it is not as accurate as performing a full reliability correction.  We get the same result when performing a reliability correction with and without a score cut.  Therefore removing low-score planets removes data without providing any advantage, and should be avoided when possible.  We make our alternative catalogs publicly available, and propose that these should be used as a test of occurrence rate methods, with the requirement that a method should provide statistically consistent occurrence rates for all these catalogs.

\end{abstract}


\section{Introduction}

The \Kepler\ space telescope \citep{Borucki2010,Koch2010} has delivered unique data that enables the characterization of exoplanet population statistics, from hot Jupiters in short-period orbits to terrestrial-size rocky planets in orbits with periods up to $\sim$one year\footnote{\url{https://exoplanetarchive.ipac.caltech.edu/docs/occurrence_rate_papers.html}}. 
By observing $>$150,000 stars nearly continuously for four years looking for transiting exoplanets, \Kepler\ detected over 4,000 planet candidates (PCs) \citep{Thompson2018}, leading to the confirmation or statistical validation of over 2,300 exoplanets.  This rich trove of exoplanet data has delivered many insights into exoplanet structure and formation, and promises deeper insights with further analysis. One of the most exciting insights to be gained from \Kepler\ data is $\eta_{\oplus}$, the occurrence rate of temperate, terrestrial-size planets orbiting Sun-like stars.  $\eta_{\oplus}$ is also a critical input to the design of future space telescopes for the characterization of habitable exoplanets.

Fully utilizing \Kepler\ data to calculate accurate occurrence rates requires a thorough understanding of how well it reflects the underlying exoplanet population.  There are several ways in which the \Kepler\ planet candidate catalog does not directly measure the real planet population.  In this paper we focus on the following:

\begin{itemize}
    \item The catalog is {\it incomplete}, missing real planets
    \item The catalog is {\it unreliable}, being polluted with false positives (FPs)
\end{itemize}

Low completeness and low reliability are particularly acute at the \Kepler\ detection limit, which happens to coincide with the period and radius of Earth-Sun analog exoplanets.  We therefore focus our attention on a period and radius range $50 \leq$ period $\leq 400$ days and $0.75 \leq$ radius $\leq 2.5$ $R_{\oplus}$, which spans the \Kepler\ detection limit.  We refer to this range as our domain of analysis.

\citet{Bryson2020} developed a method for using data provided in the final \Kepler\ data release (DR25) to characterize the completeness and reliability of the DR25 planet candidate catalog, and use this characterization in occurrence rate calculations.  The occurrence rates presented in \citet{Bryson2020} were not considered definitive, and several issues with the occurrence rate calculations were discussed in detail.  In this paper we address two issues: reliance on the specific balance between completeness and reliability of the DR25 planet candidate catalog, and the assumption that planet occurrence is described by a Poisson likelihood.

We address possible dependence on the balance of the DR25 PC catalog between completeness and reliability by creating new PC catalogs that give greater emphasis to either completeness or reliability.  As described in \S\ref{section:thresholdVariation}, these new catalogs provide lists of planet candidates in Kepler data that are just as valid as the DR25 PC catalog, so we would expect a good occurrence rate measurement to give the same result for all of these catalogs.  We will find that, when correcting for completeness and reliability, the method of \citet{Bryson2020} computes the same occurrence rates for all the catalogs.  

We address the possible dependence on the assumption of a Poisson likelihood by computing the occurrence rates using Approximate Bayesian Computation (ABC) as described in \citet{Kunimoto2020a} and \citet{kunimoto2020b}, following methods developed in \citet{Mulders2018} and \citet{He2019}.  The ABC method treats completeness and reliability, as well as the statistics of the planet population, very differently from the Poison likelihood method used by \citet{Bryson2020}.  We find that both methods result in essentially the same occurrence rate results.  

We also examine the use of DR25 Robovetter Disposition Score \citep{Thompson2018} to provide a high-reliability planet candidate catalog, potentially making a correction for reliability unnecessary.  We use the consistency of results for the PC catalogs presented in this paper as the diagnostic criterion.  We find that, while using the disposition score significantly improves consistency,  correction for reliability is still indicated.  Not using the disposition score and correcting for reliability gives the most consistent results.  

This paper is structured as follows: In \S\ref{section:compRelProducts} and \ref{section:stellarCatalogs} we review the DR25 catalog and stellar properties used in \citet{Bryson2020}.  We describe our alternative planet candidate catalogs in \S\ref{section:thresholdVariation}, and the Bayesian computation of the planet population and occurrence rates in \S\ref{section:occMethods}.  We present our results in \S\ref{section:results}, and interpret these results in \S\ref{section:discussion}.  

All results reported in this paper were produced with Python code, mostly in the form of Python Jupyter notebooks, found at the paper GitHub site\footnote{\url{https://github.com/stevepur/DR25-occurrence-public/GKRobovetterVariations}}.  This site also contains the alternative PC catalogs described in \S\ref{section:thresholdVariation}.  
\section{Methodology} \label{section:methodology}
\subsection{DR25 Completeness, Reliability, and Score} \label{section:compRelProducts}
Each planet candidate catalog starts with the DR25 catalog of 34,032 {\it threshold crossing events} (TCEs) \citep{Twicken2016}, which are periodic transit-like events, identified by a matched filter \citep{Jenkins2002}, that have a combined signal strength above a threshold (set to $7.1\sigma$ for DR25). Identification of the PCs from the TCEs was performed by a fully automated {\it Robovetter} \citep{Coughlin2017}.  The Robovetter applies a variety of tests to each TCE, many of which were tuned on the synthetic test datasets described below.  When a TCE passes tests that indicate a resemblance to a planetary transit or eclipsing binary, it is elevated to a \Kepler\ Objects of Interest (KOI). If the KOI passes further tests, it is elevated to planet candidate (PC) status.  Such automated vetting (and transit detection) is critical for the production of a statistically uniform catalog that is amenable to statistical correction for completeness and reliability.   The DR25 planet candidate catalog \citep{Thompson2018} contains 4034 identified PCs out of 8054 KOIs.  

The DR25 Robovetter uses a number of metrics to identify instrumental false alarms, and the inverted and scrambled data sets were used to tune their pass/fail thresholds.  For an extensive discussion, see \citet{Thompson2018}.

\subsubsection{Detection and Vetting Completeness} \label{section:completeness}

The DR25 completeness products are based on {\it injected data} --- a ground-truth of transiting planets obtained by injecting artificial transit signals with known characteristics on all observed stars at the pixel level \citep{Christiansen2017}.  A large number of transits were also injected on a small number of target stars to measure the dependence of completeness on transit parameters and stellar properties. This data is then analyzed by the \Kepler\ detection pipeline to produce a catalog of detections at the injected ephemerides called {\it injected and recovered TCEs}, which are then sent through the same Robovetter used to identify planet candidates.  

{\bf Detection completeness} is defined as the fraction of injected transits that are recovered as TCEs by the \Kepler\ detection pipeline, regardless of whether or not those TCEs are subsequently identified as planet candidates.  We use the detection completeness of \citet{BurkeJCat2017}, which was computed for each target star as a function of period, the simulated Multiple Event Statistic (MES; a measure of the signal-to-noise ratio (S/N) that is specific to the \Kepler\ pipeline) based on stellar noise properties measured in that star's \Kepler\ light curve.  The result is referred to as {\it completeness detection contours}.

{\bf Vetting completeness} is defined as the fraction of detected injected transits that were identified as planet candidates by the Robovetter.  Vetting completeness is computed for a population of stars based on the simulated MES and orbital period of injected transits.  We use the method of \citet{Bryson2020}, which models vetting completeness as a binomial problem with a rate given by a product of rotated logistic functions of MES and orbital period.  The specific shape of the logistic functions depend on the Robovetter thresholds.  The detection completeness contours are multiplied by this vetting completeness function for each star.

The product of vetting and detection completeness as a function of period and MES is converted to a function of period and planet radius for each star.  This product is further multiplied by the geometrical transit probability for each, which is a function of planet period and radius, given that star's radius.  The final result is a completeness contour for each star that includes detection and vetting completeness, and geometrical transit probability.

\subsubsection{Vetting Reliability} \label{section:reliability}

Vetting reliability is the fraction of planet candidates that are true planets \citep{Thompson2018}.  Reliability is estimated by determining the rate of two types of misidentified planet candidates: {\it astrophysical false positives} and non-transit-like {\it false alarms}.  Astrophysical false positives are periodic transit-like signals that are not caused by planets, most often eclipsing binaries.  We evaluate the probability that a PC is an astrophysical false positive using the False Positive Probability (FPP) of \citet{Morton2016}.  False alarms can be caused by stellar variability, but, especially at longer periods, are dominated by instrumental artifacts \citep{Thompson2018}.  False alarm reliability $R_{\mathrm{FA}}$ is defined as the fraction of PCs that are not false alarms.  The final reliability for each planet candidate is the product $(1-\mathrm{FPP}) R_{\mathrm{FA}}$

False alarm reliability $R_{\mathrm{FA}}$ is decomposed into two rates, the  {\it false alarm effectiveness} $E_{\mathrm{FA}}$, the fraction of true false alarms correctly identified as false alarms by the Robovetter, and the observed false-alarm rate $F_{\mathrm{FA}}$, the fraction of TCEs classified as false alarms.  $E_{\mathrm{FA}}$ is measured using observed data manipulated to eliminate true periodic transit-like signals, called inverted and scrambled data, described in \citet{Thompson2018}.  Any TCE found in the inverted and scrambled data is a false alarm.  Then the reliability against false alarms is given by \citep{Thompson2018}
\begin{equation} \label{eqn:reliabilityRate}
    R_{\mathrm{FA}} = 1 - \frac{F_{\mathrm{FA}}}{1-F_{\mathrm{FA}}} 
        \left( \frac{1-E_{\mathrm{FA}}}{E_{\mathrm{FA}}} \right).
\end{equation}
$R_{\mathrm{FA}}$ is computed as a two-dimensional function of MES and period, based on what PCs were found in the scrambled and inverted data, and the observed false alarms.  The reliability of a PC is computed by evaluating $R_{\mathrm{FA}}$ at that PC's MES and period.

In \citet{Bryson2020} $E_{\mathrm{FA}}$ and $F_{\mathrm{FA}}$ are characterized separately.  However, we have found that this separate characterization resulted in negative values for $R_{\mathrm{FA}}$ when applied to the high-completeness catalog described in \S\ref{section:thresholdVariation}, which is not meaningful.  Therefore in this paper we characterize $E_{\mathrm{FA}}$ and $F_{\mathrm{FA}}$ in a joint MCMC inference using a likelihood formed from the product of the individual likelihoods for $E_{\mathrm{FA}}$ and $F_{\mathrm{FA}}$.  The requirement of $0 \leq R_{\mathrm{FA}} \leq 1$ is enforced by making this requirement part of the prior for the joint inference.  We performed this joint inference of $E_{\mathrm{FA}}$ and $F_{\mathrm{FA}}$ on all catalogs discussed in this paper, including DR25.

\begin{figure*}[ht]
  \centering
  \includegraphics[width=0.48\linewidth]{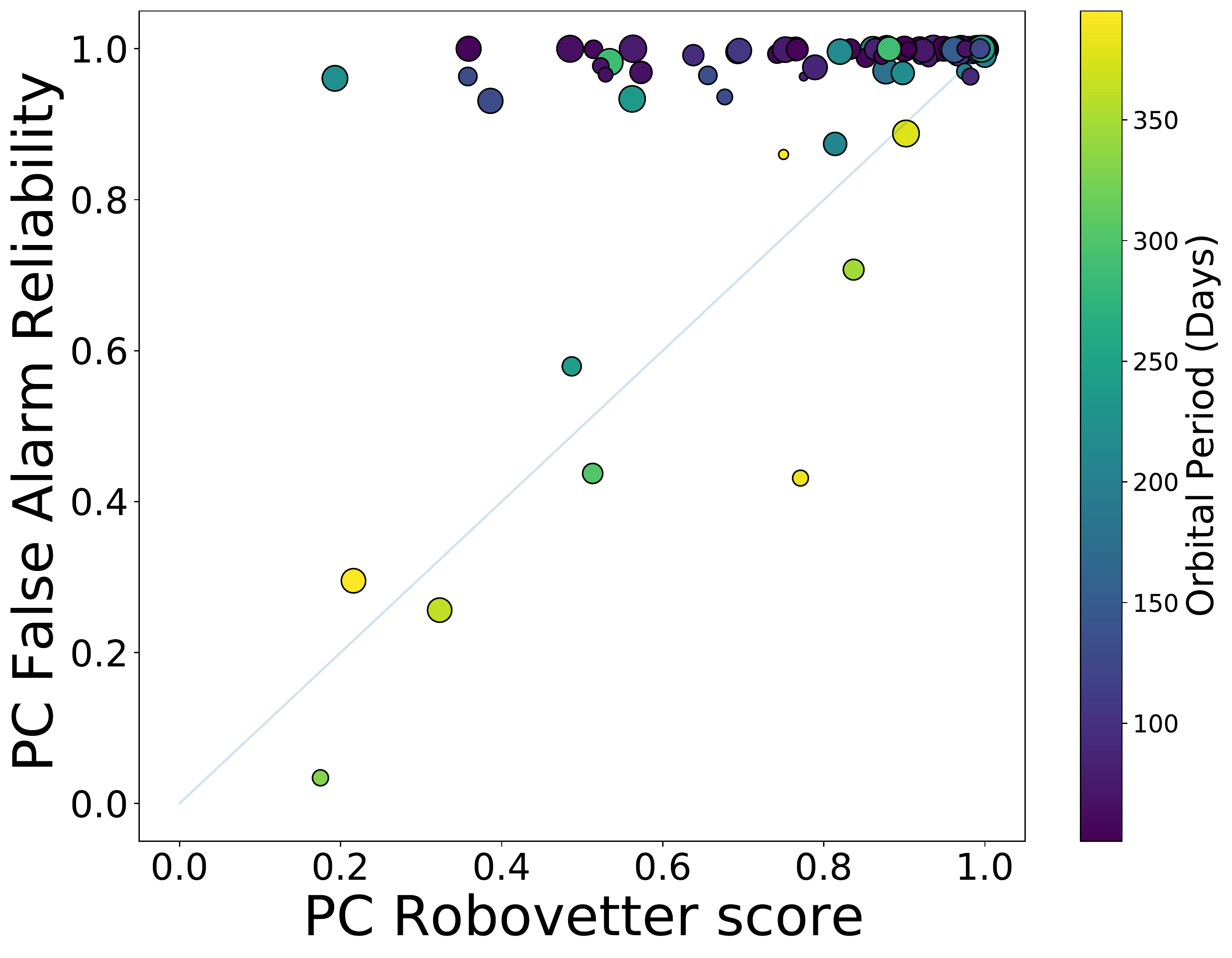} 
  \includegraphics[width=0.48\linewidth]{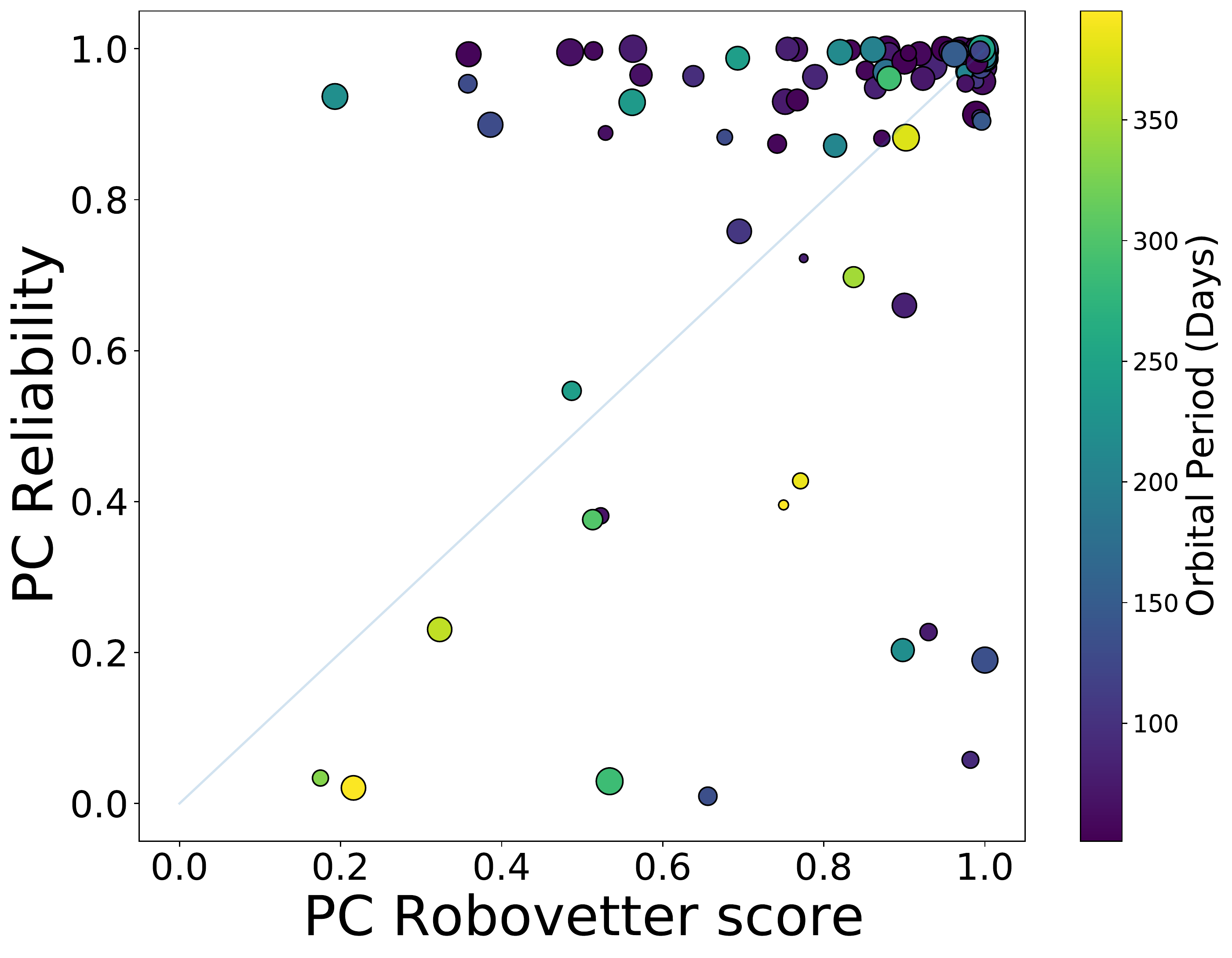} 
\caption{Left: Robovetter disposition score plotted against false alarm reliability for the DR25 GK PC population with $50 \leq$ period $\leq 400$ days and $0.75 \leq$ radius $\leq 2.5$ $R_{\oplus}$, with the color of each planet showing its orbital period and marker size showing its radius.  Most PCs, with short orbital periods, have high reliability but strongly varying score, while long-period PCs have a strong correlation between score and reliability.  Right: the same PC population showing the combined false alarm and astrophysical false positive reliability.  In this case a high score cut does not remove PCs which have low reliability due to a high false positive probability.} \label{figure:scoreVsReliability}
\end{figure*}

\begin{figure*}[ht]
  \centering
  \includegraphics[width=0.48\linewidth]{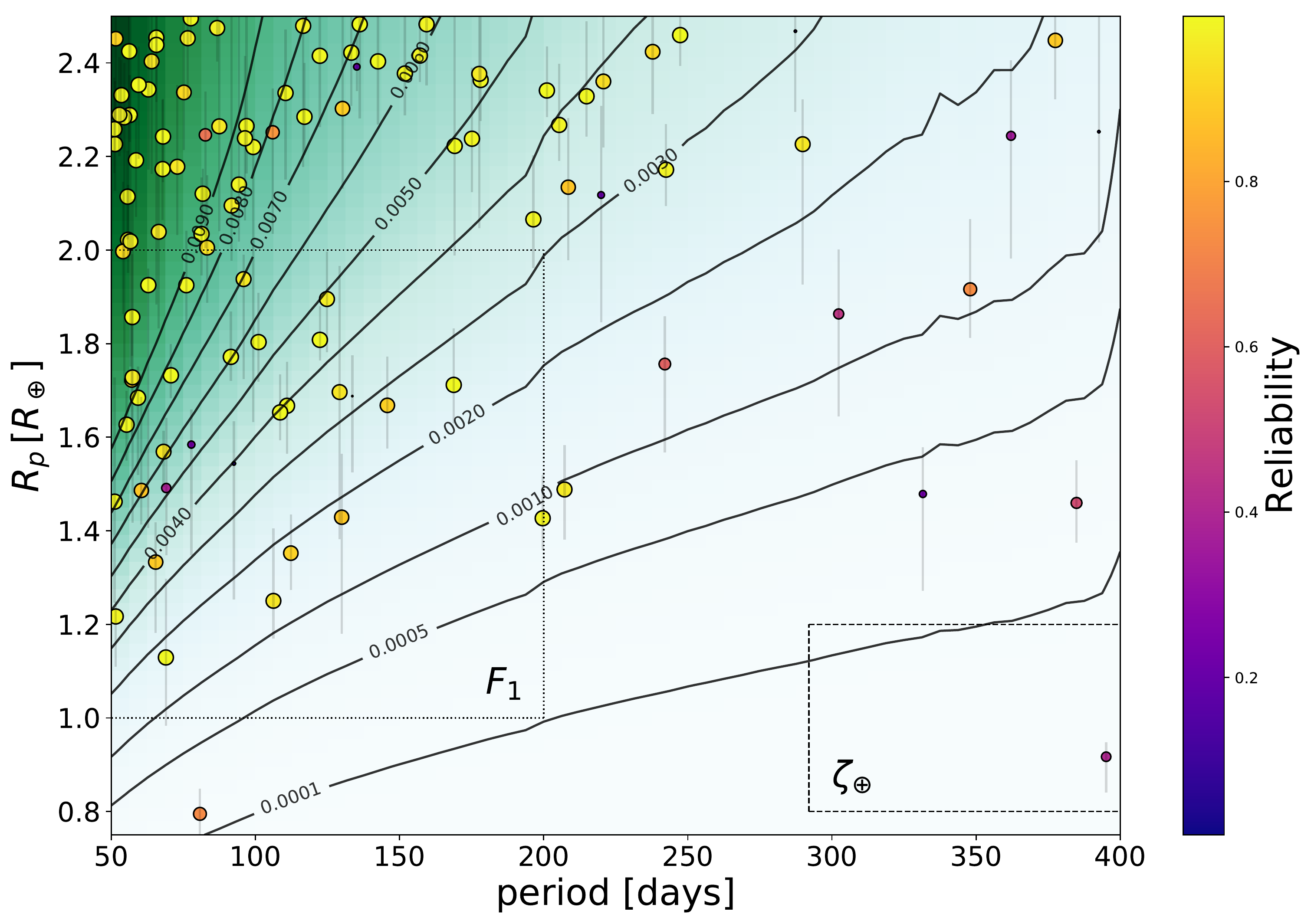} 
  \includegraphics[width=0.48\linewidth]{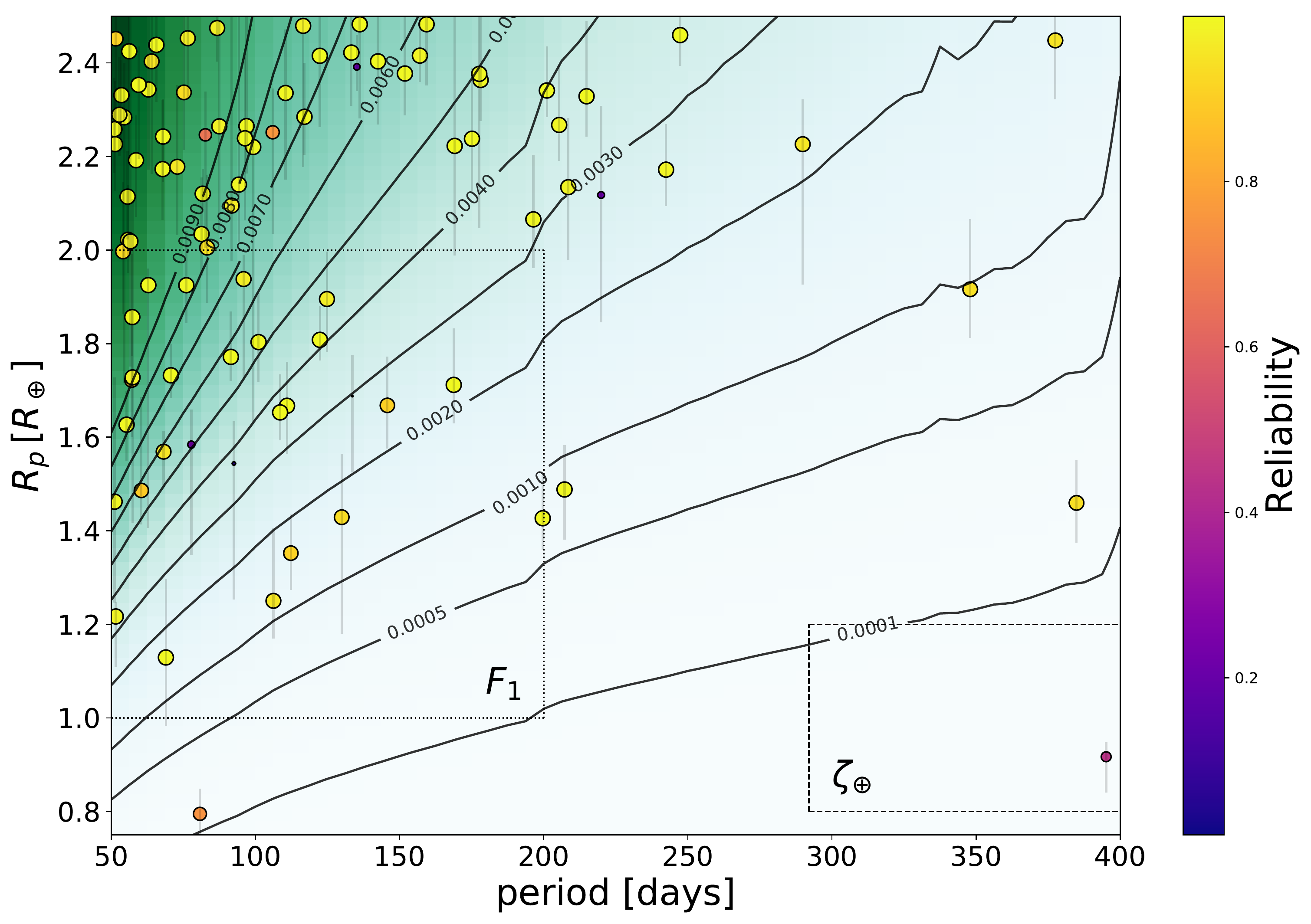} \\
  \includegraphics[width=0.48\linewidth]{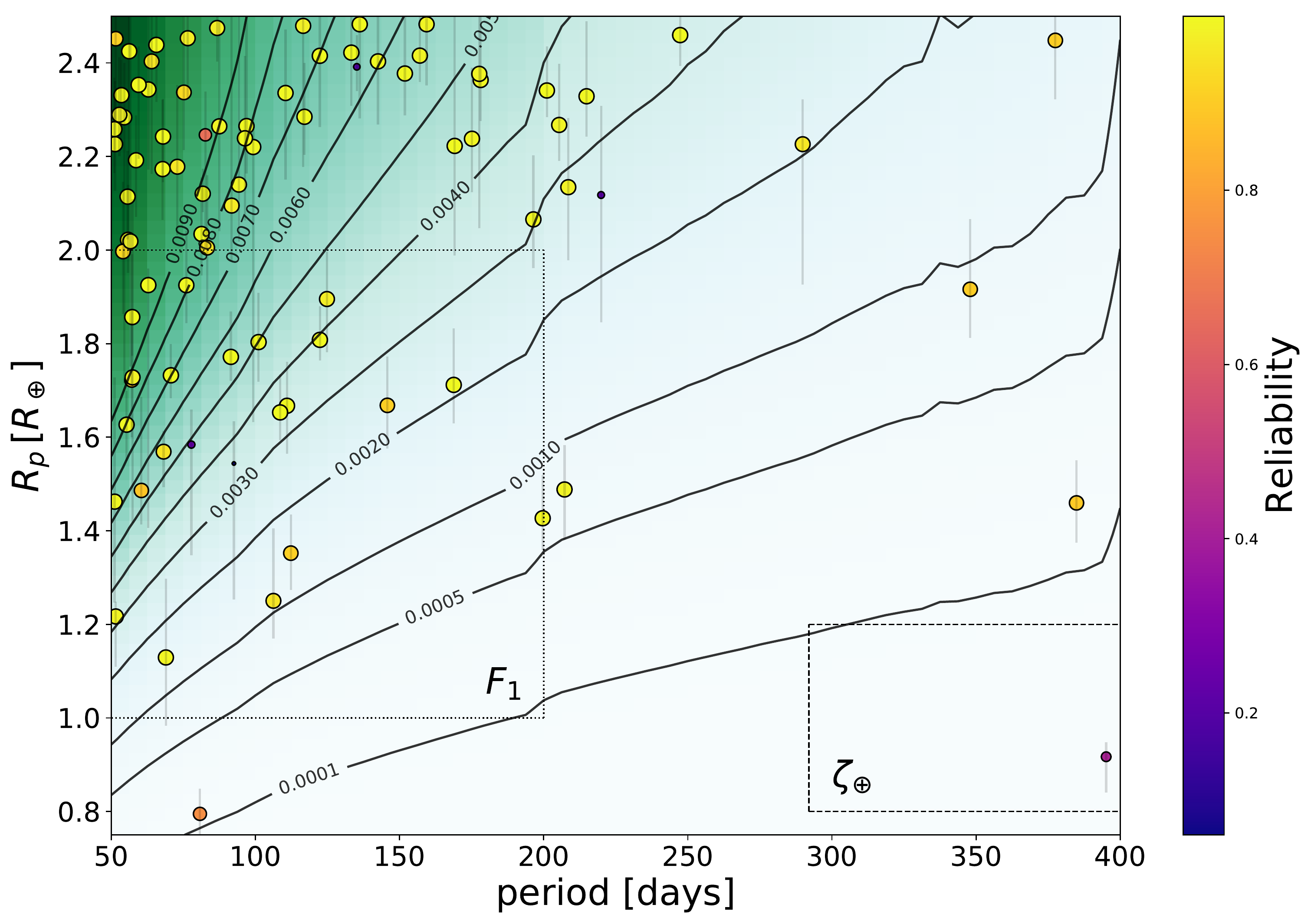}
  \includegraphics[width=0.48\linewidth]{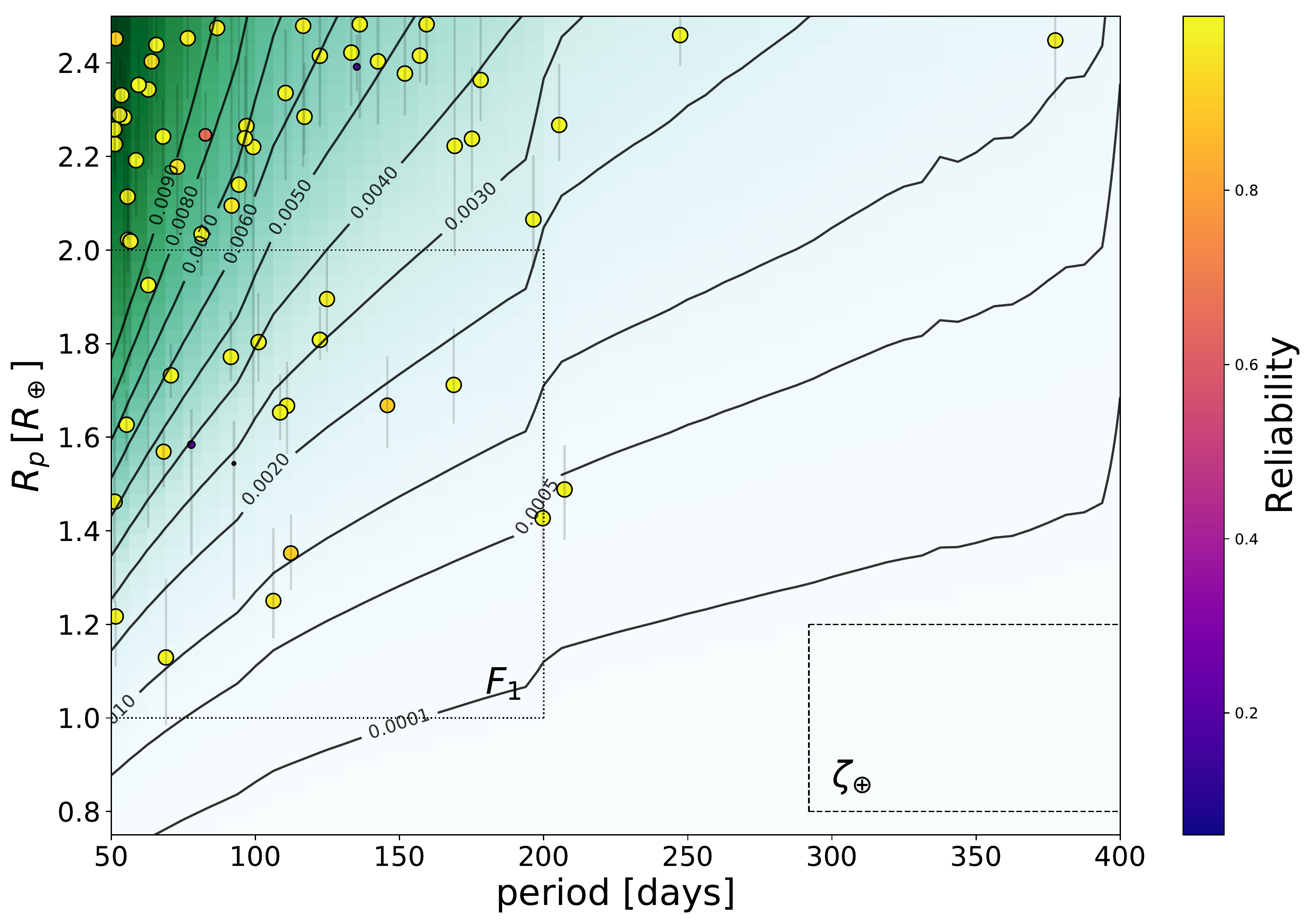}
  \caption{The DR25 planet candidate population after imposing various score cuts. Top Left: score cut = 0.  Top right: score cut = 0.6. Bottom Left: score cut = 0.7.  Bottom Right: score cut = 0.9. The planet candidates are colored and sized by reliability with planet radius error bars.  The background color map and contours indicate the summed completeness function $\eta(p, r)$.  The box on the left indicates the region integrated to obtain the occurrence rate $F_1$, while the box on the right indicates the integration region for the occurrence rate $\zeta_{\oplus}$.  The $\zeta_{\oplus}$ box extends out to 438 days. The occurrence rate $F_0$ is the integral over the range of the figure.}   \label{figure:planetPopulationScoreCut}
\end{figure*}

\subsubsection{Robovetter Disposition Score} \label{section:score}

As described in \citet{Thompson2018}, the Robovetter disposition score is a measure of the confidence of the Robovetter's classification of a TCE into a PC or FP.  This score is measured by varying the Robovetter metrics according to their uncertainties, and the score of a TCE is the fraction of those variations for which the TCE is classified as a PC.  The score can be thought of as the propagation of the uncertainty of the Robovetter metrics for each TCE.  A high-score PC (near 1.0) is almost always classified as a PC, while a low-score PC (near 0.0) is almost always classified as a FP.

Robovetter disposition score and false alarm reliability are often conflated, but are conceptually very different.  The score of a PC is determined by the Robovetter metrics for that PC based on that PC's observed data.  False alarm reliability is determined by the rate of the Robovetter's identification of PCs in the inverted and scrambled data, and the observed rate of false alarms, at the PC's MES and period.  

The relationship between score and reliability is shown in Figure~\ref{figure:scoreVsReliability} for PCs in our domain of analysis $50 \leq$ period $\leq 400$ days and $0.75 \leq$ radius $\leq 2.5$ $R_{\oplus}$.  The left panel shows false alarm reliability.  PCs with shorter periods have high false alarm reliability but strongly varying score, indicating no relationship.  Long-period PCs, on the other hand, show a strong correlation between score and false alarm reliability.  This correlation provides some justification for the use of score as a proxy for false-alarm reliability, producing a ``high-reliability'' PC population by removing those PCs below a score threshold (\citet{Mulders2018}, for example).  But such a score cut will remove many high-reliability planets.  The right panel of Figure~\ref{figure:scoreVsReliability} shows the PC reliability including astrophysical false positives.  We see that even a high score cut such as 0.9 does not remove all low-reliability PCs due to their having a high false positive probability.  This indicates that reliability correction is still useful when using a score cut.  

Figure~\ref{figure:planetPopulationScoreCut} shows the impact of various score cuts on the PC population in our domain of analysis.  In this paper we will study the impact score cuts and address the possibility that using high-score PCs may provide more accurate occurrence rates.

\subsection{Input Stellar and Planet Catalogs} \label{section:stellarCatalogs}

As in \citet{Bryson2020}, our stellar catalog uses the Gaia-based stellar properties from \citet{Berger2020a} combined with the DR25 stellar catalog at the exoplanet archive, with the cuts described in the baseline case of \citet{Bryson2020}.  This gives us 57,015 GK stars whose noise properties and observational coverage make them appropriate for a statistical exoplanet survey.

We use planet properties from the \Kepler\ Threshold Crossing Events (TCE) catalog, with planet radii corrected using the Gaia-based stellar radii from \citet{Berger2020a} as in \citet{Bryson2020}.  These radii differ from those in \citet{Berger2020b} by a constant factor $ = 1.00226$, due to a small difference in the value of $R_\oplus / R_\odot$.

\subsection{Varying the Robovetter Vetting Thresholds} \label{section:thresholdVariation}

We produce and compare different planet candidate catalogs, with differing balances between completeness and reliability, by varying the thresholds used by the Robovetter to identify planet candidates (PCs) from the transit signals identified as Transit Crossing Events (TCEs) by the \Kepler\ data analysis pipeline.  We produce four PC catalogs for the stellar population described in \S\ref{section:stellarCatalogs}:
\begin{itemize}
    \item {\bf DR25,} the original \Kepler\ planet candidate catalog, which was analyzed in detail in \citet{Bryson2020}.
    \item {\bf High Reliability,} which uses more restrictive Robovetter thresholds and rejects more borderline transit detections compared to the original DR25 catalog, resulting in higher reliability and lower completeness.
    \item {\bf High Completeness,} which uses less restrictive Robovetter thresholds and accepts more borderline transit detections compared to the original DR25 catalog, resulting in lower reliability and higher completeness.
    \item {\bf FPWG PC,} which attempts to tune the Robovetter thresholds to pass DR25 false positive KOIs that are identified as possible planets by the \Kepler\ False Positive Working Group (FPWG) \citep{Bryson2015}.
\end{itemize}
The high-reliability and high-completeness catalogs use the alternative Robovetter thresholds described in \S5.2 of \citet{Thompson2018}.  Details of the alternative vetting thresholds are given in Appendix~\ref{appendix}, as well as the new planet candidates that appear in the high-reliability and FPWG PC catalogs.  We believe that these four catalogs are equally valid, imperfect, catalogs of planet candidates in the \Kepler\ data, each with differing reliability and completeness.

\begin{figure*}[ht]
  \centering
  \includegraphics[width=0.48\linewidth]{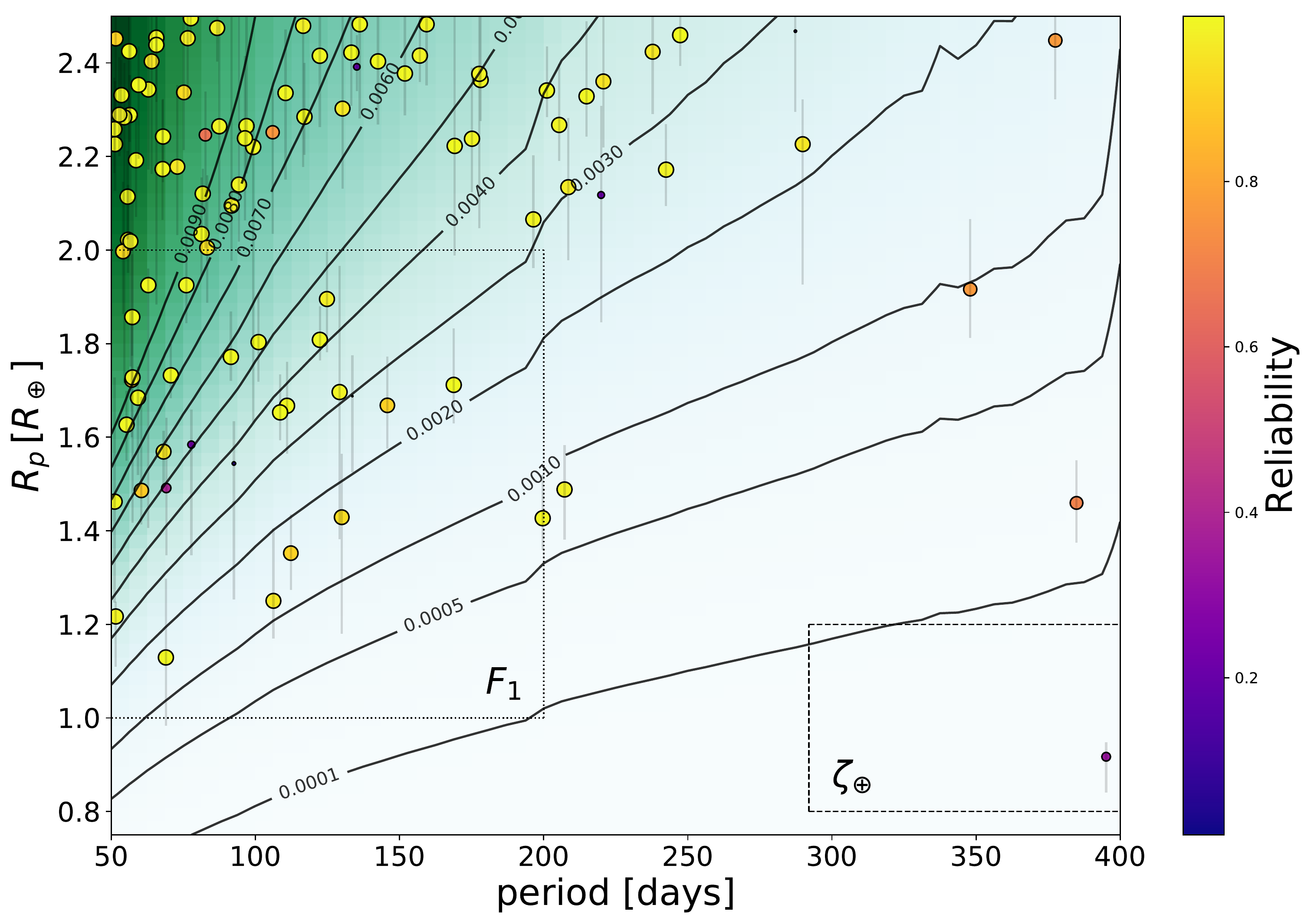} 
  \includegraphics[width=0.48\linewidth]{summedCompletenessDR25.pdf} \\
  \includegraphics[width=0.48\linewidth]{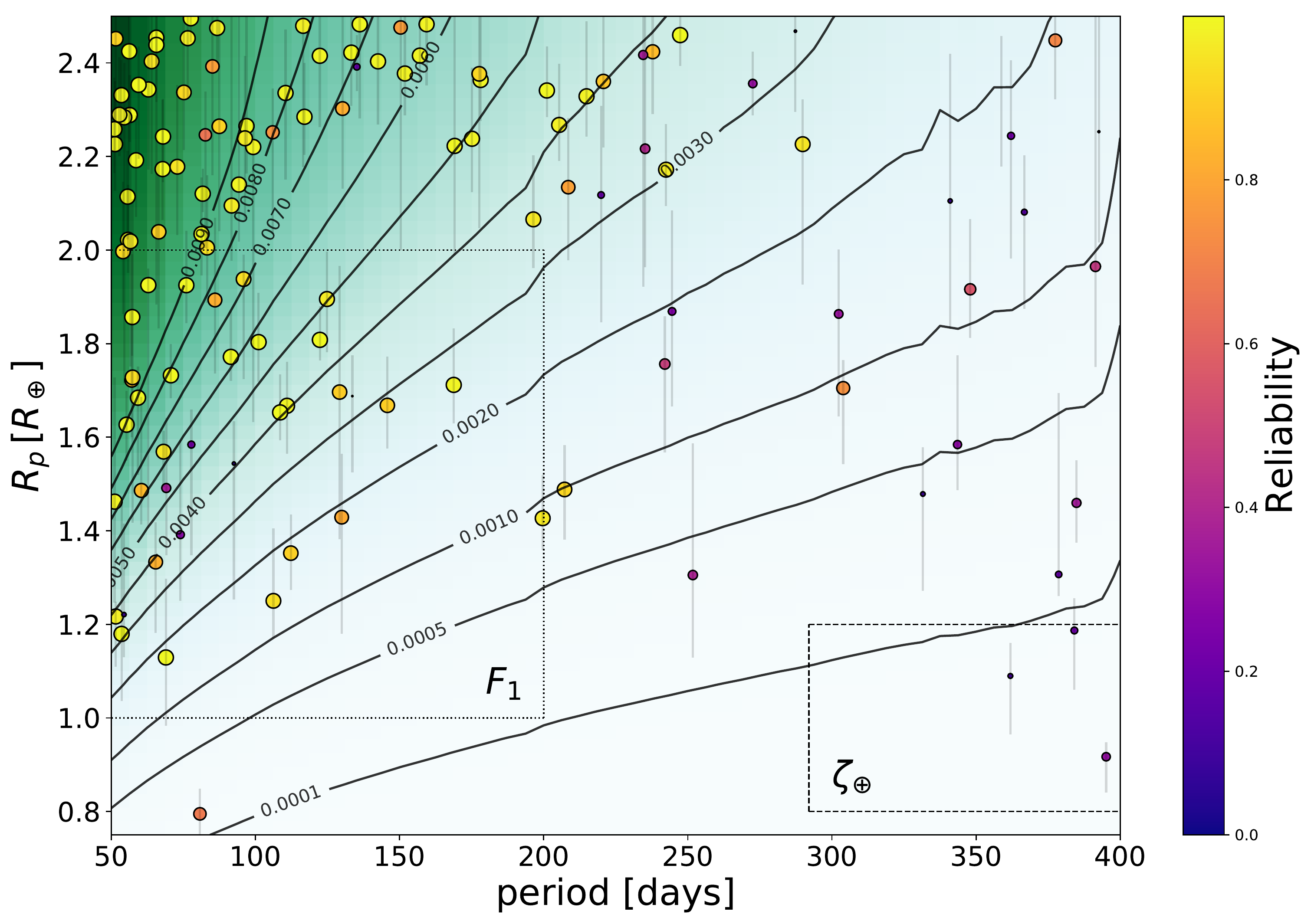}
  \includegraphics[width=0.48\linewidth]{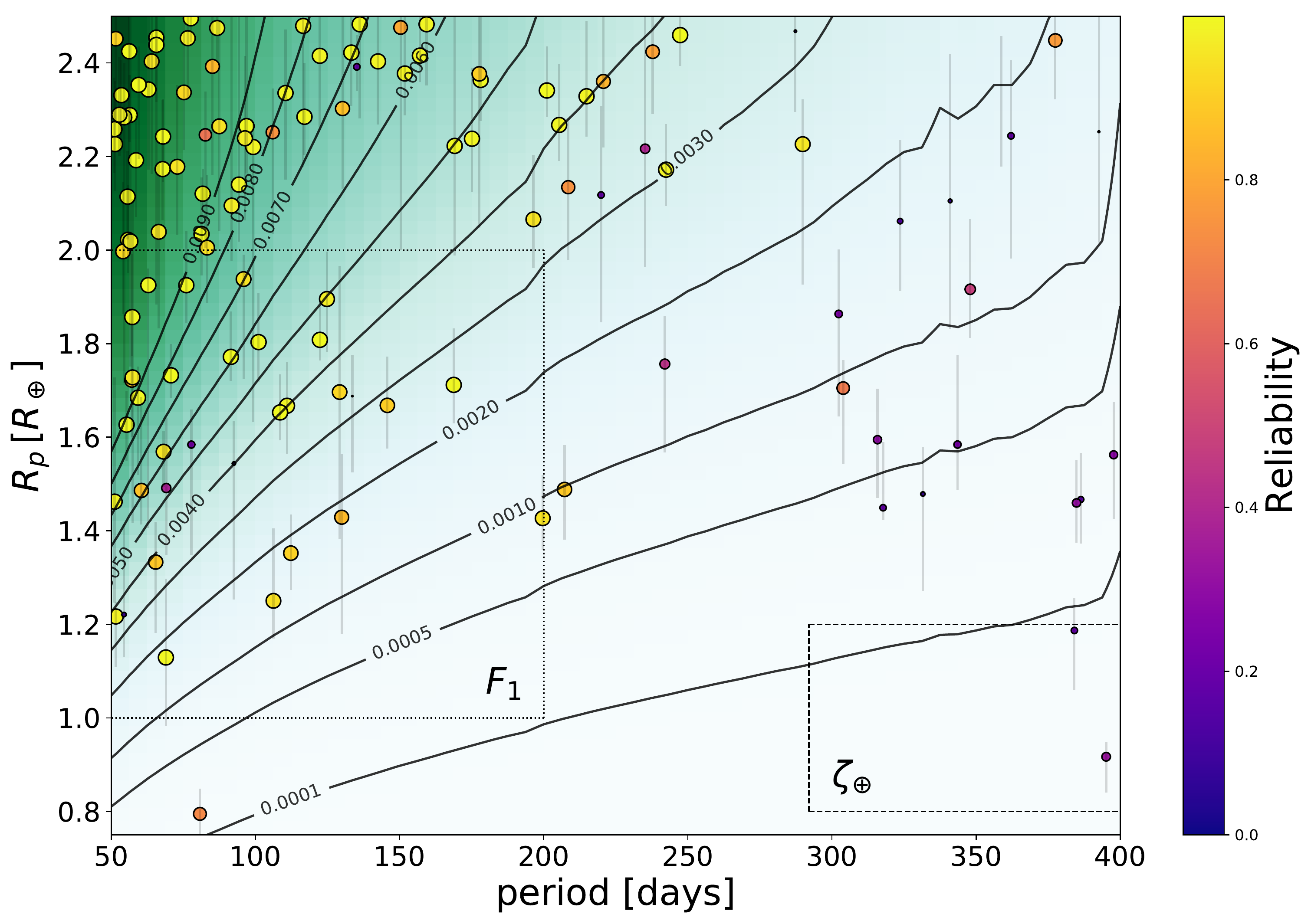}
  \caption{The planet candidate population for the four catalogs described in \S\ref{section:thresholdVariation}.  Top Left: high reliability.  Top Right: DR25. Bottom Left: FPWG PC. Bottom Right: high completeness.  PCs are colored and sized by reliability with planet radius error bars.  The background color map and contours indicate the summed completeness function $\eta(p, r)$.  The box on the left indicates the region integrated to obtain $F_1$, while the box on the right indicates the integration region for $\zeta_{\oplus}$.  The $\zeta_{\oplus}$ box extends out to 438 days.} \label{figure:planetPopulation}
\end{figure*}

These catalogs are shown for our domain of analysis $50 \leq$ period $\leq 400$ days and $0.75 \leq$ radius $\leq 2.5$ $R_{\oplus}$ in Figure~\ref{figure:planetPopulation}.  We see that there is a strong variation between the cases in the number of planets with orbital periods $> 200$ days.  

In the high completeness case, 14 planet candidates appear within our domain of analysis that were vetted as false positives in DR25.  We manually inspected these 14 new PCs using the TCERT Vetting Reports\footnote{\url{https://exoplanetarchive.ipac.caltech.edu/docs/Kepler_TCE_docs.html}}, and found that only 5 presented plausible planetary transit signals.  The other 9 are likely false alarms due to instrumental artifacts or stellar variability.  

For each catalog and its corresponding Robovetter thresholds, we run the Robovetter on the injected, inverted and scrambled data, producing the data required to compute the vetting completeness and reliability of each catalog.  

\begin{figure}[ht]
  \centering
  \includegraphics[width=\linewidth]{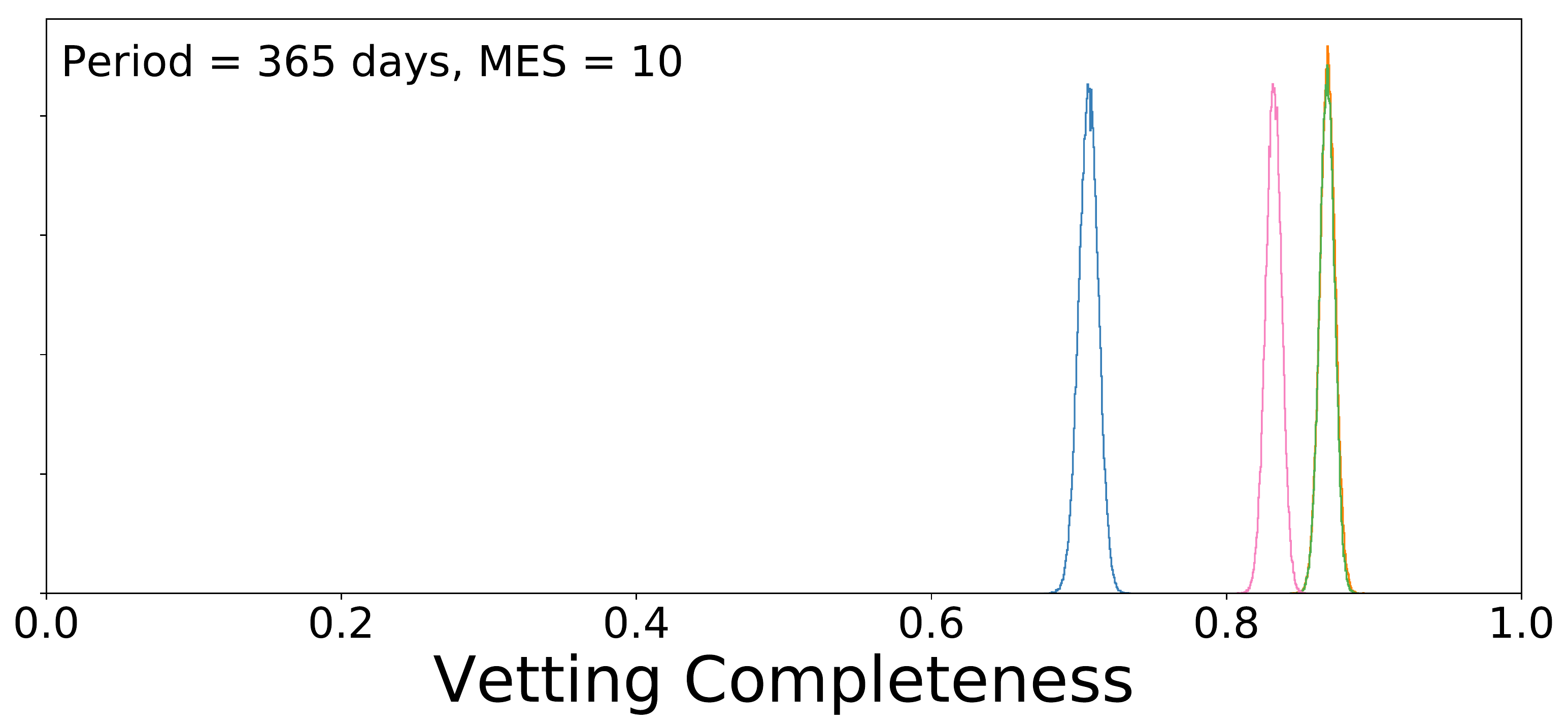} 
   \caption{Vetting completeness distributions evaluated at period = 365 days and expected MES = 10 for high reliability (blue), DR25 (pink), FPWG PC (green) and high completeness (orange) vetting.  } \label{figure:vetCompExamples}
\end{figure}

Figure~\ref{figure:vetCompExamples} shows example distributions of vetting completeness for our planet candidate catalogs for the long-period, low MES case of period 365 days and MES = 10.  In this case there is a large variation in the vetting completeness between the catalogs.  As expected, the high-reliability case has lower vetting completeness.

\begin{figure}[ht]
  \centering
  \includegraphics[width=\linewidth]{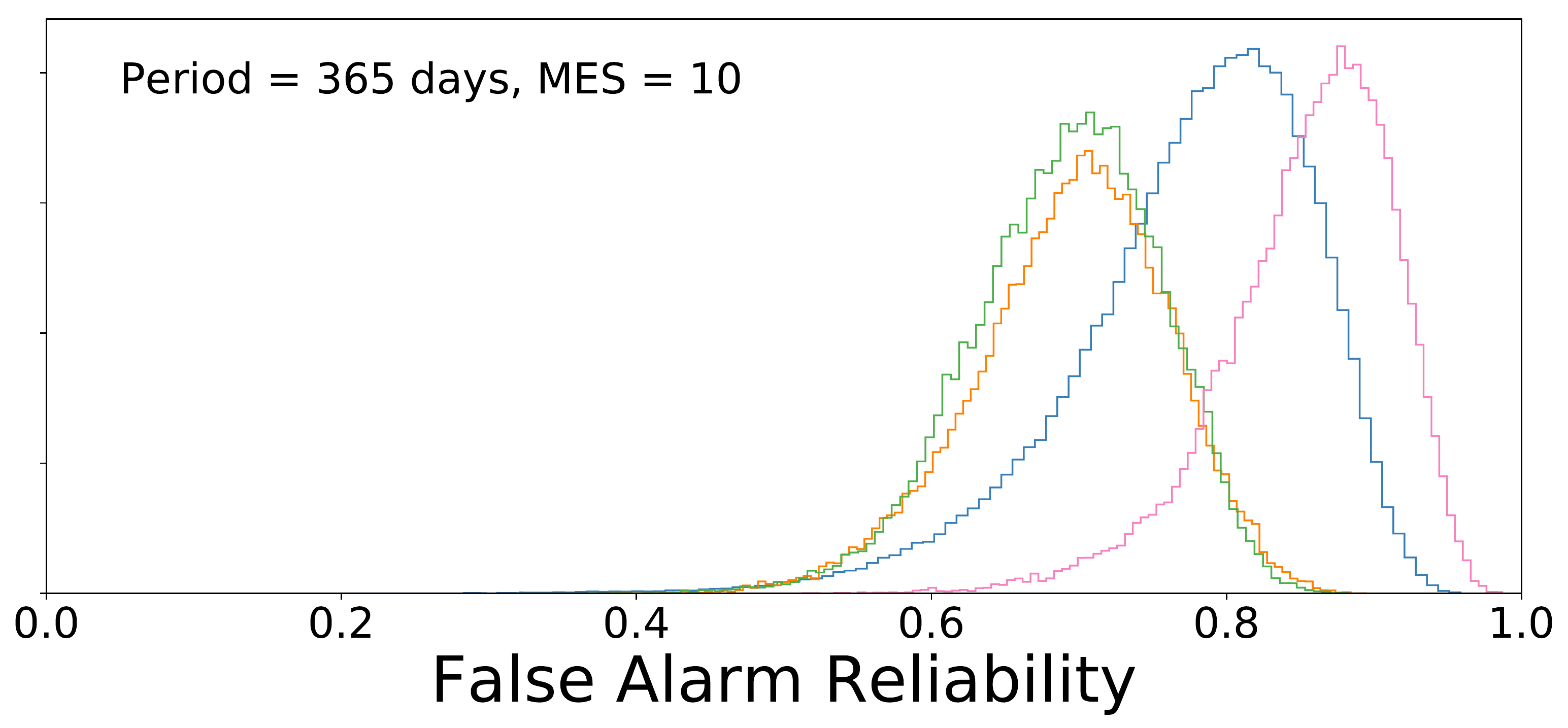} \\
  \includegraphics[width=\linewidth]{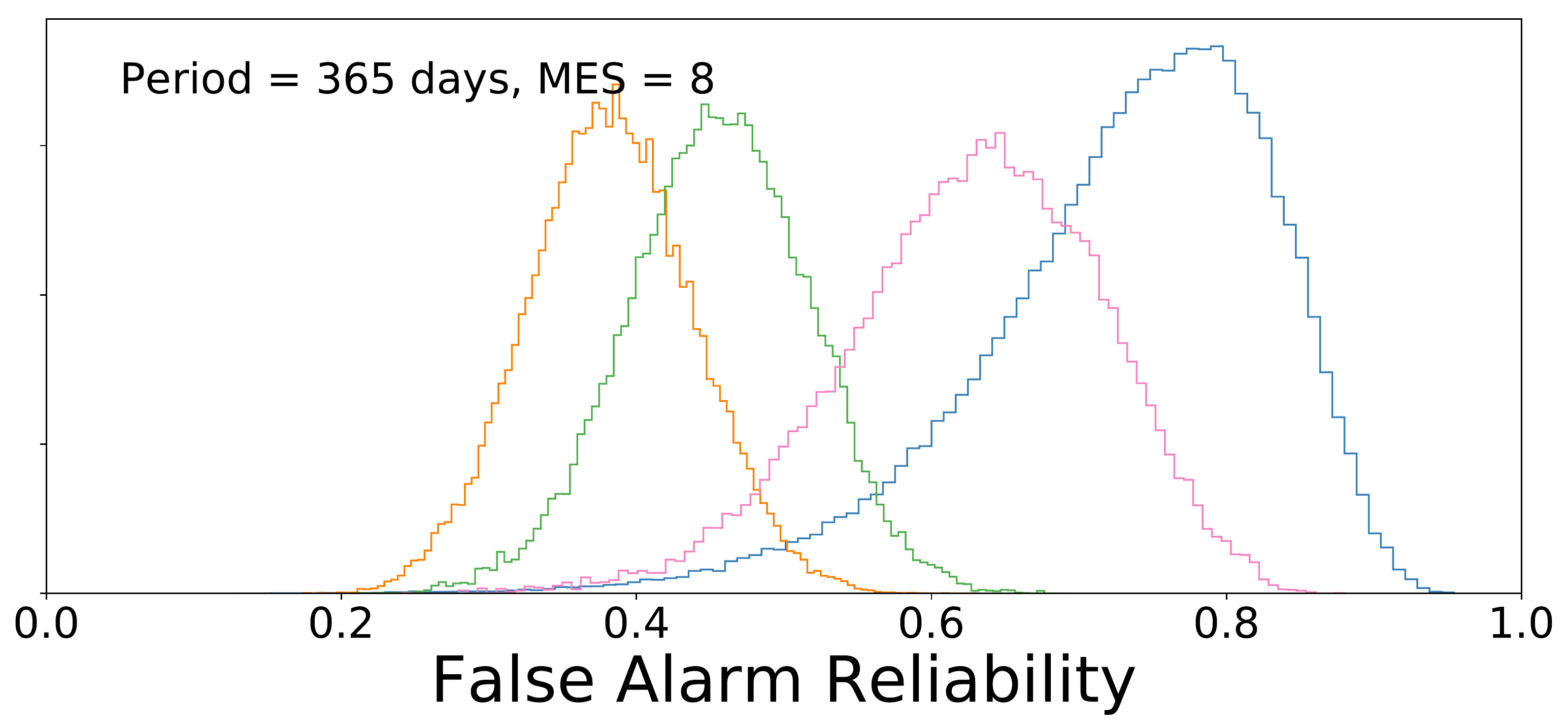} \\
   \caption{Reliability distributions for high reliability (blue), DR25 (pink), FPWG PC (green) and high completeness (orange) vetting, evaluated with the posterior distributions. Top: period = 365 days and MES = 10.  Bottom: period = 365 days and MES = 8. } \label{figure:reliabilityCompExamples}
\end{figure}

Figure~\ref{figure:reliabilityCompExamples} shows the resulting false alarm reliability distributions in two cases near the detection limit.  Again we see a large amount of variation in the reliability between the catalogs we consider.  As expected, the high-completeness catalog has lower false alarm reliability at a given period and MES.

\subsection{Occurrence Rate Methods} \label{section:occMethods}
We compute occurrence rates as the average of the number of planets per star $f(p, r)$ over a specified stellar population and range of planet period $p$ and radius $r$.  We do this in two major steps: 1) the determination of  $\frac{\mathrm{d}^2 f}  {\mathrm{d} p \, \mathrm{d} r}$, the differential rate of planets per star for the specified stellar population and catalog of planet candidates on those stars, and 2) the integration of that rate over the specified planet period and radius range.  Step 1, described in this section, is performed for a planet and period range $50 \leq$ period $\leq 400$ days and $0.75 \leq$ radius $\leq 2.5$ $R_{\oplus}$.  Using the results of step 1, step 2 computes several occurrence rates by integrating the differential rate over various ranges described in \S\ref{section:results}.  

To explore the dependence of our occurrence rates on the specific Bayesian inference method, we compute all occurrence rates for the four catalogs from \S\ref{section:thresholdVariation} using the Poisson-Likelihood MCMC method of \citet{Burke2015} and the Approximate Bayesian Computation method of \citet{Kunimoto2020a}. Both methods are modified to account for vetting completeness and reliability, with the Poisson-likelihood method described in \citet{Bryson2020} and the ABC method described in \citet{kunimoto2020b}.  Both methods use a standard power-law model for the planet population differential rate $\lambda$: for $\boldsymbol{\theta} = \left(F_0, \alpha, \beta \right)$,
\begin{equation} \label{eqn:powerLaw}
\begin{split}
    \lambda(p, r, \boldsymbol{\theta}) 
    &\equiv \frac{\mathrm{d}^2 f}  {\mathrm{d} p \, \mathrm{d} r} \\
    &=  F_0 \frac{(\alpha + 1) r^{\alpha}}{r_{\max}^{\alpha+1}-r_{\min}^{\alpha+1}} \frac{(\beta + 1) p^{\beta}}{p_{\max}^{\beta+1}-p_{\min}^{\beta+1}}
\end{split}
\end{equation}
where $f$ is the number of planets per star.  Given $\boldsymbol{\theta}$, we find $f$ for a particular period and radius range by integrating $\lambda$ over that range.

The normalization in Equation~(\ref{eqn:powerLaw}) is chosen so that the integral of $\lambda$ from $r_{\min}$ to $r_{\max}$ and from $p_{\min}$ to $p_{\max}$ $= F_0$, so $F_0$ is the number of planets per star in our domain of analysis.
The occurrence rates we present in this paper are the integral of $\lambda$ over various period and radius ranges.

Both inference methods use the same stellar and planet population, and the same characterization of completeness and reliability computed using the approach \citet{Bryson2020}. These steps are as follows:
\begin{itemize}
    \item Select a subset of the target star population, which will be our parent population of stars that are searched for planets.  We apply various cuts intended to select well-behaved and well-observed stars, and we restrict our analysis to GK dwarfs.
    \item Use the injected data to characterize vetting completeness.
    \item Compute the detection completeness, incorporating vetting completeness and geometrical probability for each star and sum over the stars, as described in \S\ref{section:completeness}.
    \item Use observed, inverted, and scrambled data to characterize false alarm reliability, as described in \S\ref{section:reliability}.
    \item Assemble the collection of planet candidates, including computing the reliability of each candidate from the false alarm reliability and false positive probability.
\end{itemize}
Because choice of score cut and catalog change both the vetting completeness and reliability of the PC population, all of steps except stellar parent sample selection are computed for each catalog and score cut.  
For the Poisson likelihood inference of the parameters in Equation (\ref{eqn:powerLaw}), reliability is implemented by running the MCMC computation many times, with the planets removed with a probability given by their reliability.  For details see \citet{Bryson2020}.  For the Poisson likelihood case, we compute infer the coefficients of Equation (\ref{eqn:powerLaw}) for four choices of score cut for all four populations.

The ABC-based inference of the parameters in Equation (\ref{eqn:powerLaw}) is computed using the approach of \citet{kunimoto2020b}. In summary, the underlying \Kepler\ planet population is forward modeled by drawing $N_{p} = F_{0}N_{s}$ planets according to Equation (\ref{eqn:powerLaw}), where $N_{s} = 57,015$ is the total number of stars in the sample. Following the procedure of \citet{Mulders2018}, this involves assigning each planet a period between 50 and 400 days from the cumulative distribution function of $p^{\beta}$, and a radius between 0.75 and 2.5 $R_{\oplus}$ from the cumulative distribution function of $r^{\alpha}$. The detectable planet sample is then simulated from the underlying population by drawing from a Bernoulli distribution with the star-averaged detection probability. We compare the detected planets to the observed PC population using a distance function, which quantifies agreement between the period distributions, radius distributions, and sample sizes of the catalogs. For the former two distances, we chose the two-sample Anderson-Darling (AD) statistic, which has been shown to be more powerful than the commonly used Kolmogorov-Smirnoff test \citep{Engmann2011}. With each iteration of the ABC algorithm, model parameters are accepted when the resulting population's distance from the observed population is less than 75th quantile of the previous iteration's accepted distances. Following the guidance of \citet{prangle2017}, we confirmed that our algorithm converged by observing that the distances between simulated and observed catalogues approached zero with each iteration. These simulations are repeated within a Population Monte Carlo ABC algorithm to infer the parameters that give the closest match between simulated and observed catalogs. We perform the ABC inference on the four catalogs, without a score cut.

This forward model is appropriate for estimating the average number of planets per star in a given period and radius range, as is achieved by the Poisson likelihood function method. However, forward modeling has the advantage of being more versatile, especially in the face of increasingly complicated population models. Notably, \citet{Mulders2018} first used forward modeling to explore planetary architectures by taking into account correlations in planet properties, while \citet{He2019} used ABC to describe exoplanet periods and sizes using a clustered point process model. Another advantage is the ease with which ABC can implement reliability. Rather than requiring many inferences on different catalogues, we modify the distance function to accept weighted data.

\section{Results} \label{section:results}

We test our occurrence rate methods described in \S\ref{section:occMethods} on the four catalogs described in \S\ref{section:thresholdVariation}, using both the Poisson likelihood and ABC inference methods.  For the Poisson likelihood method, we study the impact of score cut.  As explained in \S\ref{section:thresholdVariation}, we treat these catalogs as equally valid, alternative measurements of the exoplanet population by \Kepler, and will take the consistency of results using these catalogs as a diagnostic of the quality of the occurrence rate method.

For all cases, we present both the population model parameters $\boldsymbol{\theta}$ for the differential occurrence rate $\lambda(\mathrm{period}, \mathrm{radius}, \boldsymbol{\theta})$ given by Equation~(\ref{eqn:powerLaw}), as well as several occurrence rates obtained by integrating $\lambda(\mathrm{period}, \mathrm{radius}, \boldsymbol{\theta})$ over various planet period and radius ranges.  The occurrence rates we compute are 
\begin{itemize}
    \item {\bf The log-differential rate of planets per star evaluated at Earth's period and radius} $\Gamma_{\oplus} \equiv \mathrm{d}^2 f / \mathrm{d} \log p \, \mathrm{d} \log r = p_{\oplus} r_{\oplus} \lambda \left(p_{\oplus}, r_{\oplus}, \boldsymbol{\theta} \right)$.
    \item {\bf $F_0$, the number of planets per star in our domain of analysis,} given by the integrating $\lambda$ over $50 \leq \mathrm{period} \leq 400$ days and $0.75 \leq \mathrm{radius} \leq 2.5$ $R_{\oplus}$.
    \item {\bf $F_1$, defined in \citet{Burke2015}}, given by integrating $\lambda$ over $50 \leq \mathrm{period} \leq 200$ days and $1 \leq \mathrm{radius} \leq 2$ $R_{\oplus}$.
    \item {\bf The number of planets per star with period and radius within $20\%$ of Earth, $\zeta_{\oplus}$ \citep{Burke2015}}, given by integrating $\lambda$ over $0.8p_{\oplus} \leq \mathrm{period} \leq 1.2 p_{\oplus}$ days and $0.8 r_{\oplus} \leq \mathrm{radius} \leq 1.2 r_{\oplus}$.
    \item {\bf The SAG13\footnote{\label{footnote:sag13}\url{https://exoplanets.nasa.gov/exep/exopag/sag/\#sag13}} definition of $\eta_{\oplus}$}  as the integral of $\lambda(\mathrm{period}, \mathrm{radius}, \boldsymbol{\theta})$ over $237 \leq \mathrm{period} \leq 860$ days and $0.5 \leq \mathrm{radius} \leq 1.5$ $R_{\oplus}$.
\end{itemize}
All these occurrence rates probe the \Kepler\ detection limit, with $F_1$ being furthest from the detection limit.  $F_0$, $F_1$, and part of $\zeta_{\oplus}$ are shown in Figure~\ref{figure:planetPopulation}.

\begin{figure*}[ht]
  \centering
  \Large No Reliability Correction \hspace{1 in} Corrected for Reliability\\
  \includegraphics[width=0.48\linewidth]{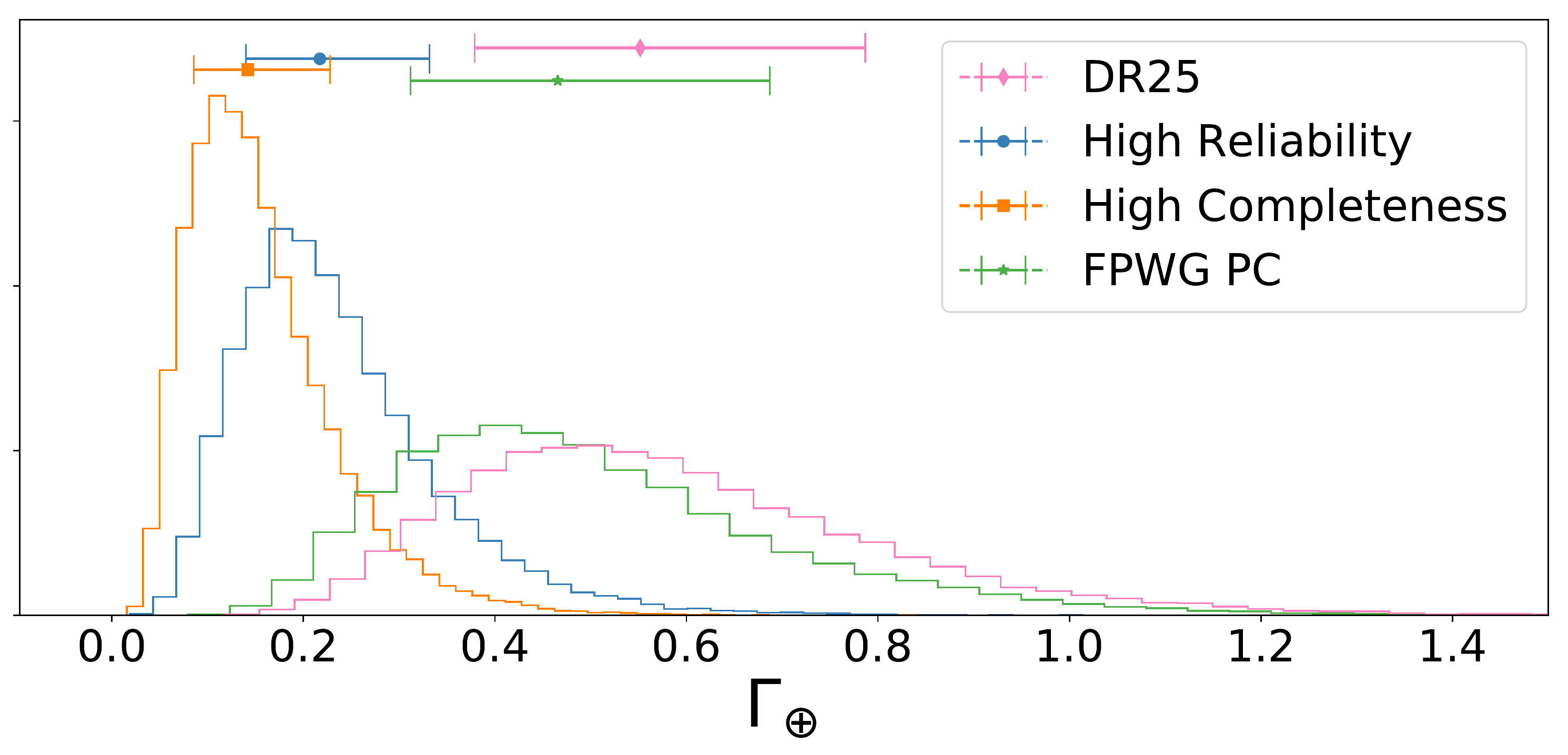} 
  \includegraphics[width=0.48\linewidth]{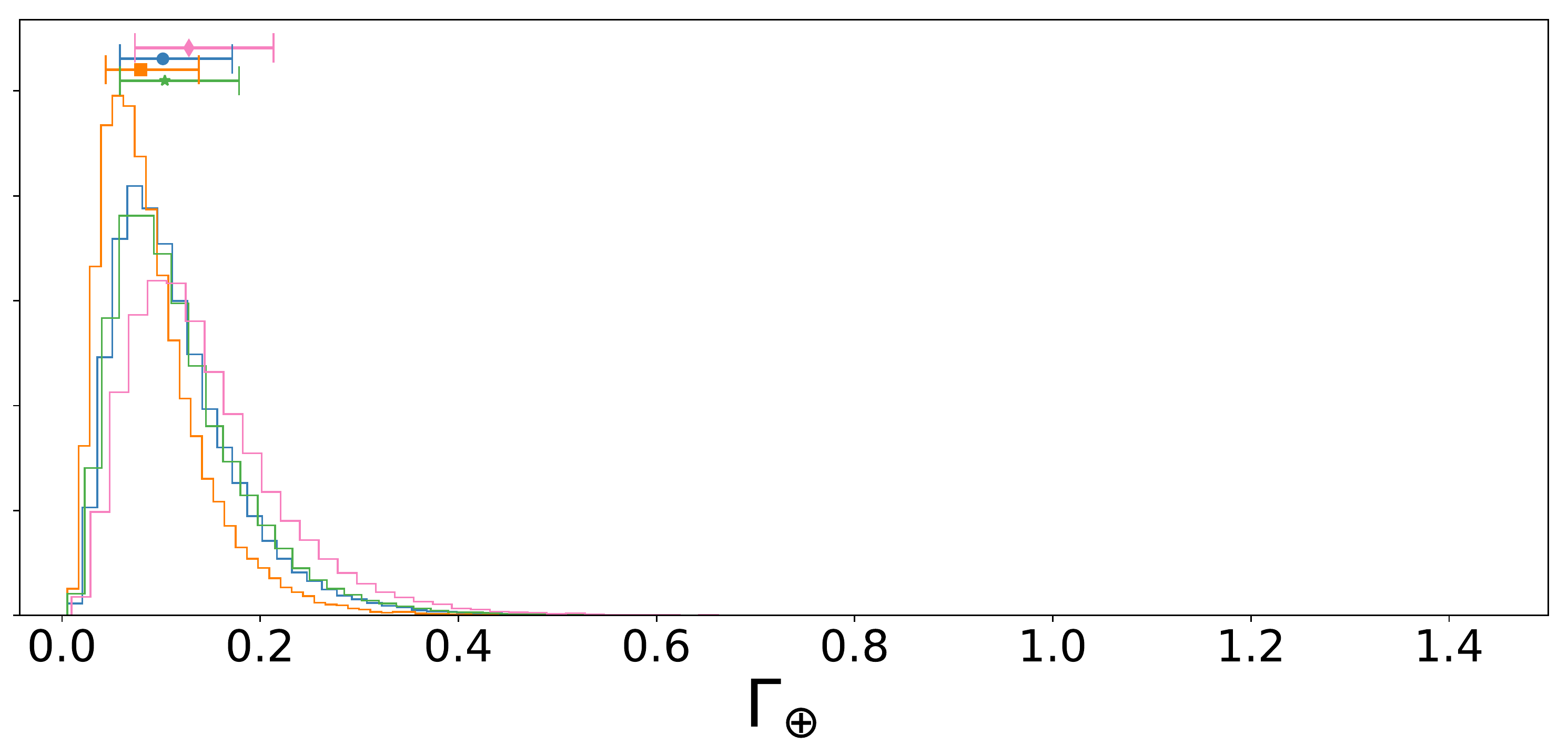} \\
  \includegraphics[width=0.48\linewidth]{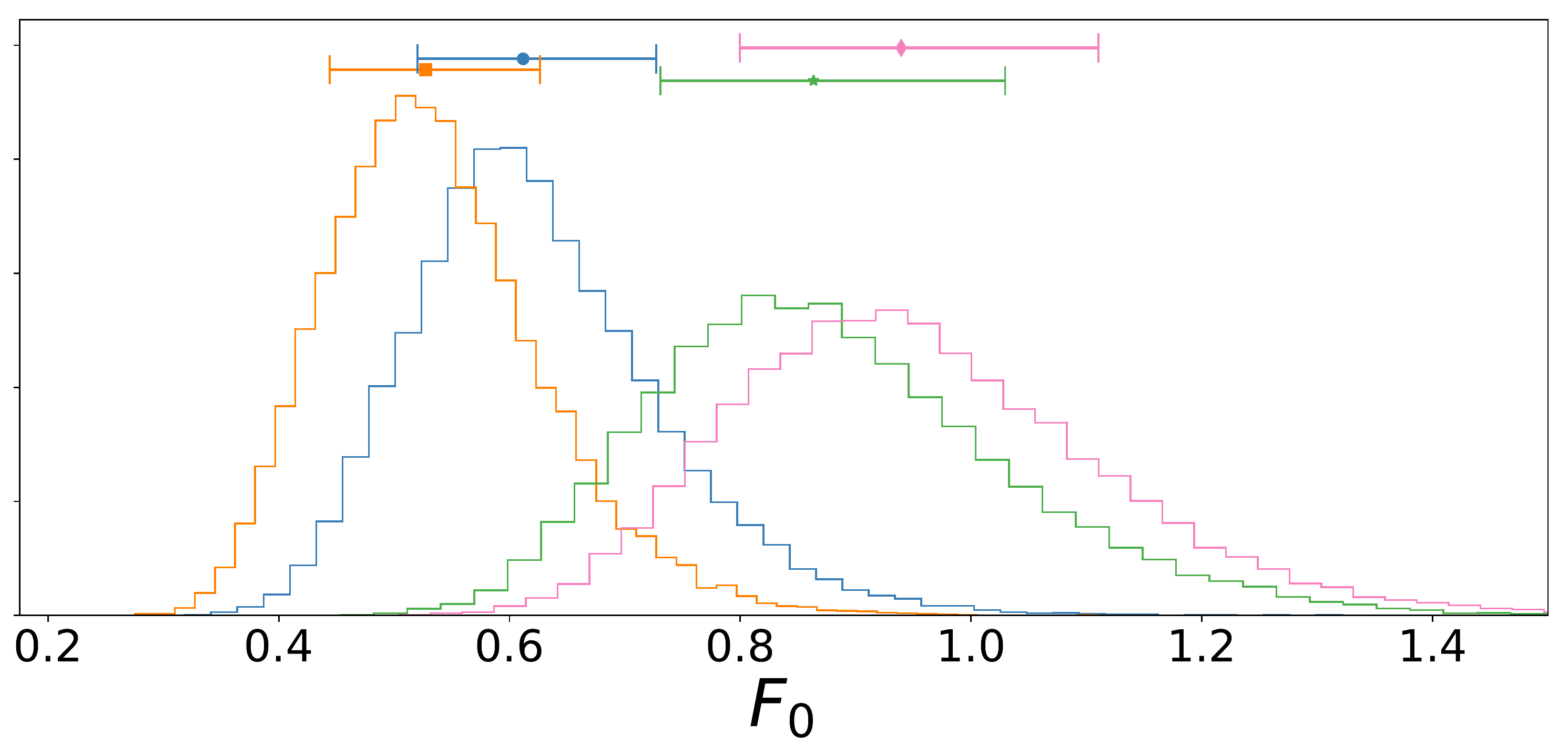}
  \includegraphics[width=0.48\linewidth]{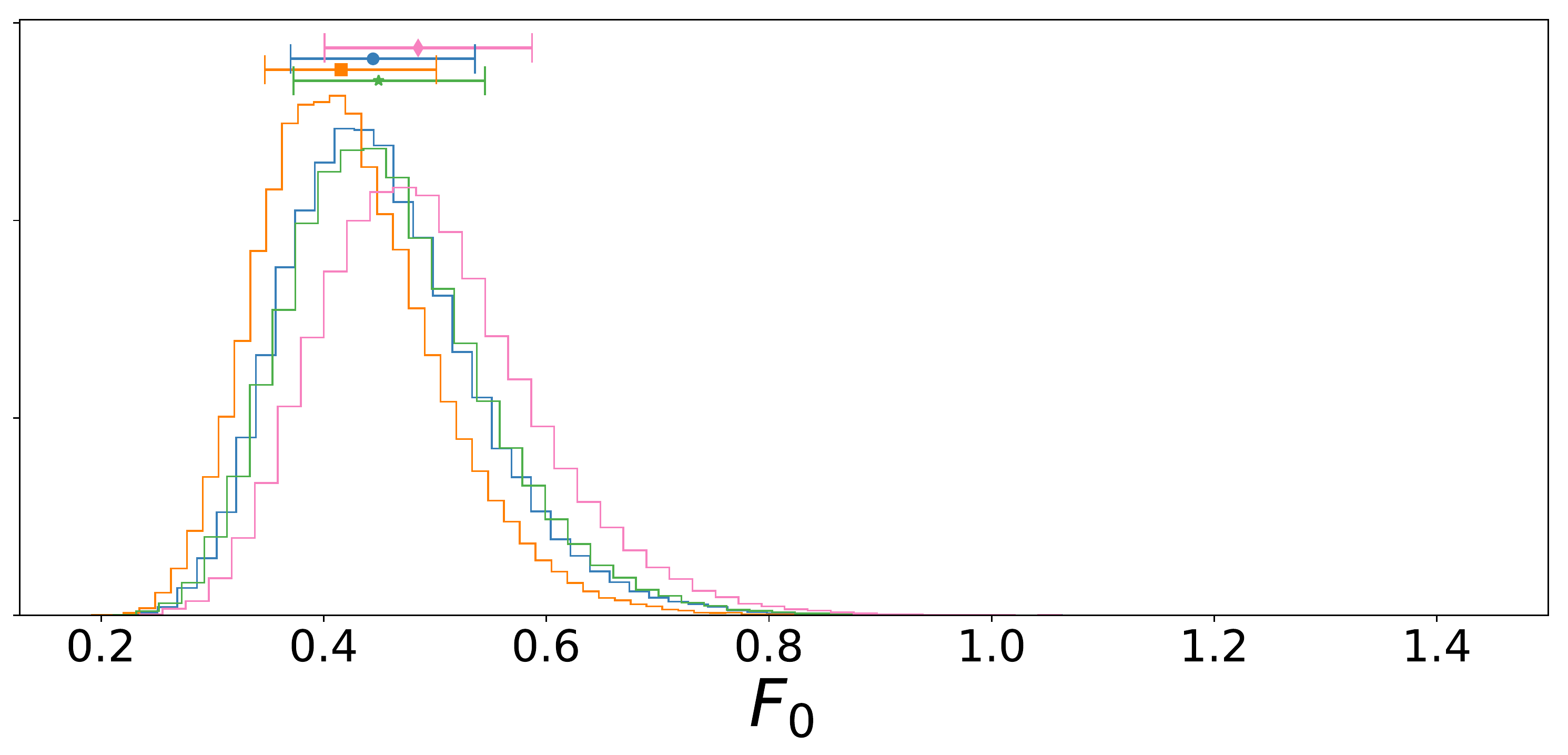} \\
  \includegraphics[width=0.48\linewidth]{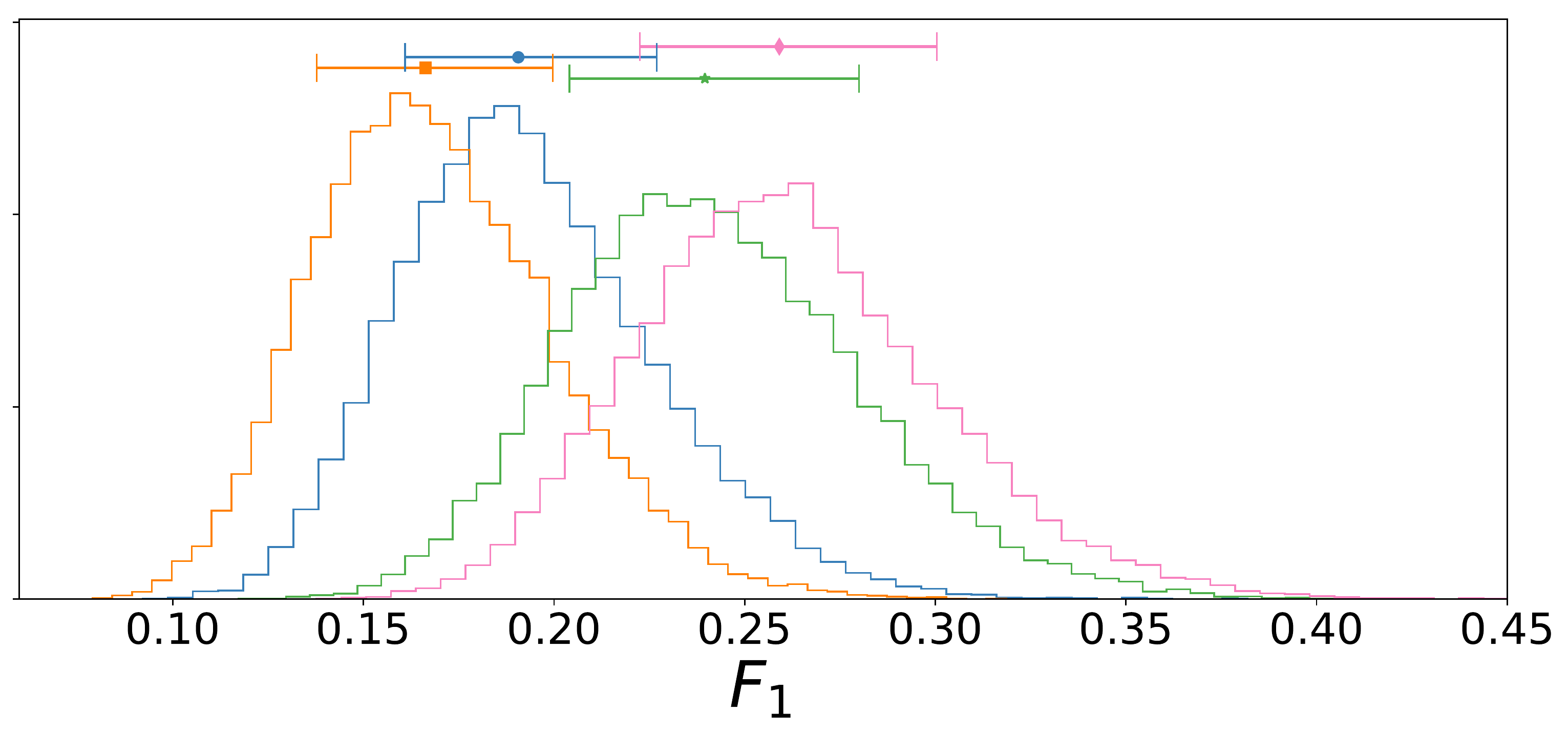}
  \includegraphics[width=0.48\linewidth]{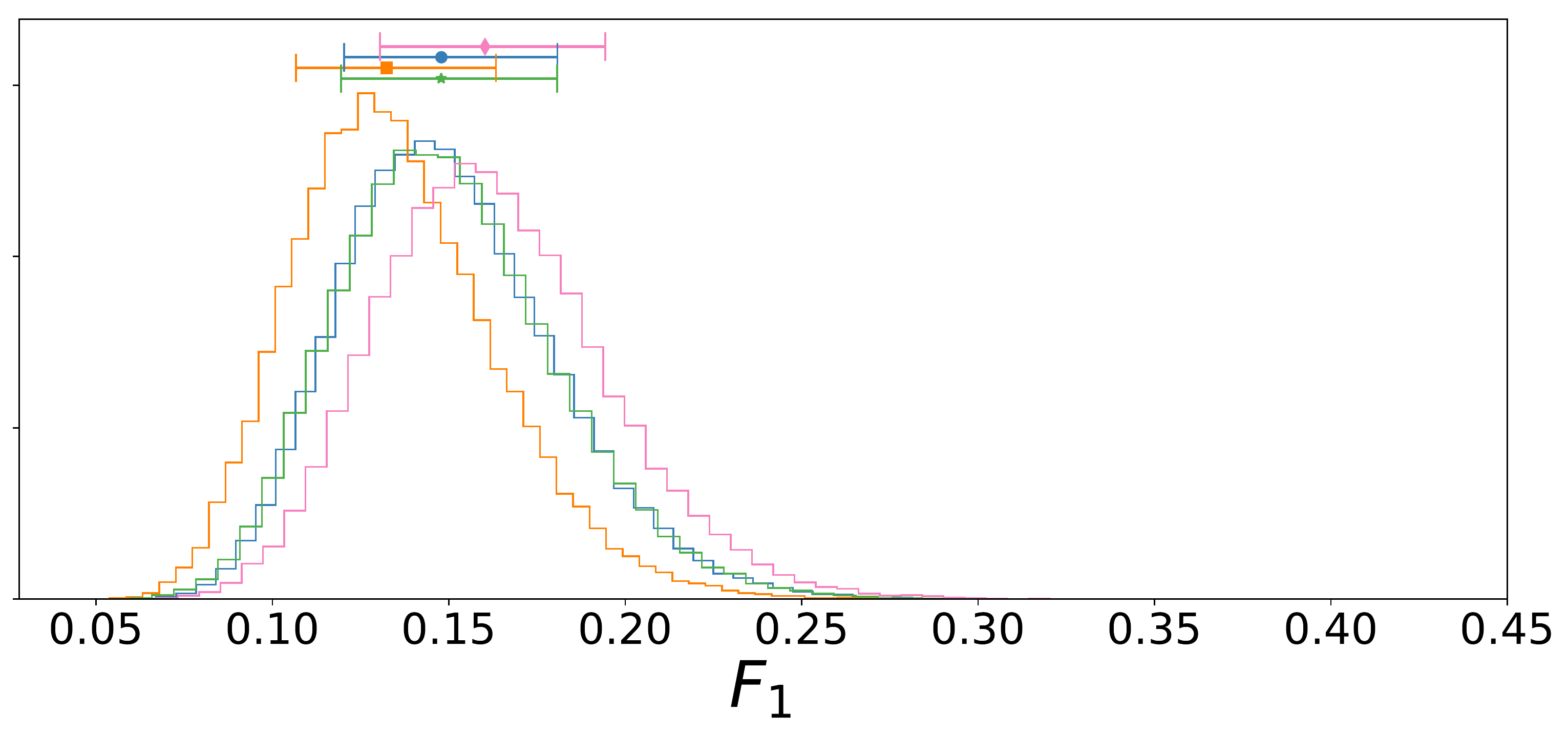} \\
  \includegraphics[width=0.48\linewidth]{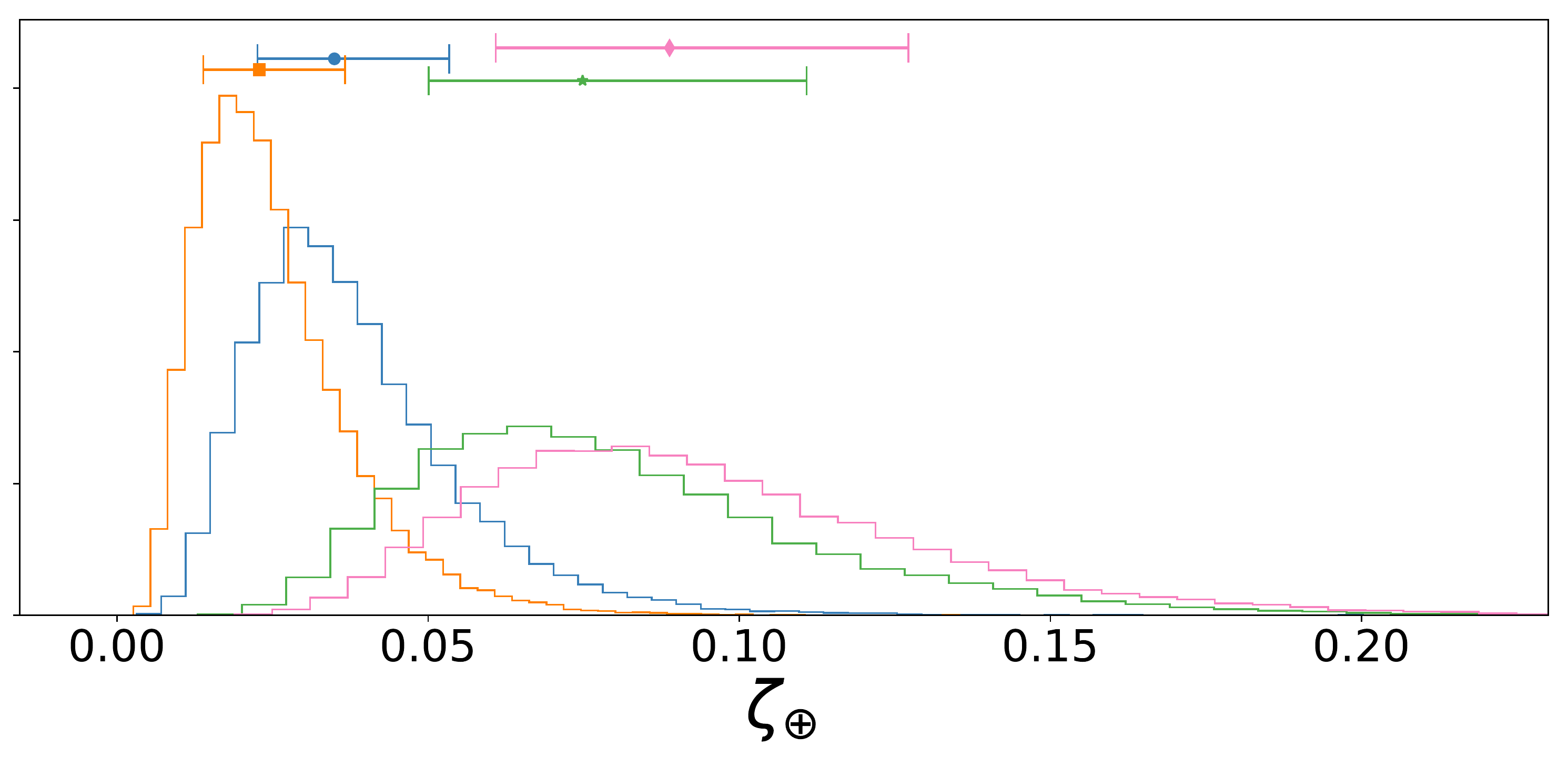}
  \includegraphics[width=0.48\linewidth]{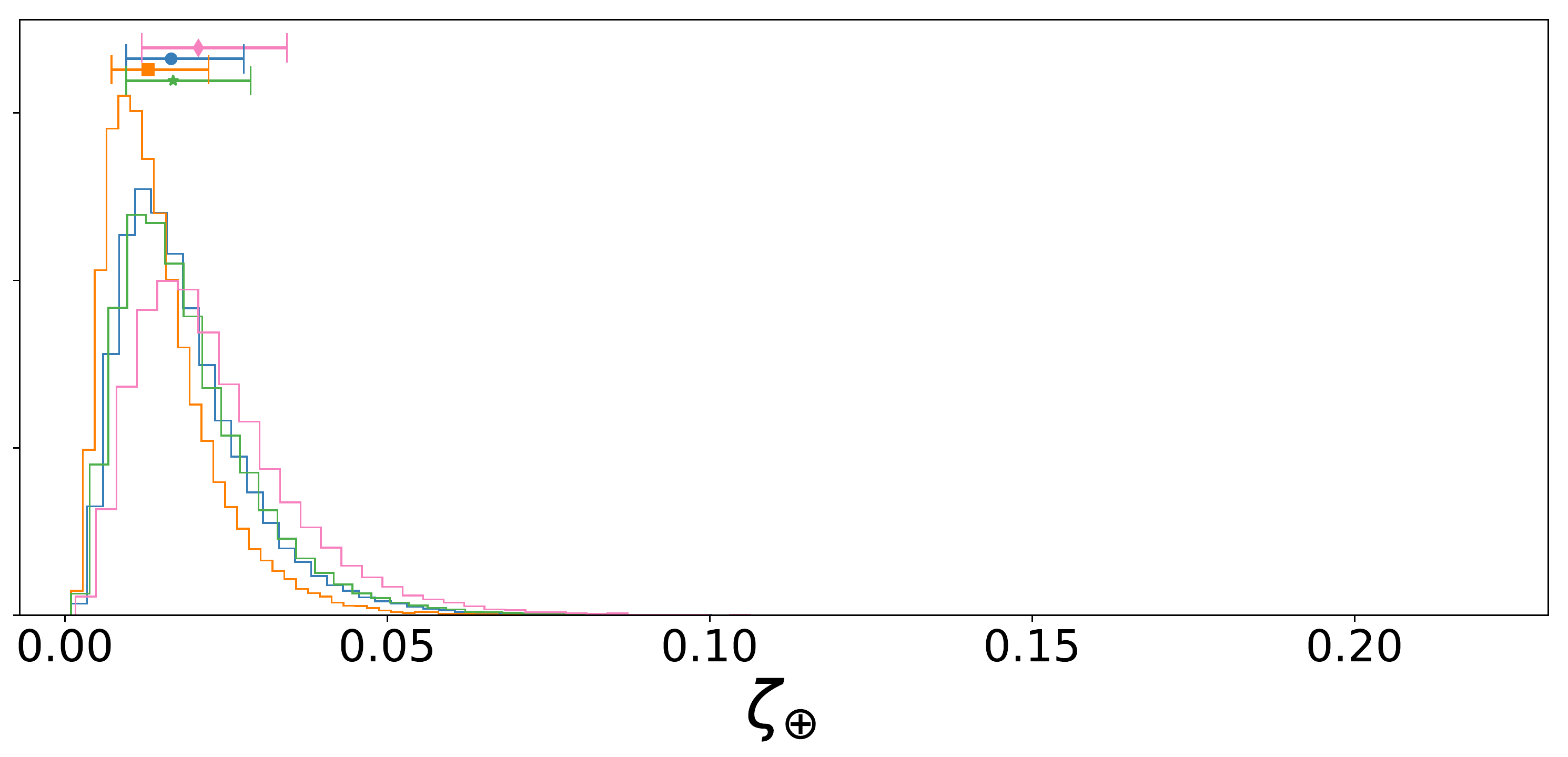} \\
  \includegraphics[width=0.48\linewidth]{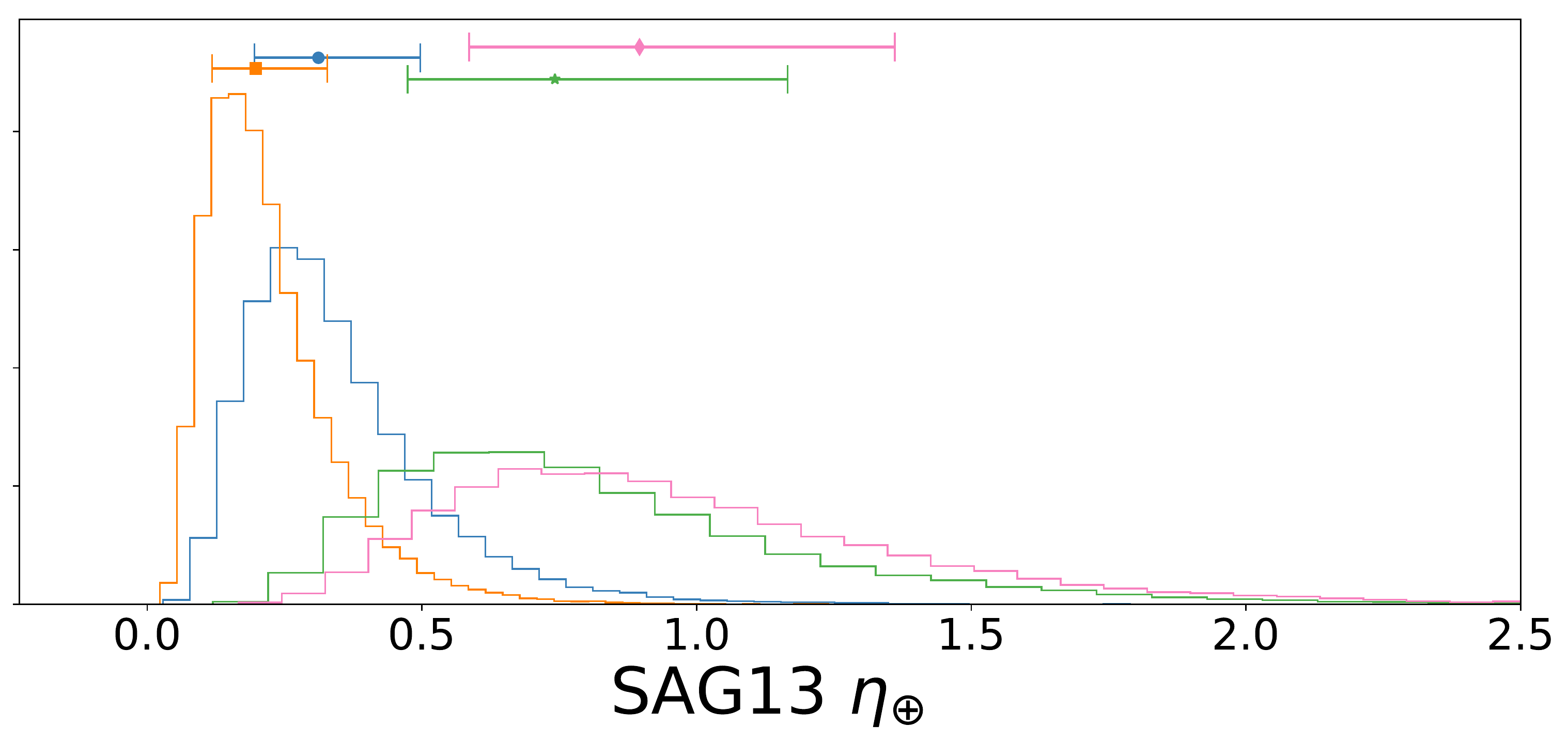}
  \includegraphics[width=0.48\linewidth]{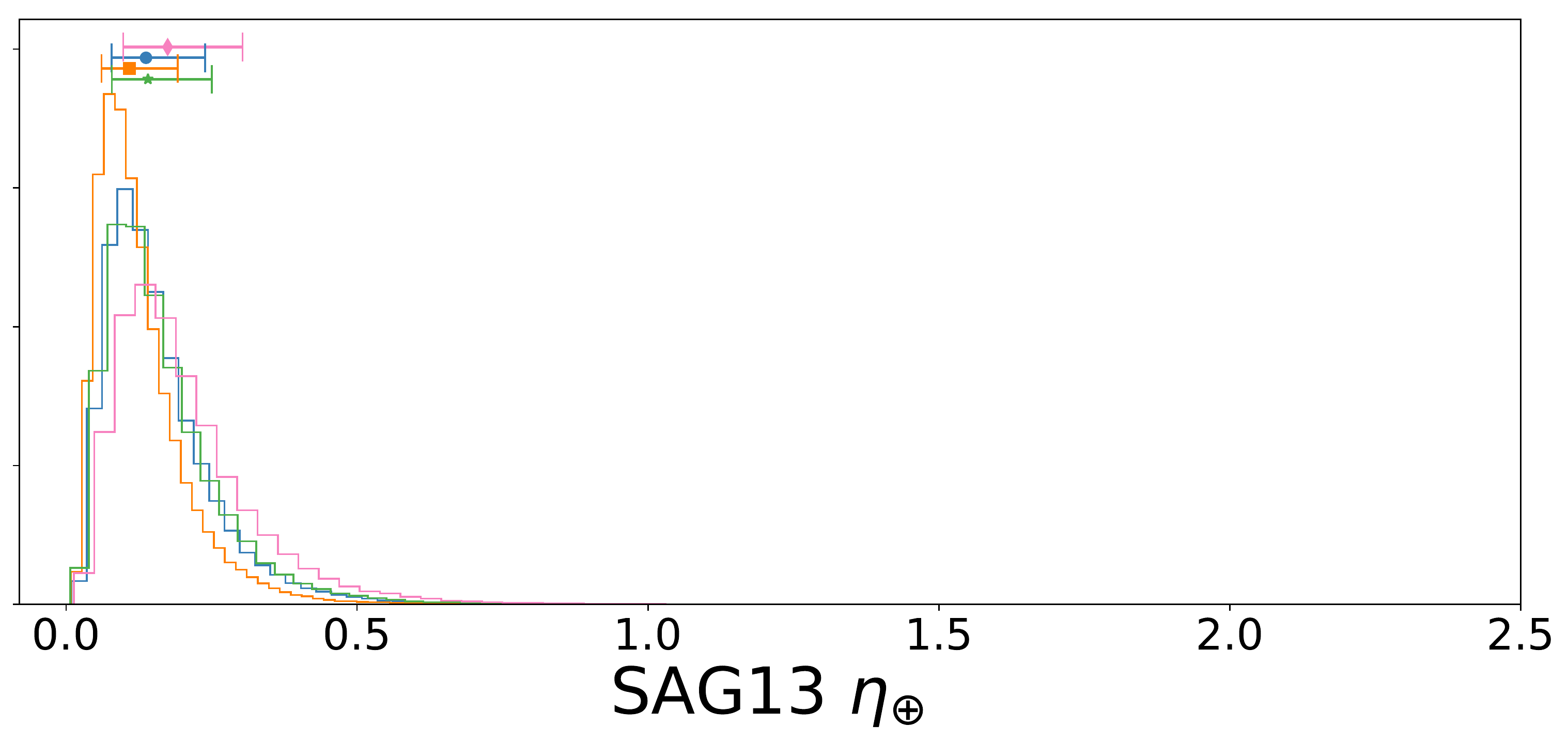}
   \caption{Distributions of various occurrence rates for the  high reliability (blue), DR25 (pink), FPWG PC (green) and high completeness (orange) vetting, computed with the Poisson method.  Left: without correcting for reliability.  Right: corrected for reliability.} \label{figure:allDist}
\end{figure*}

\begin{figure*}[ht]
  \centering
  \Large No Reliability Correction \hspace{1 in} Corrected for Reliability\\
  \includegraphics[width=0.48\linewidth]{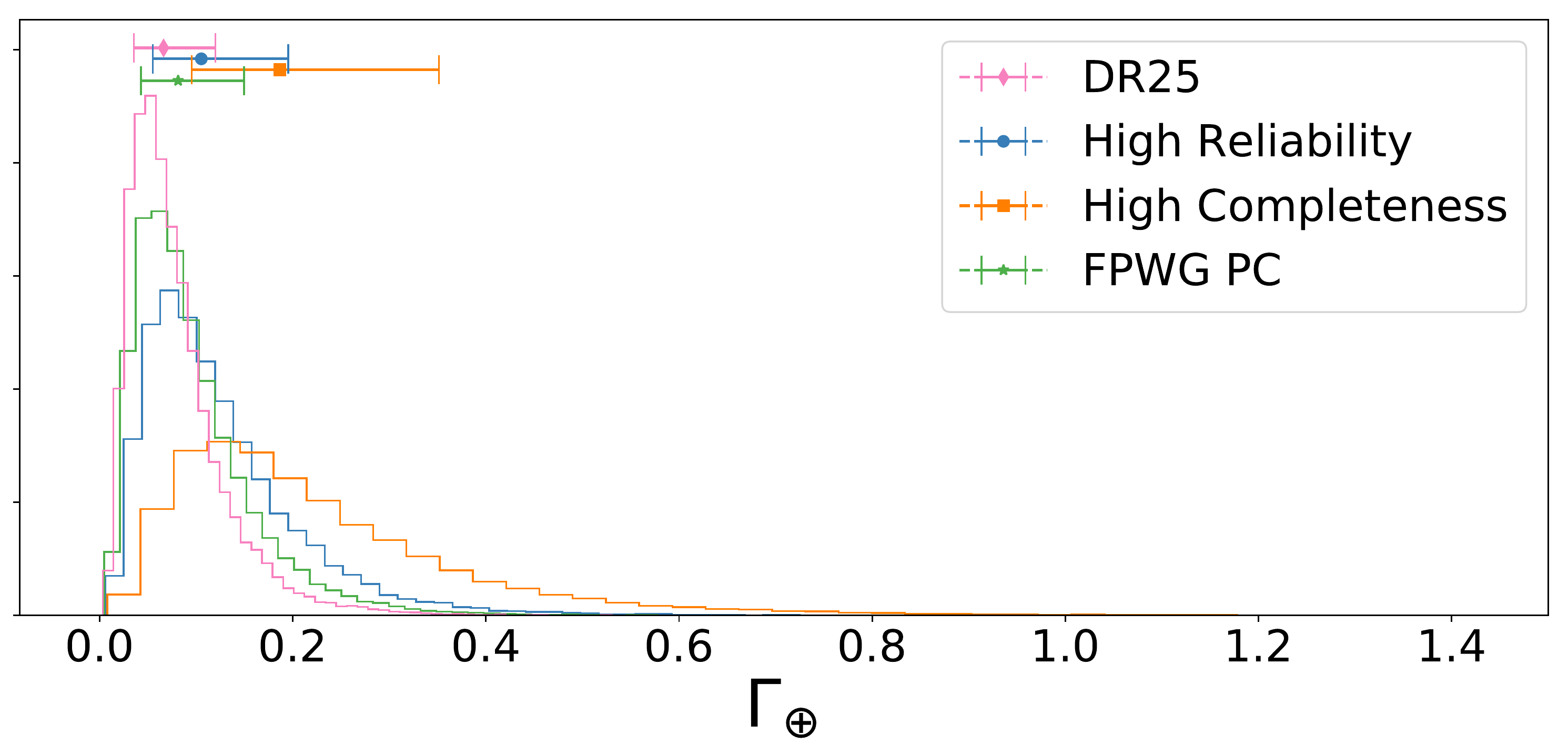} 
  \includegraphics[width=0.48\linewidth]{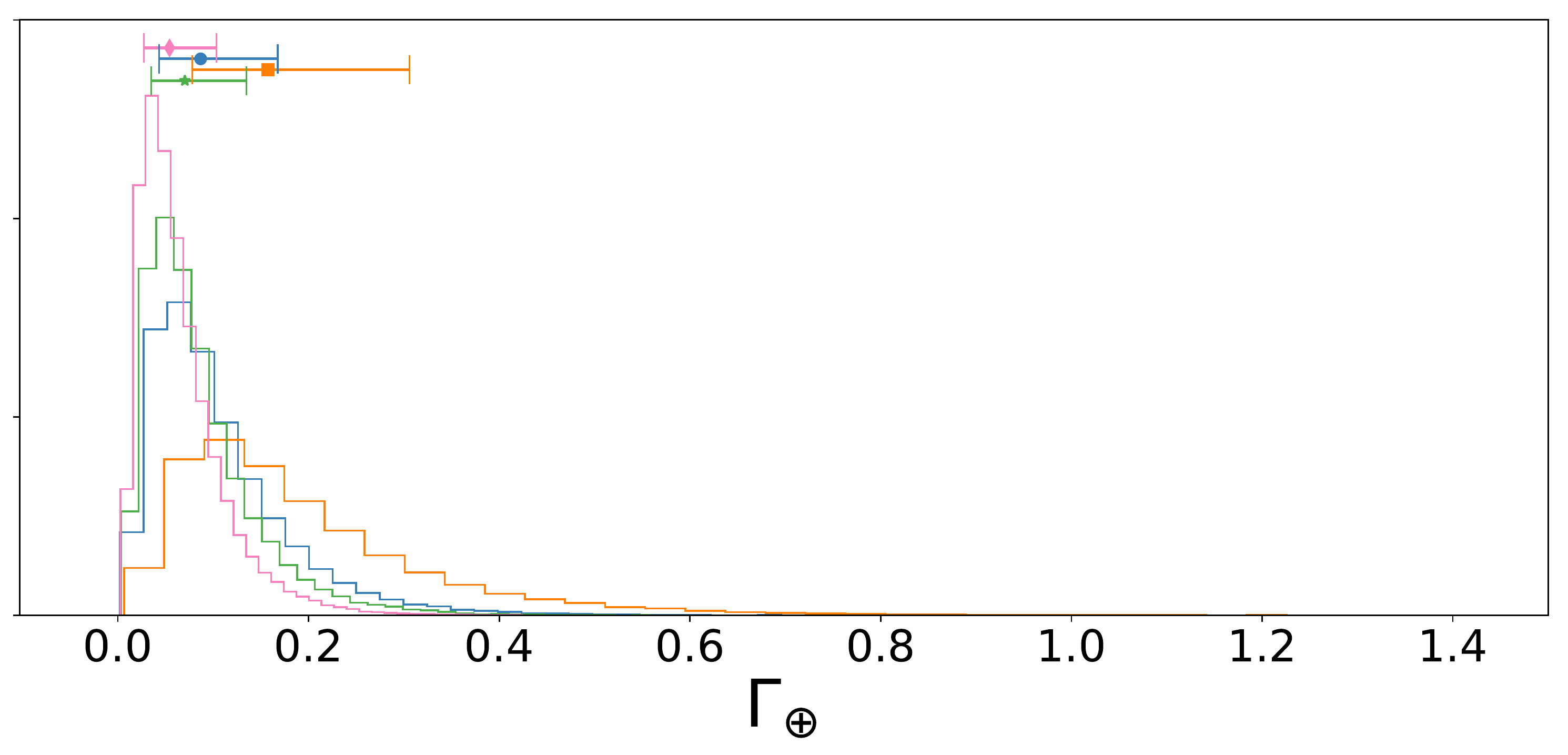} \\
  \includegraphics[width=0.48\linewidth]{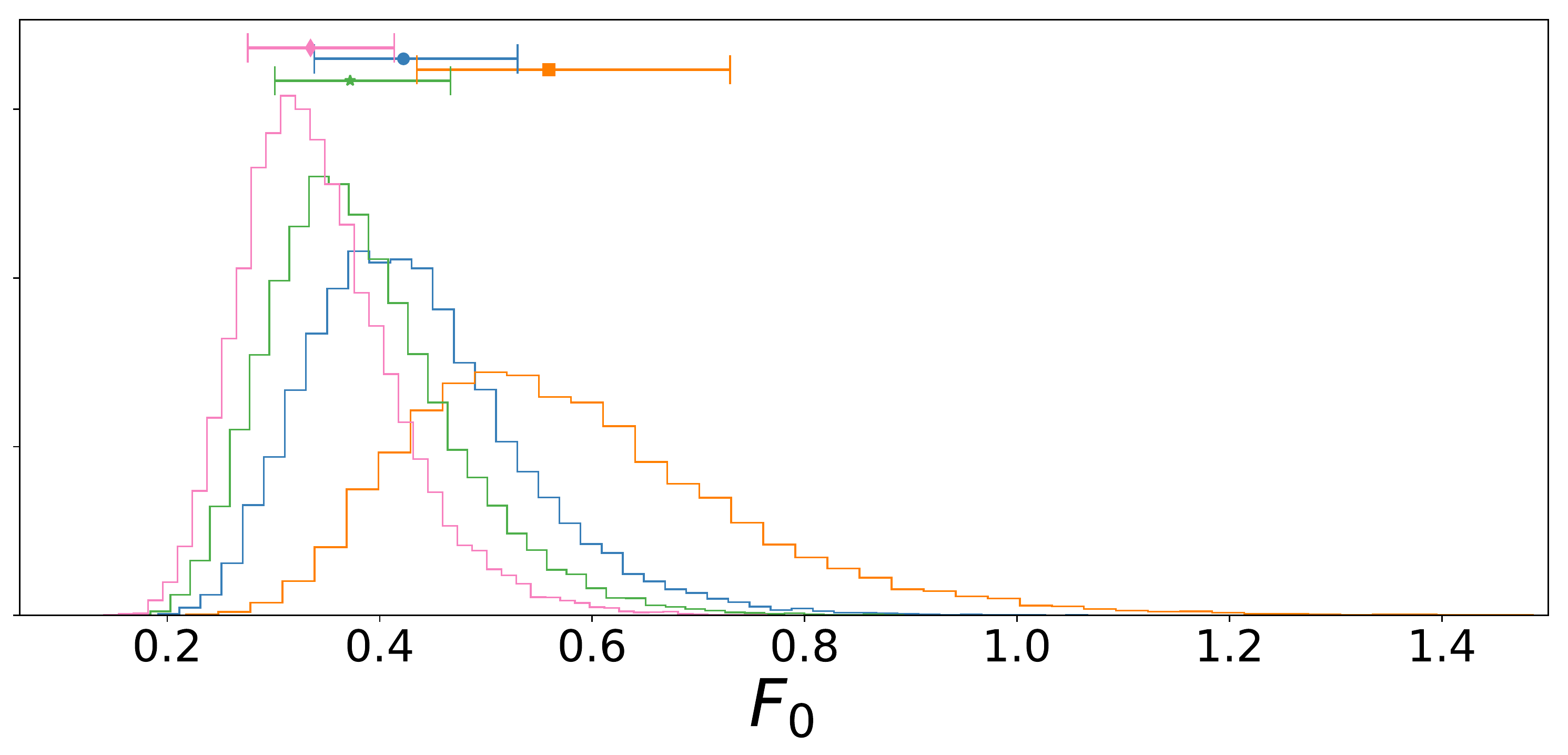}
  \includegraphics[width=0.48\linewidth]{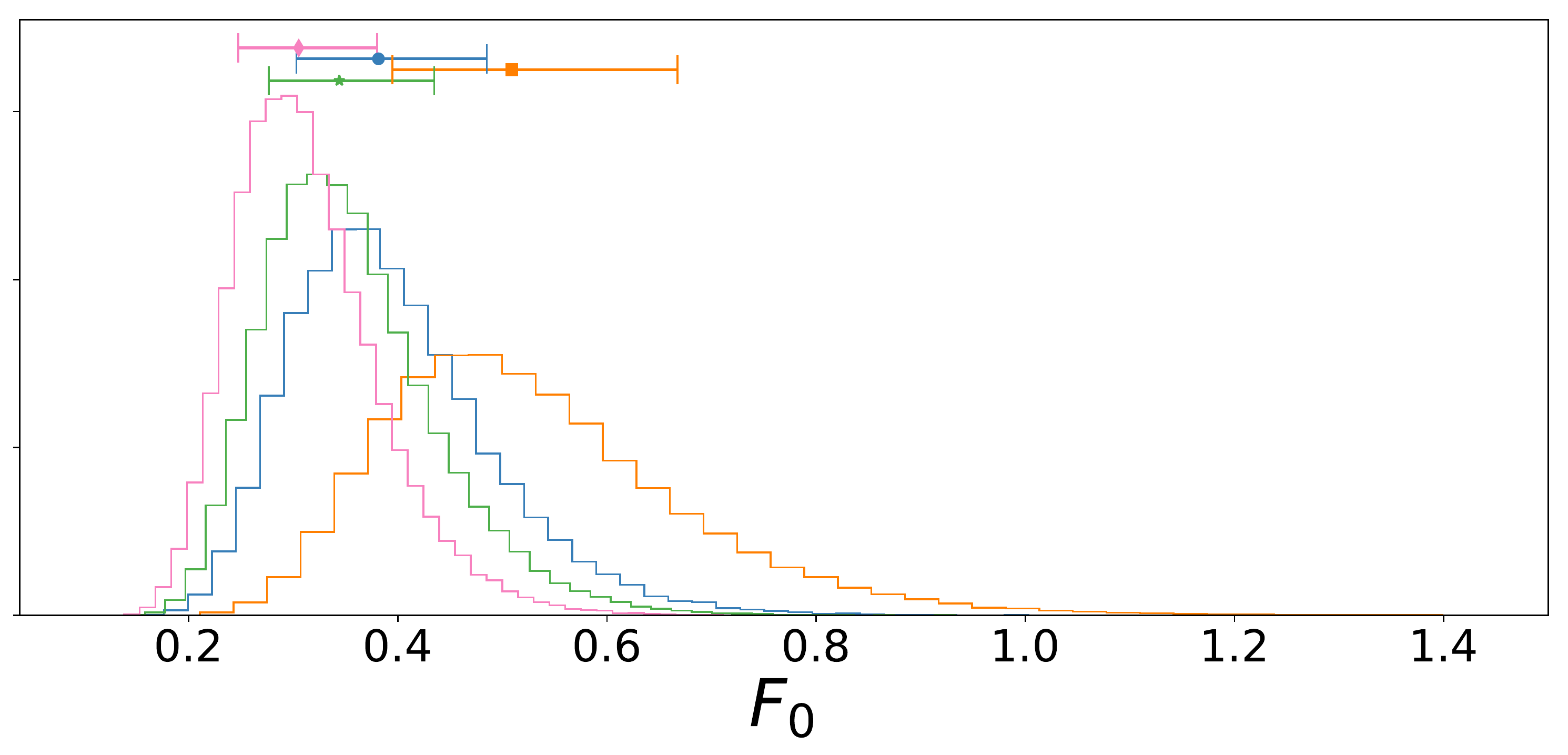} \\
  \includegraphics[width=0.48\linewidth]{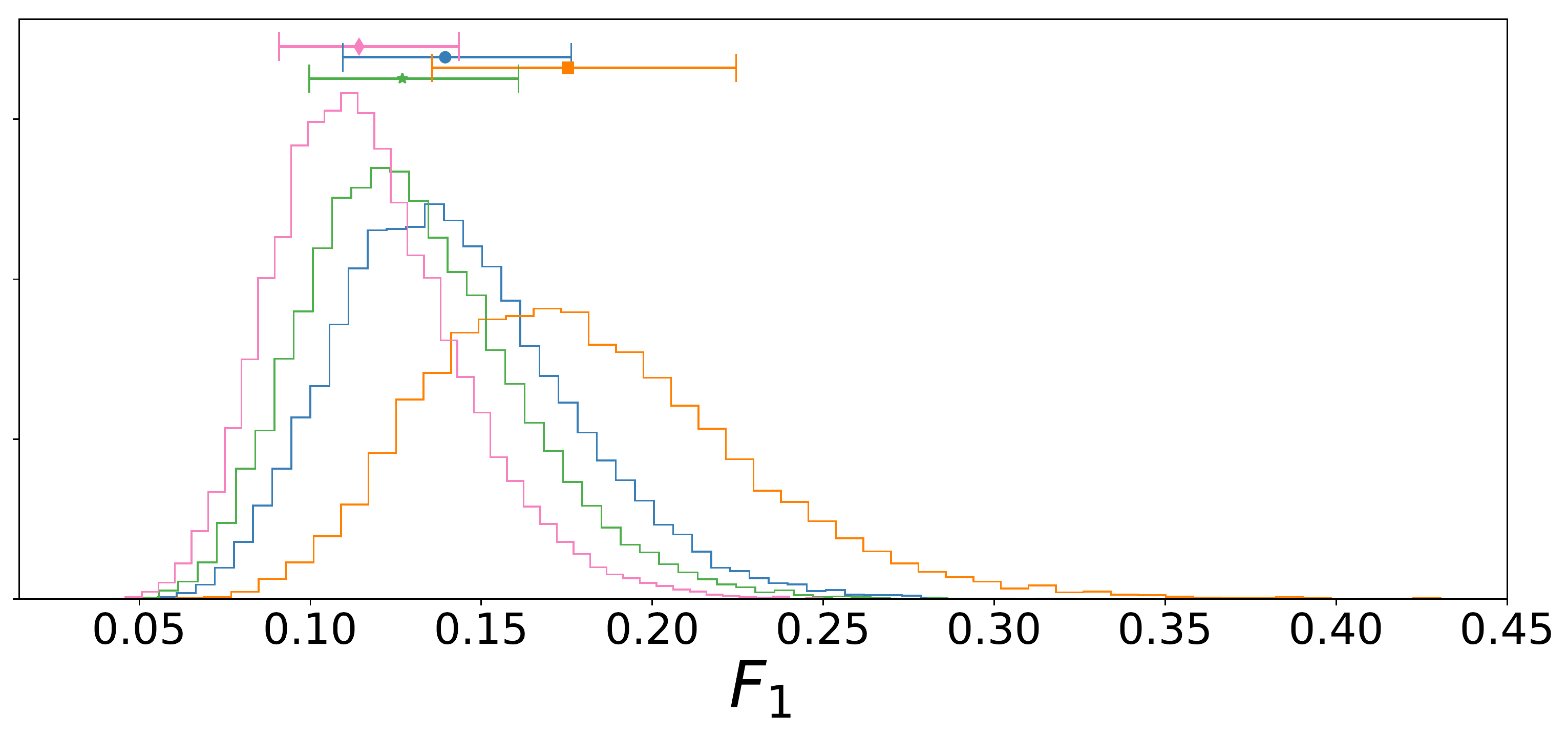}
  \includegraphics[width=0.48\linewidth]{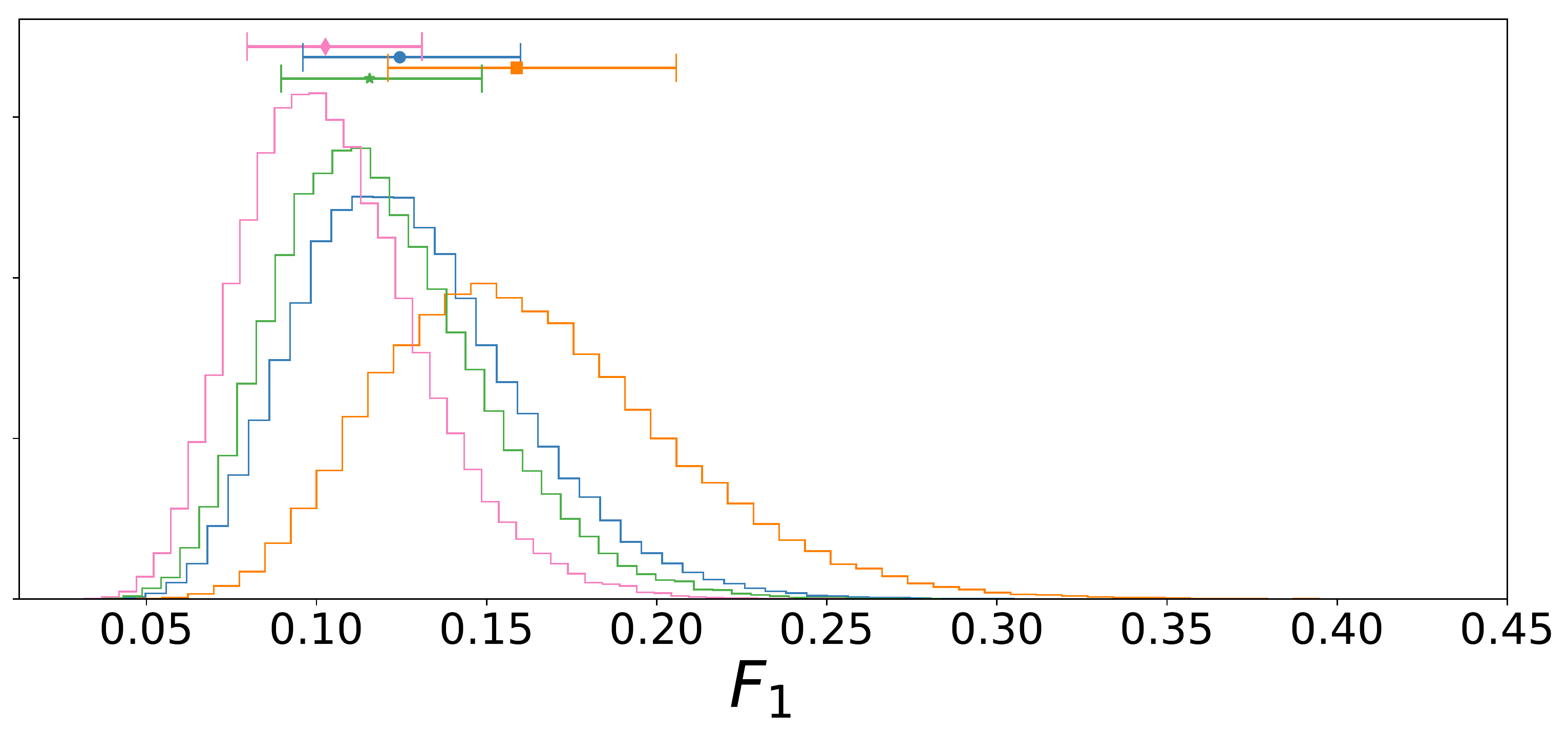} \\
  \includegraphics[width=0.48\linewidth]{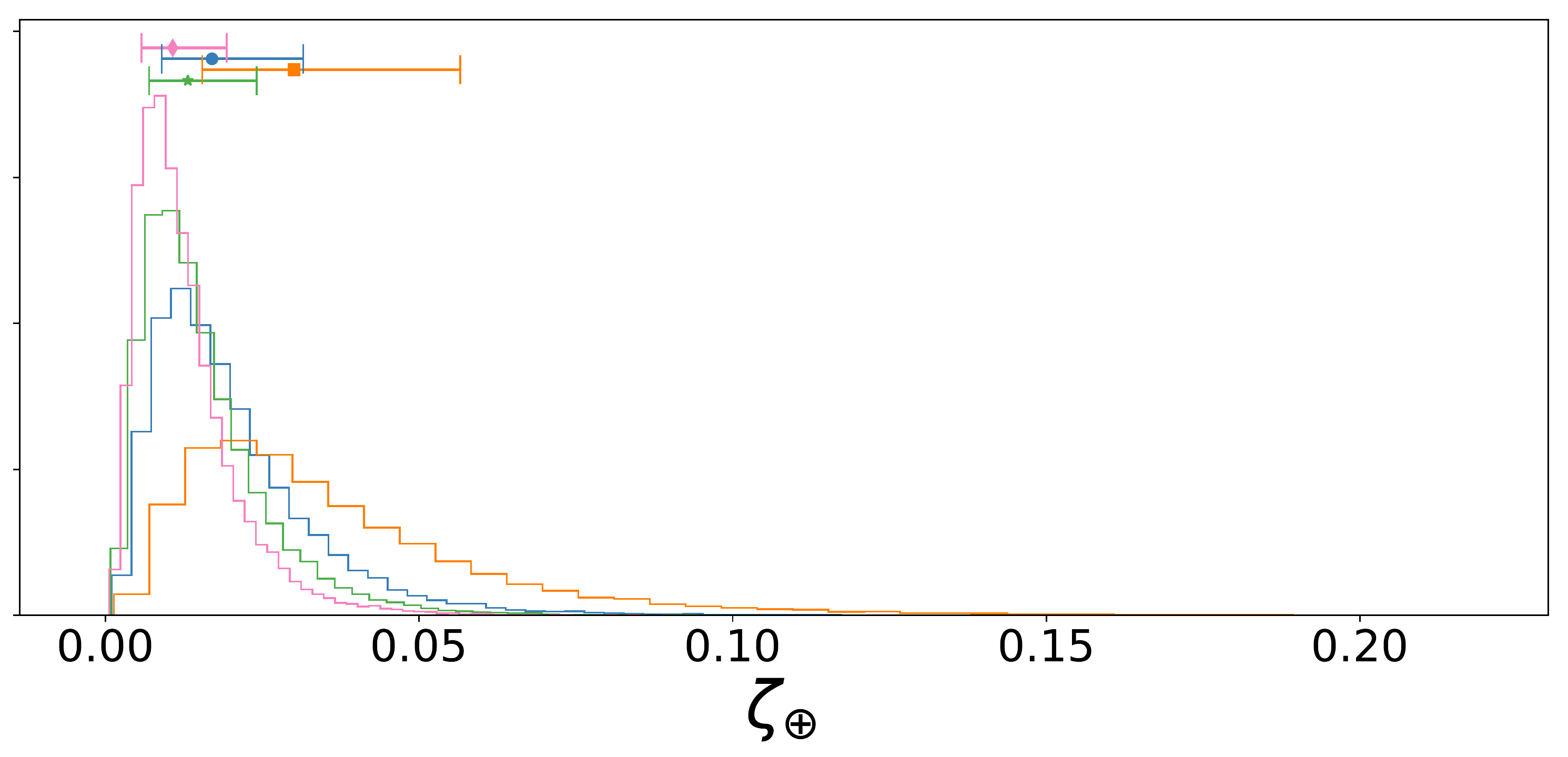}
  \includegraphics[width=0.48\linewidth]{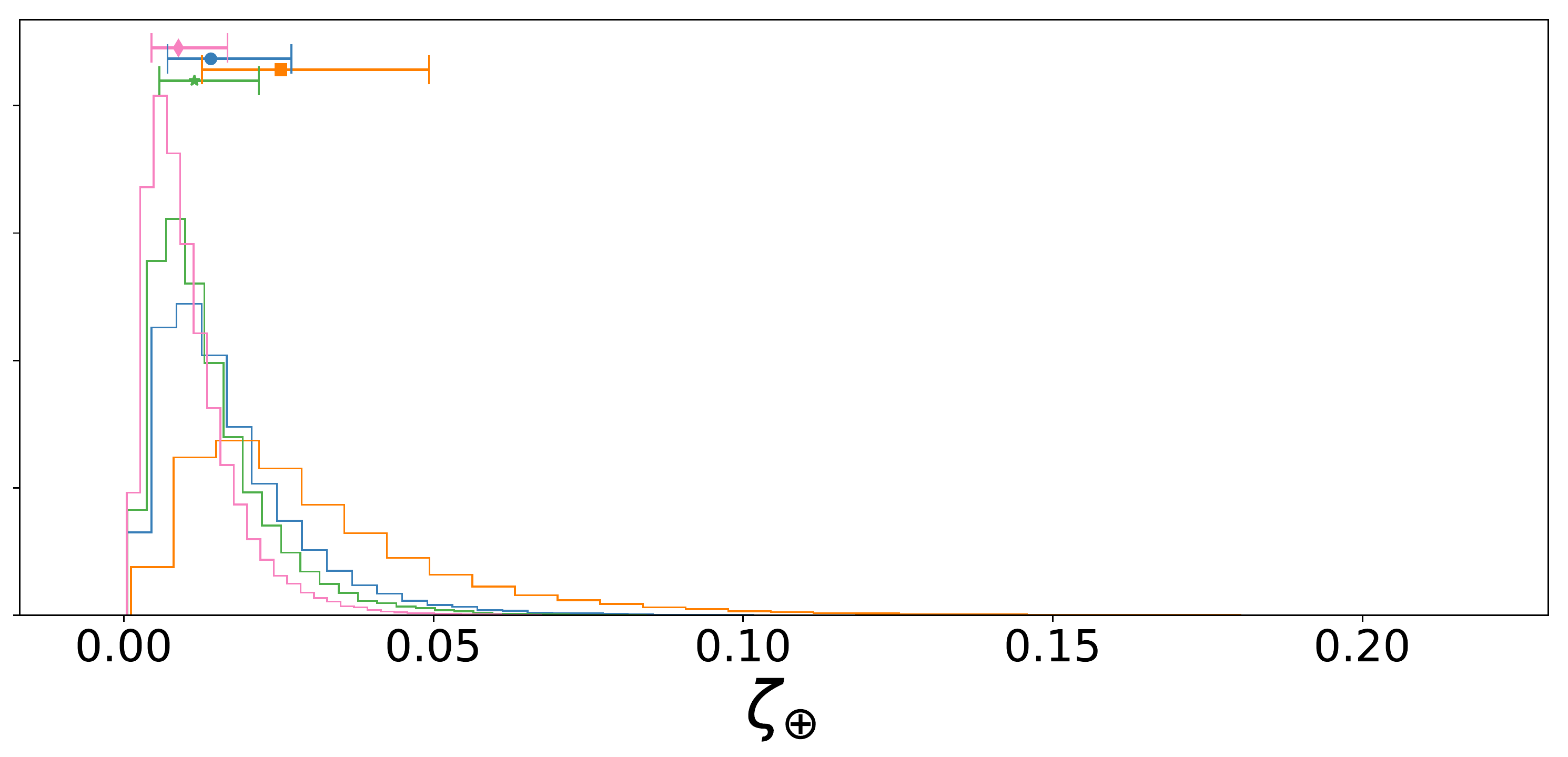} \\
  \includegraphics[width=0.48\linewidth]{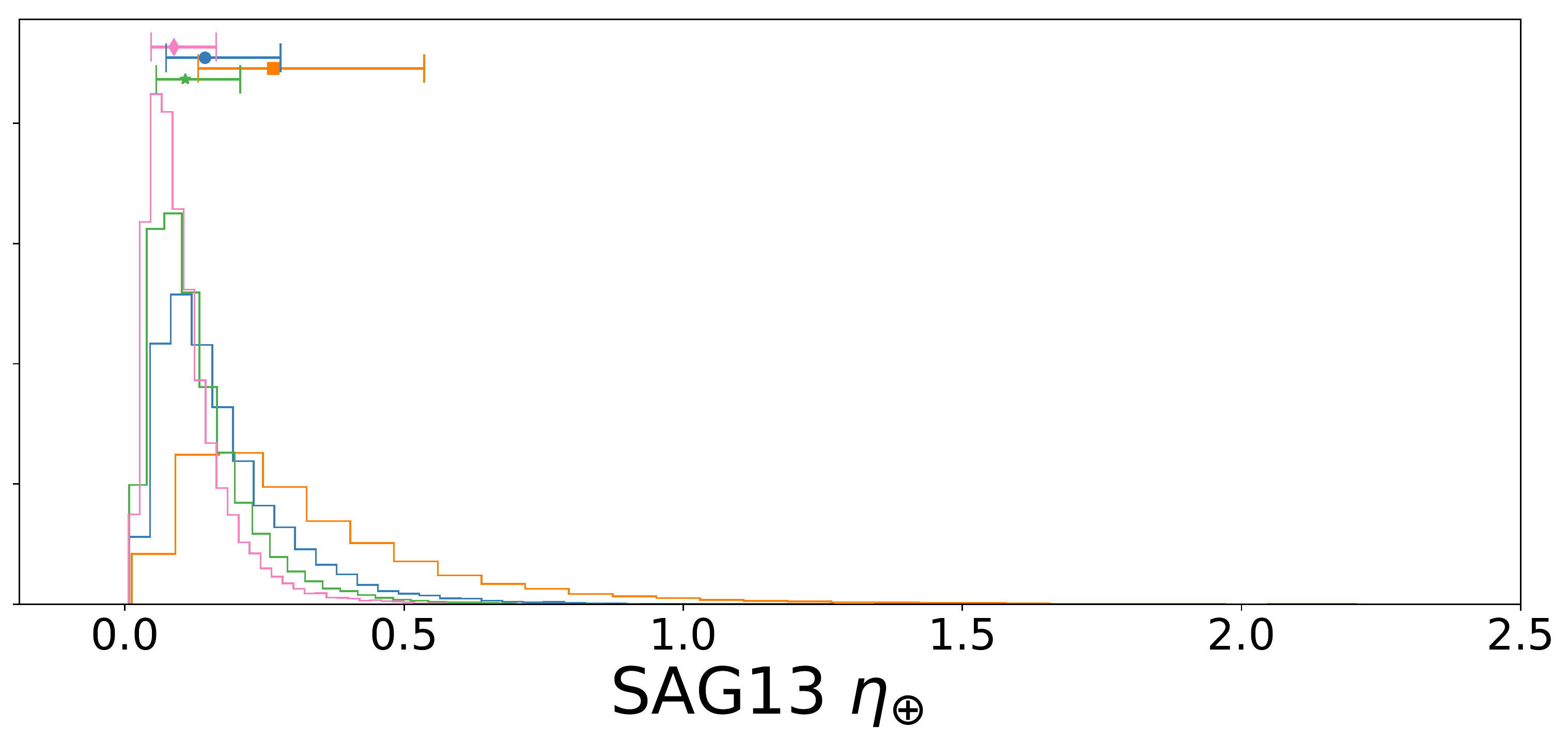}
  \includegraphics[width=0.48\linewidth]{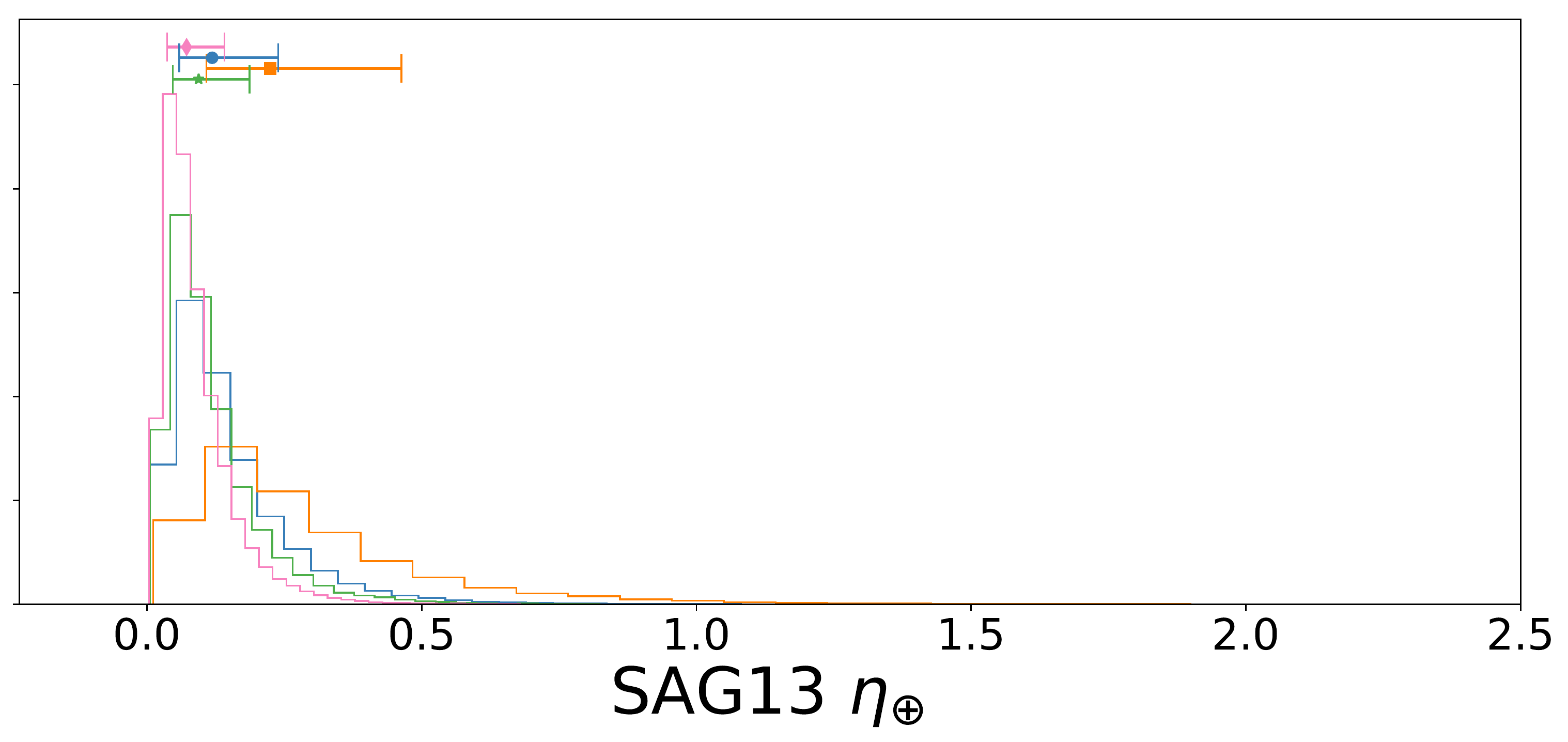}
   \caption{Distributions of various occurrence rates using a score cut of 0.9 for the  high reliability (blue), DR25 (pink), FPWG PC (green) and high completeness (orange) vetting, computed with the Poisson method.  Left: without correcting for reliability.  Right: corrected for reliability.} \label{figure:sc0p9Dist}
\end{figure*}

\subsection{Poisson Likelihood Method}
 
Figure~\ref{figure:allDist} shows the distribution of the five occurrence rates resulting from the posteriors of $\boldsymbol{\theta}$, all near the \Kepler\ detection limit, for the four catalogs described in \S\ref{section:thresholdVariation}.  The left panels show the occurrence rates without correcting for reliability, while the right panels are corrected for reliability.  For all occurrence rates, if we do not correct for reliability the different choices for Robovetter thresholds yield different occurrence rates, in some cases varying by more than a factor of 3.  When we correct for reliability, we find that the different choices for Robovetter thresholds yield very similar occurrence rates, with closely-overlapping distributions.  We take this to indicate the methods we use for completeness and reliability characterization and correction are working as intended, and correction for both completeness and reliability are required.  This is the case even when the PC population is polluted by a significant number of false alarms, as is the case for the high-completeness catalog as described in \S\ref{section:thresholdVariation}.

Figure~\ref{figure:sc0p9Dist} shows the same occurrence rate posteriors as Figure~\ref{figure:allDist}, but with a score cut applied to the planet catalog that removes planets with Robovetter disposition score below 0.9. This score cut is expected to yield a higher-reliability population in all the catalog.  This population for the DR25 catalog is shown in the lower right panel of  Figure~\ref{figure:planetPopulationScoreCut}.  We see that, without correcting for reliability (left panels), the distributions for the different catalogs are more consistent than in the left panels of Figure~\ref{figure:allDist}.  As discussed in \S\ref{section:score}, even a high score cut of 0.9 passes planets with low reliability due to astrophysical false positive probability, so correction for reliability is appropriate with a score cut.  But there are relatively few such low-reliability planets with this high score cut, so the impact of reliability correction on the distributions in Figure~\ref{figure:sc0p9Dist} (right panels) is minor.  Comparing Figures~\ref{figure:sc0p9Dist} and \ref{figure:allDist} shows, however, that whether corrected for reliability or not, the spread of occurrence rates from the four catalogs using a score cut is notably larger than the spread when correcting for reliability not using a score cut.

\renewcommand{\arraystretch}{1.25}
\begin{table*}[ht]
\centering
\caption{Fit coefficients for various score cuts}\label{table:fitResults}
\begin{tabular}{ r r r r r r c}
\hline
\hline
         & \multicolumn{5}{c}{No Reliability Correction}  \\
Parameter & Score Cut & DR25 & High Reliability & High Completeness & FPWG PC & Max Separation $(\sigma)$ \\
\hline
$F_0$
& 0.0
& $0.612^{+0.115}_{-0.091}$
& $0.527^{+0.099}_{-0.083}$
& $0.863^{+0.166}_{-0.133}$
& $0.939^{+0.171}_{-0.140}$
& 2.40
\\
& 0.6
& $0.524^{+0.110}_{-0.087}$
& $0.590^{+0.131}_{-0.105}$
& $0.511^{+0.104}_{-0.085}$
& $0.490^{+0.097}_{-0.080}$
& 0.70
\\
& 0.7
& $0.511^{+0.111}_{-0.090}$
& $0.587^{+0.133}_{-0.107}$
& $0.477^{+0.106}_{-0.080}$
& $0.455^{+0.098}_{-0.077}$
& 0.92
\\
& 0.9
& $0.422^{+0.107}_{-0.084}$
& $0.559^{+0.171}_{-0.124}$
& $0.372^{+0.094}_{-0.071}$
& $0.335^{+0.079}_{-0.059}$
& 1.52
\\
$\alpha$
& 0.0
& $0.285^{+0.499}_{-0.495}$
& $0.670^{+0.570}_{-0.538}$
& $-0.161^{+0.479}_{-0.450}$
& $-0.323^{+0.443}_{-0.418}$
& 1.42
\\
& 0.6
& $0.237^{+0.539}_{-0.530}$
& $0.302^{+0.567}_{-0.570}$
& $0.232^{+0.554}_{-0.524}$
& $0.286^{+0.556}_{-0.520}$
& 0.09
\\
& 0.7
& $0.229^{+0.575}_{-0.527}$
& $0.346^{+0.581}_{-0.552}$
& $0.309^{+0.554}_{-0.544}$
& $0.379^{+0.573}_{-0.549}$
& 0.19
\\
& 0.9
& $0.613^{+0.717}_{-0.675}$
& $0.333^{+0.734}_{-0.682}$
& $0.719^{+0.696}_{-0.678}$
& $0.838^{+0.703}_{-0.680}$
& 0.51
\\
$\beta$
& 0.0
& $-0.546^{+0.173}_{-0.176}$
& $-0.653^{+0.192}_{-0.191}$
& $-0.241^{+0.161}_{-0.162}$
& $-0.218^{+0.155}_{-0.158}$
& 1.75
\\
& 0.6
& $-0.679^{+0.191}_{-0.199}$
& $-0.565^{+0.201}_{-0.204}$
& $-0.700^{+0.191}_{-0.195}$
& $-0.702^{+0.195}_{-0.200}$
& 0.49
\\
& 0.7
& $-0.707^{+0.203}_{-0.206}$
& $-0.569^{+0.200}_{-0.216}$
& $-0.751^{+0.204}_{-0.205}$
& $-0.751^{+0.202}_{-0.210}$
& 0.61
\\
& 0.9
& $-0.761^{+0.244}_{-0.250}$
& $-0.580^{+0.253}_{-0.254}$
& $-0.859^{+0.234}_{-0.254}$
& $-0.899^{+0.237}_{-0.244}$
& 0.92
\\
& \multicolumn{5}{c}{With Reliability Correction}  \\
Parameter & Score Cut & DR25 & High Reliability & High Completeness & FPWG PC & Max Separation $(\sigma)$ \\
\hline
$F_0$
& 0.0
& $0.444^{+0.092}_{-0.074}$
& $0.416^{+0.085}_{-0.069}$
& $0.449^{+0.096}_{-0.076}$
& $0.485^{+0.102}_{-0.084}$
& 0.58
\\
& 0.6
& $0.436^{+0.099}_{-0.078}$
& $0.485^{+0.116}_{-0.089}$
& $0.407^{+0.092}_{-0.072}$
& $0.379^{+0.083}_{-0.066}$
& 0.87
\\
& 0.7
& $0.422^{+0.100}_{-0.077}$
& $0.498^{+0.124}_{-0.094}$
& $0.403^{+0.093}_{-0.073}$
& $0.367^{+0.085}_{-0.066}$
& 1.03
\\
& 0.9
& $0.382^{+0.104}_{-0.078}$
& $0.509^{+0.158}_{-0.114}$
& $0.344^{+0.091}_{-0.067}$
& $0.305^{+0.075}_{-0.058}$
& 1.49
\\
$\alpha$
& 0.0
& $0.736^{+0.618}_{-0.589}$
& $1.100^{+0.647}_{-0.618}$
& $0.766^{+0.630}_{-0.599}$
& $0.553^{+0.610}_{-0.565}$
& 0.63
\\
& 0.6
& $0.579^{+0.656}_{-0.609}$
& $0.666^{+0.665}_{-0.627}$
& $0.638^{+0.672}_{-0.617}$
& $0.757^{+0.670}_{-0.625}$
& 0.20
\\
& 0.7
& $0.671^{+0.684}_{-0.647}$
& $0.631^{+0.686}_{-0.639}$
& $0.618^{+0.666}_{-0.632}$
& $0.791^{+0.702}_{-0.662}$
& 0.18
\\
& 0.9
& $0.783^{+0.769}_{-0.725}$
& $0.467^{+0.770}_{-0.720}$
& $0.829^{+0.768}_{-0.721}$
& $1.008^{+0.770}_{-0.720}$
& 0.51
\\
$\beta$
& 0.0
& $-0.800^{+0.202}_{-0.211}$
& $-0.787^{+0.207}_{-0.213}$
& $-0.774^{+0.208}_{-0.211}$
& $-0.741^{+0.202}_{-0.207}$
& 0.20
\\
& 0.6
& $-0.753^{+0.210}_{-0.215}$
& $-0.637^{+0.212}_{-0.217}$
& $-0.817^{+0.213}_{-0.220}$
& $-0.844^{+0.214}_{-0.224}$
& 0.68
\\
& 0.7
& $-0.777^{+0.217}_{-0.224}$
& $-0.634^{+0.220}_{-0.227}$
& $-0.829^{+0.221}_{-0.228}$
& $-0.871^{+0.225}_{-0.232}$
& 0.74
\\
& 0.9
& $-0.769^{+0.256}_{-0.258}$
& $-0.588^{+0.260}_{-0.267}$
& $-0.861^{+0.248}_{-0.259}$
& $-0.909^{+0.246}_{-0.262}$
& 0.89
\\
\end{tabular}
\tablecomments{Median values and 16th and 84th percentile error bars of the posteriors of $\boldsymbol{\theta}$ in Equation (\ref{eqn:powerLaw}) for the high reliability, DR25, FPWG PC and high catalogs $F_0$ is the occurrence rate of planets with $50 \leq$ period $\leq 400$ days and $0.75 \leq$ radius $\leq 2.5$ $R_{\oplus}$.  These values were computed using the Poisson Likelihood MCMC method.  The maximum separation in each row is the maximum over each row of the difference in medians divided by the propagated uncertainty of that distance.
}
\end{table*}

\renewcommand{\arraystretch}{1.25}
\begin{table*}[ht]
\centering
\caption{Occurrence rates for various score cuts}\label{table:occurrenceResults}
\begin{tabular}{ r r r r r r c}
\hline
\hline
         & \multicolumn{5}{c}{No Reliability Correction}  \\
Parameter & Score Cut & DR25 & High Reliability & High Completeness & FPWG PC & Max Separation $(\sigma)$ \\
\hline
$\Gamma_\oplus$ & 0.0 & $0.217^{+0.114}_{-0.077}$ & $0.142^{+0.086}_{-0.056}$ & $0.466^{+0.222}_{-0.154}$ & $0.552^{+0.235}_{-0.173}$ & 2.12 \\
& 0.6 & $0.170^{+0.103}_{-0.066}$ & $0.204^{+0.131}_{-0.084}$ & $0.164^{+0.094}_{-0.064}$ & $0.152^{+0.090}_{-0.059}$ & 0.42 \\
& 0.7 & $0.162^{+0.101}_{-0.065}$ & $0.196^{+0.130}_{-0.080}$ & $0.141^{+0.090}_{-0.056}$ & $0.130^{+0.082}_{-0.053}$ & 0.58 \\
& 0.9 & $0.105^{+0.090}_{-0.050}$ & $0.186^{+0.165}_{-0.091}$ & $0.081^{+0.068}_{-0.039}$ & $0.066^{+0.054}_{-0.031}$ & 1.14 \\
$F_1$ & 0.0 & $0.191^{+0.036}_{-0.030}$ & $0.166^{+0.033}_{-0.029}$ & $0.240^{+0.040}_{-0.036}$ & $0.259^{+0.041}_{-0.037}$ & 1.87 \\
& 0.6 & $0.173^{+0.035}_{-0.029}$ & $0.185^{+0.039}_{-0.034}$ & $0.171^{+0.033}_{-0.030}$ & $0.163^{+0.032}_{-0.028}$ & 0.47 \\
& 0.7 & $0.171^{+0.036}_{-0.031}$ & $0.184^{+0.040}_{-0.034}$ & $0.162^{+0.034}_{-0.028}$ & $0.153^{+0.033}_{-0.027}$ & 0.66 \\
& 0.9 & $0.139^{+0.037}_{-0.030}$ & $0.175^{+0.049}_{-0.040}$ & $0.127^{+0.034}_{-0.027}$ & $0.114^{+0.029}_{-0.023}$ & 1.24 \\
$\zeta_\oplus$ & 0.0 & $0.035^{+0.018}_{-0.012}$ & $0.023^{+0.014}_{-0.009}$ & $0.075^{+0.036}_{-0.025}$ & $0.089^{+0.038}_{-0.028}$ & 2.12 \\
& 0.6 & $0.027^{+0.017}_{-0.011}$ & $0.033^{+0.021}_{-0.013}$ & $0.026^{+0.015}_{-0.010}$ & $0.025^{+0.015}_{-0.010}$ & 0.42 \\
& 0.7 & $0.026^{+0.016}_{-0.011}$ & $0.032^{+0.021}_{-0.013}$ & $0.023^{+0.015}_{-0.009}$ & $0.021^{+0.013}_{-0.008}$ & 0.57 \\
& 0.9 & $0.017^{+0.015}_{-0.008}$ & $0.030^{+0.027}_{-0.015}$ & $0.013^{+0.011}_{-0.006}$ & $0.011^{+0.009}_{-0.005}$ & 1.14 \\
SAG13 $\eta_\oplus$ & 0.0 & $0.312^{+0.185}_{-0.117}$ & $0.197^{+0.130}_{-0.079}$ & $0.742^{+0.423}_{-0.268}$ & $0.896^{+0.465}_{-0.310}$ & 2.08 \\
& 0.6 & $0.236^{+0.163}_{-0.097}$ & $0.292^{+0.212}_{-0.126}$ & $0.226^{+0.148}_{-0.092}$ & $0.209^{+0.141}_{-0.085}$ & 0.44 \\
& 0.7 & $0.224^{+0.158}_{-0.095}$ & $0.279^{+0.211}_{-0.119}$ & $0.192^{+0.138}_{-0.080}$ & $0.177^{+0.125}_{-0.074}$ & 0.59 \\
& 0.9 & $0.143^{+0.135}_{-0.070}$ & $0.266^{+0.270}_{-0.135}$ & $0.108^{+0.098}_{-0.052}$ & $0.088^{+0.075}_{-0.041}$ & 1.15 \\
\\
         & \multicolumn{5}{c}{With Reliability Correction}  \\
Parameter & Score Cut & DR25 & High Reliability & High Completeness & FPWG PC & Max Separation $(\sigma)$ \\
\hline
$\Gamma_\oplus$ & 0.0 & $0.102^{+0.070}_{-0.043}$ & $0.079^{+0.059}_{-0.035}$ & $0.104^{+0.075}_{-0.045}$ & $0.128^{+0.085}_{-0.055}$ & 0.61 \\
& 0.6 & $0.112^{+0.081}_{-0.050}$ & $0.132^{+0.101}_{-0.059}$ & $0.096^{+0.073}_{-0.044}$ & $0.082^{+0.062}_{-0.037}$ & 0.58 \\
& 0.7 & $0.101^{+0.080}_{-0.047}$ & $0.138^{+0.109}_{-0.064}$ & $0.096^{+0.072}_{-0.043}$ & $0.076^{+0.062}_{-0.036}$ & 0.69 \\
& 0.9 & $0.087^{+0.081}_{-0.044}$ & $0.157^{+0.148}_{-0.079}$ & $0.070^{+0.065}_{-0.035}$ & $0.054^{+0.049}_{-0.027}$ & 1.10 \\
$F_1$ & 0.0 & $0.148^{+0.033}_{-0.028}$ & $0.132^{+0.031}_{-0.026}$ & $0.148^{+0.033}_{-0.028}$ & $0.160^{+0.034}_{-0.030}$ & 0.65 \\
& 0.6 & $0.145^{+0.034}_{-0.029}$ & $0.152^{+0.037}_{-0.031}$ & $0.138^{+0.032}_{-0.028}$ & $0.128^{+0.030}_{-0.026}$ & 0.57 \\
& 0.7 & $0.140^{+0.034}_{-0.029}$ & $0.157^{+0.039}_{-0.033}$ & $0.137^{+0.033}_{-0.028}$ & $0.125^{+0.031}_{-0.026}$ & 0.72 \\
& 0.9 & $0.124^{+0.035}_{-0.029}$ & $0.159^{+0.047}_{-0.038}$ & $0.116^{+0.033}_{-0.026}$ & $0.103^{+0.028}_{-0.023}$ & 1.19 \\
$\zeta_\oplus$ & 0.0 & $0.016^{+0.011}_{-0.007}$ & $0.013^{+0.009}_{-0.006}$ & $0.017^{+0.012}_{-0.007}$ & $0.021^{+0.014}_{-0.009}$ & 0.61 \\
& 0.6 & $0.018^{+0.013}_{-0.008}$ & $0.021^{+0.016}_{-0.010}$ & $0.016^{+0.012}_{-0.007}$ & $0.013^{+0.010}_{-0.006}$ & 0.58 \\
& 0.7 & $0.016^{+0.013}_{-0.008}$ & $0.022^{+0.017}_{-0.010}$ & $0.015^{+0.012}_{-0.007}$ & $0.012^{+0.010}_{-0.006}$ & 0.69 \\
& 0.9 & $0.014^{+0.013}_{-0.007}$ & $0.025^{+0.024}_{-0.013}$ & $0.011^{+0.010}_{-0.006}$ & $0.009^{+0.008}_{-0.004}$ & 1.10 \\
SAG13 $\eta_\oplus$ & 0.0 & $0.137^{+0.101}_{-0.059}$ & $0.109^{+0.083}_{-0.048}$ & $0.141^{+0.110}_{-0.062}$ & $0.174^{+0.129}_{-0.076}$ & 0.58 \\
& 0.6 & $0.153^{+0.121}_{-0.069}$ & $0.185^{+0.155}_{-0.085}$ & $0.129^{+0.106}_{-0.059}$ & $0.109^{+0.089}_{-0.049}$ & 0.62 \\
& 0.7 & $0.137^{+0.118}_{-0.064}$ & $0.194^{+0.169}_{-0.091}$ & $0.128^{+0.105}_{-0.059}$ & $0.101^{+0.089}_{-0.048}$ & 0.73 \\
& 0.9 & $0.119^{+0.120}_{-0.060}$ & $0.224^{+0.239}_{-0.116}$ & $0.094^{+0.093}_{-0.047}$ & $0.072^{+0.069}_{-0.036}$ & 1.13 \\
\end{tabular}
\tablecomments{Occurrence rate results for various score cuts when not correcting for reliability resulting from high reliability, DR25, FPWG PC and high completeness vetting.  $F_1$ is the integrated planet rate over $50 \leq \mathrm{period} \leq 200$ days and $1 \leq \mathrm{radius} \leq 2$ $R_{\oplus}$,  $\zeta_{\oplus}$ is the integrated rate within 20\% of Earth's orbital period and size, and $\mathrm{SAG}13~\eta_{\oplus}$ is the integrated rate over $237 \leq \mathrm{period} \leq 860$ days and $0.5 \leq \mathrm{radius} \leq 1.5$ $R_{\oplus}$.}
\end{table*}

We compute the planet population model parameters $\boldsymbol{\theta}$ and resulting occurrence rates for the score cuts 0, 0.6, 0.7, and 0.9 for all four catalogs.  We provide results both not corrected for reliability and corrected for reliability.  The resulting $\boldsymbol{\theta}$ are given in Tables~\ref{table:fitResults}, and the occurrence rates are give in Table~\ref{table:occurrenceResults}.  These tables give the central values and 14th and 86th percentile confidence intervals.  These tables introduce the {\it maximum separation} metric, which quantifies the separation between the distributions from the catalogs.  For each pair of catalogs $\{ i,j \}$, we compute $d_{i,j} = (m_i - m_j)/\sigma_{d_{i,j}}$, the difference in medians $m_i$ and $m_j$ divided by the uncertainty in that distance propagated from $\sigma_i$ and $\sigma_j$, where $\sigma_i$ is the $68\%$ confidence interval of distribution $i$. Then the maximum separation in a row is the largest $d_{i,j}$ over all pairs $\{ i,j \}$ in that row.

Based on the maximum separation, we see in Tables~\ref{table:fitResults} and ~\ref{table:occurrenceResults} that the spread between the different catalogs can be over $2 \sigma$ with no score cut and not correcting for reliability.  Correcting for reliability reduces the separation to well under $1 \sigma$.  Applying a score cut of 0.6 or 0.7 also reduces the separation to under $1 \sigma$ with and without reliability correction.  Applying a score cut of 0.9 results in a separation in $F_0$ greater than $1 \sigma$, though the separation in $\alpha$ and $\beta$ is less than $1 \sigma$.  The larger separation in $F_0$ for a score cut of 0.9 drives a separation greater than $1 \sigma$ for the occurrence rates in Table~\ref{table:occurrenceResults}.

\begin{figure*}[ht]
  \centering
  \Large \hspace{0.2 in} No Reliability Correction \hspace{1 in} Corrected for Reliability\\
  \includegraphics[width=0.46\linewidth]{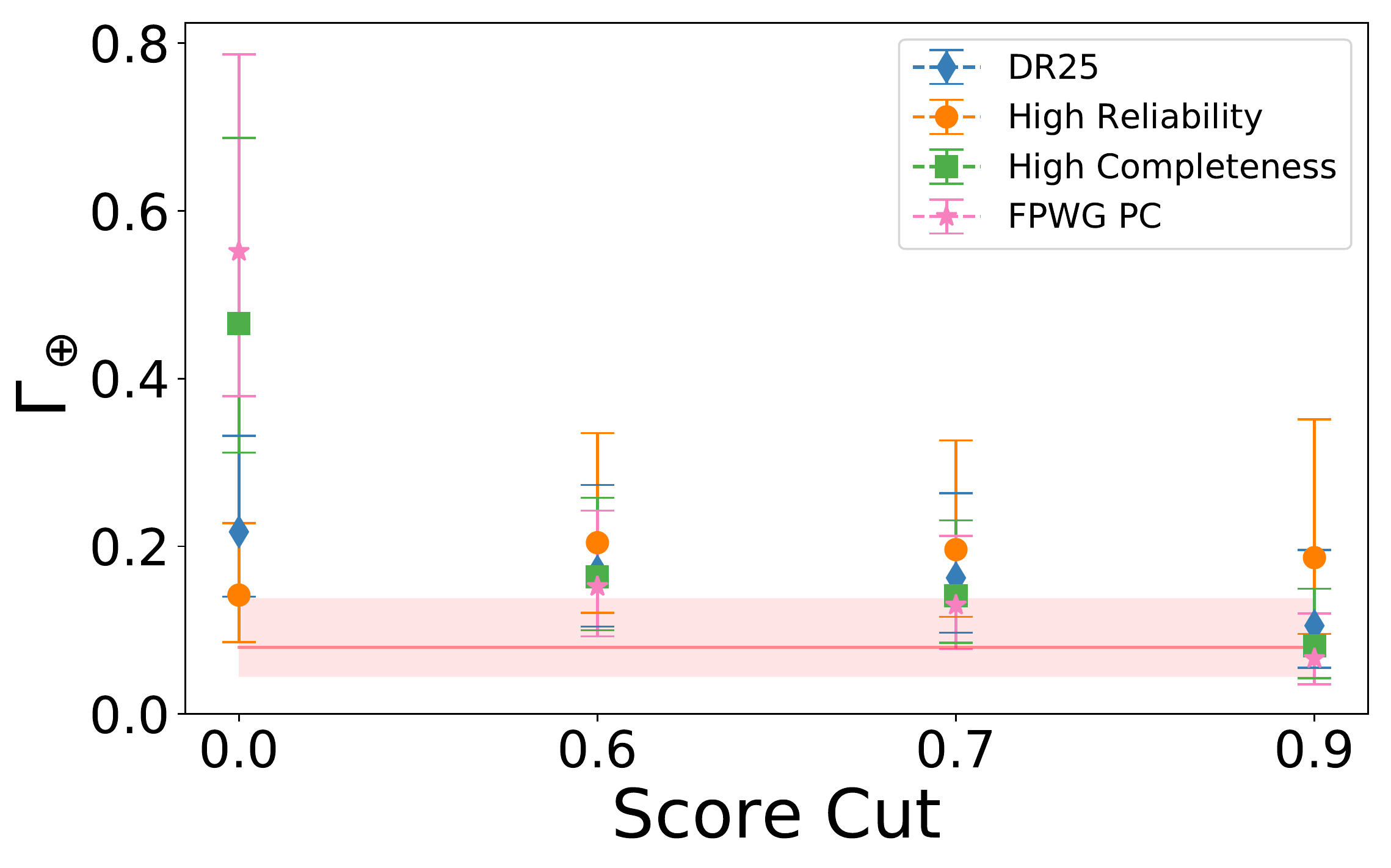} 
  \includegraphics[width=0.46\linewidth]{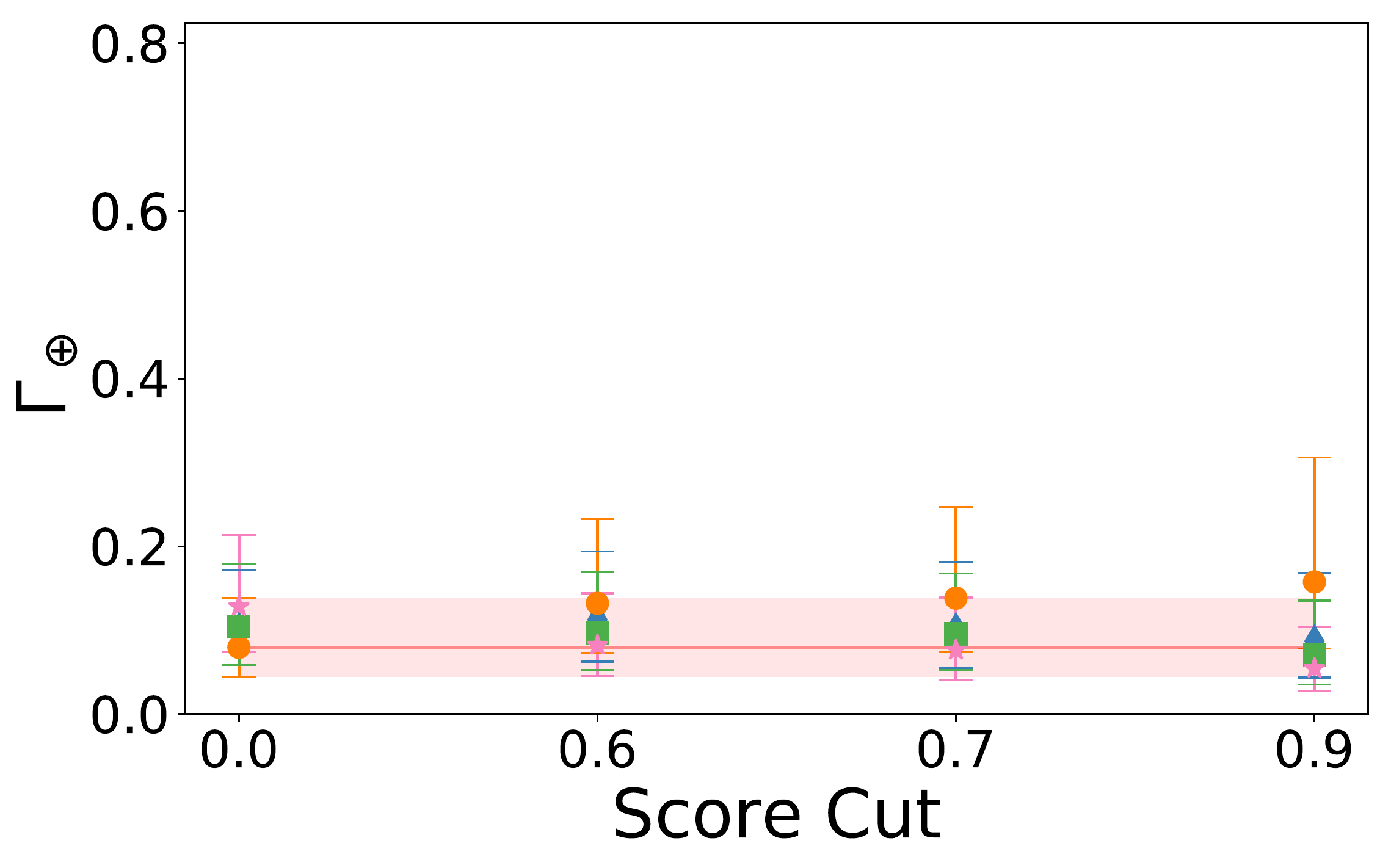} \\
  \includegraphics[width=0.46\linewidth]{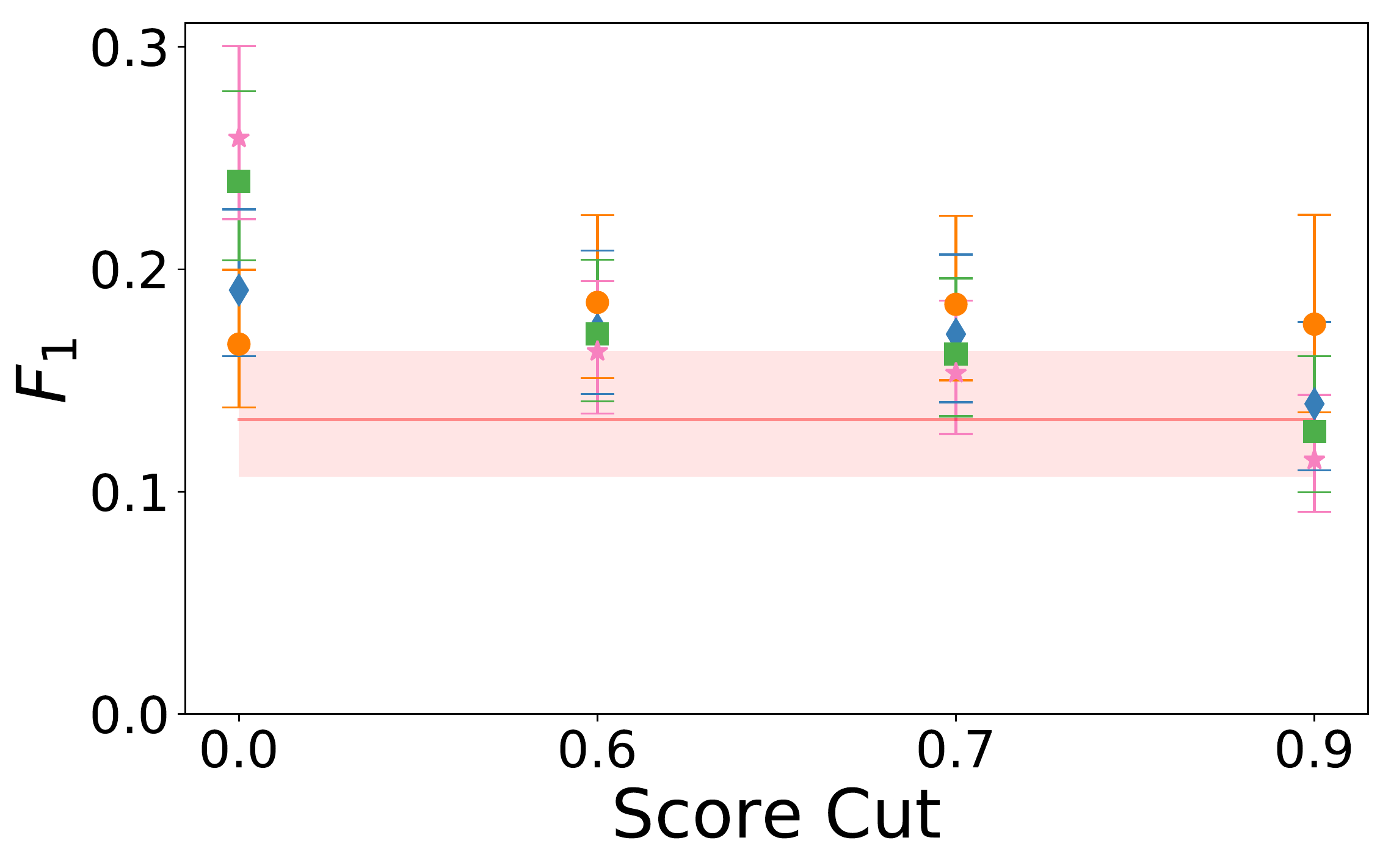}
  \includegraphics[width=0.46\linewidth]{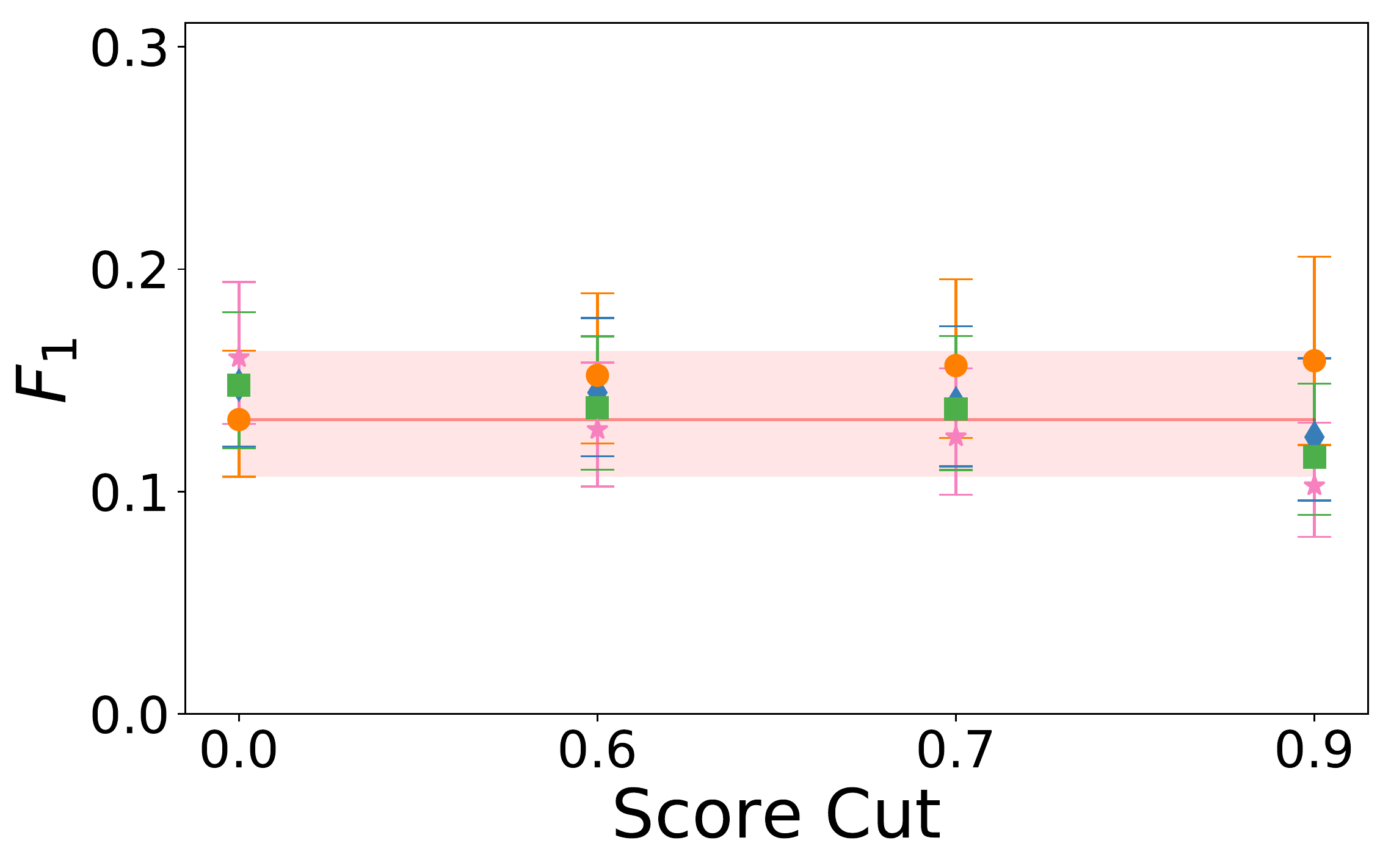} \\
  \includegraphics[width=0.46\linewidth]{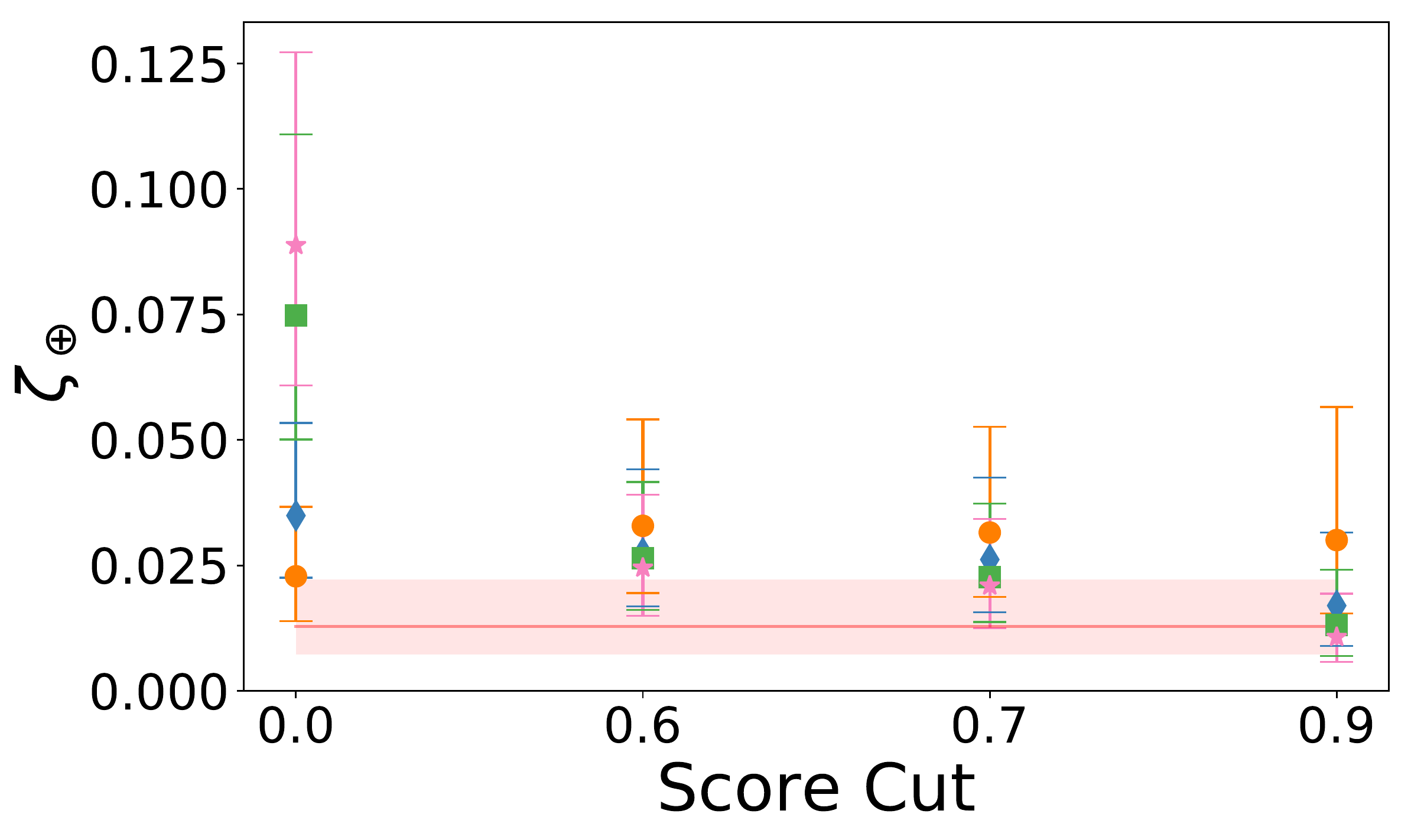}
  \includegraphics[width=0.46\linewidth]{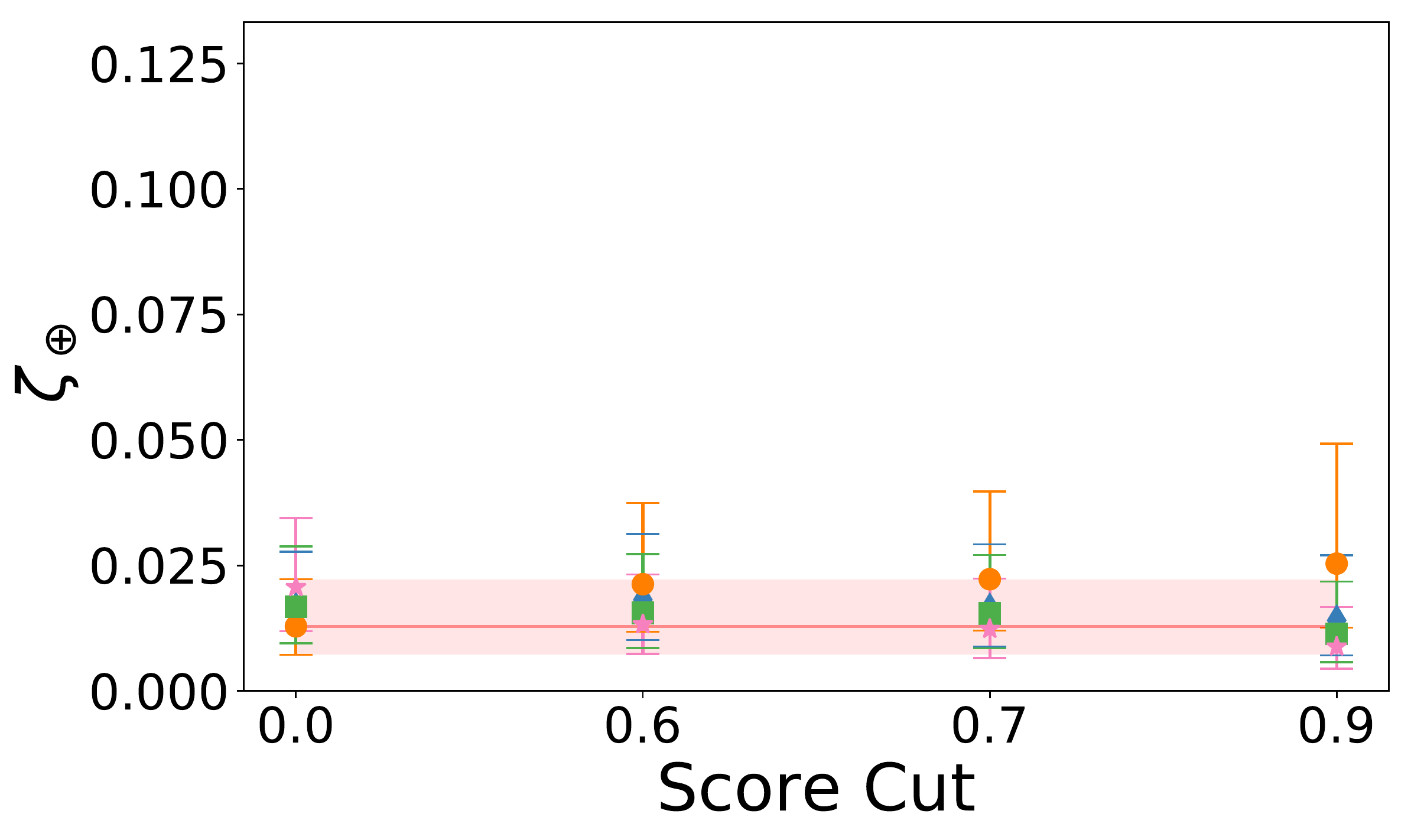} \\
  \includegraphics[width=0.46\linewidth]{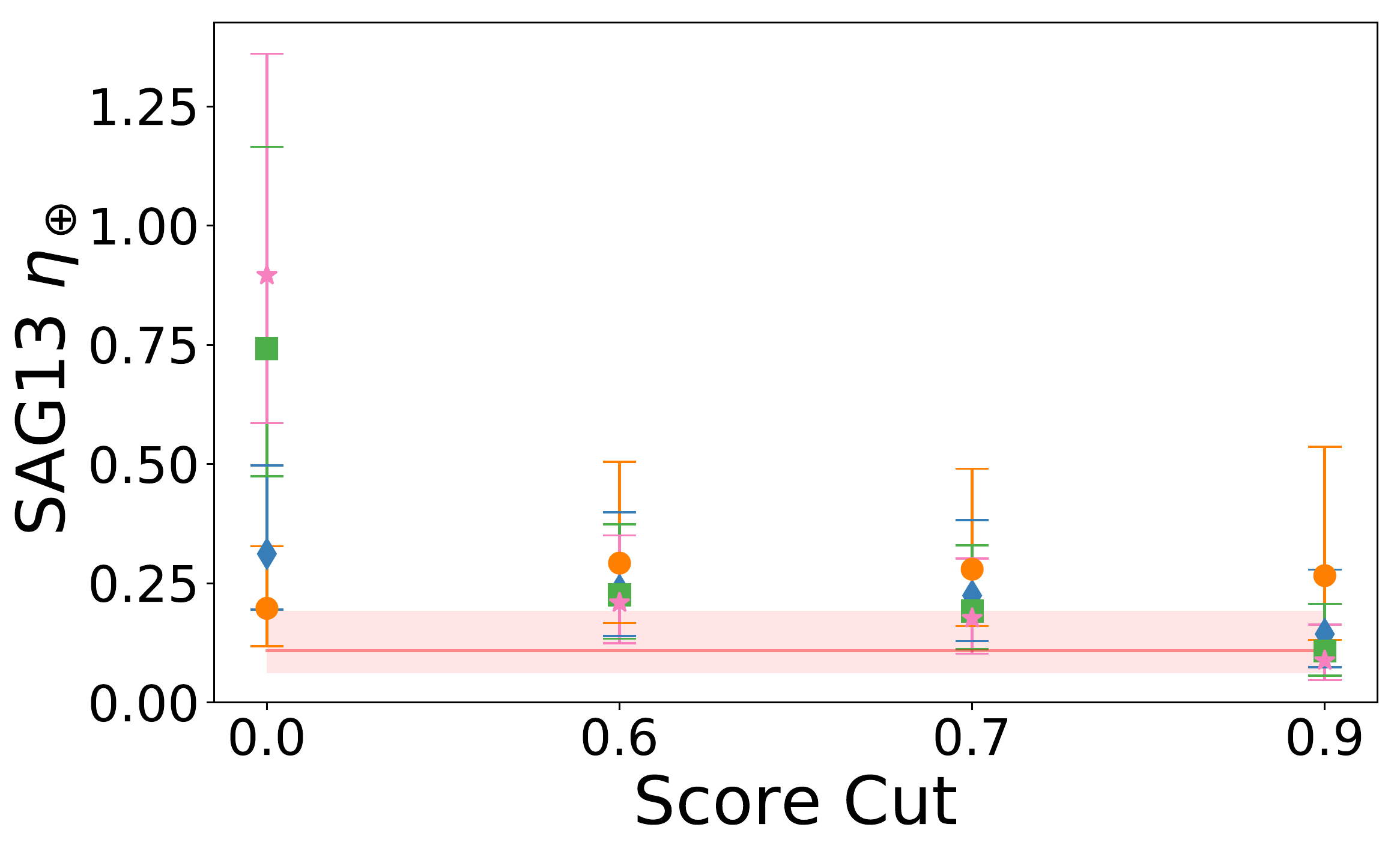}
  \includegraphics[width=0.46\linewidth]{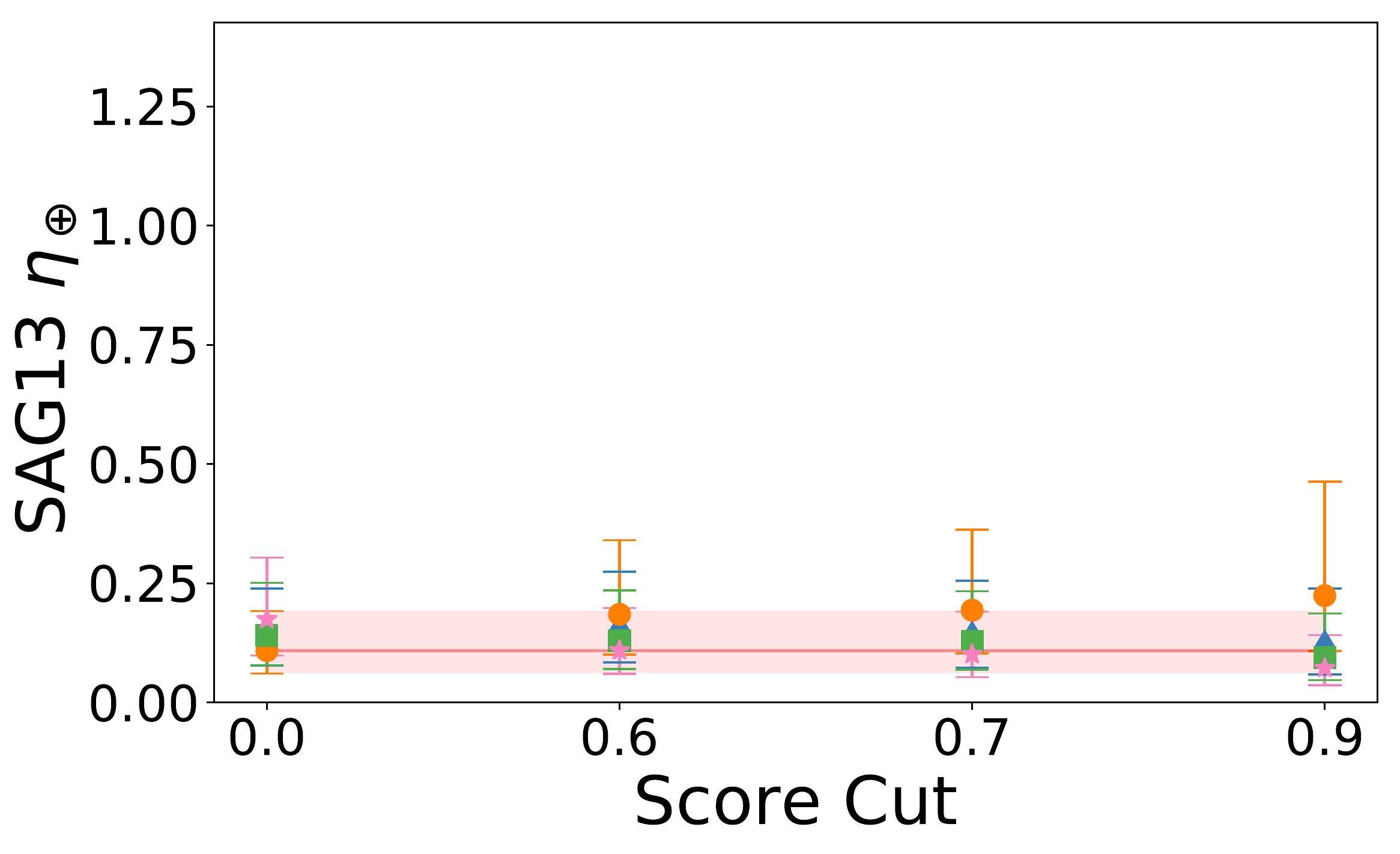}
   \caption{The median and 68\% confidence intervals for the four catalogs of various occurrence rates with various score cuts, computed with the Poisson method.  Left: without correcting for reliability.  Right: corrected for reliability.  The red horizontal line and stripe are the median and 86\% confidence intervals for the DR25 catalog with no score cut.} \label{figure:scCompare}
\end{figure*}

\begin{figure*}[ht]
  \centering
  \Large \hspace{0.2 in} No Reliability Correction \hspace{1 in} Corrected for Reliability\\
  \includegraphics[width=0.46\linewidth]{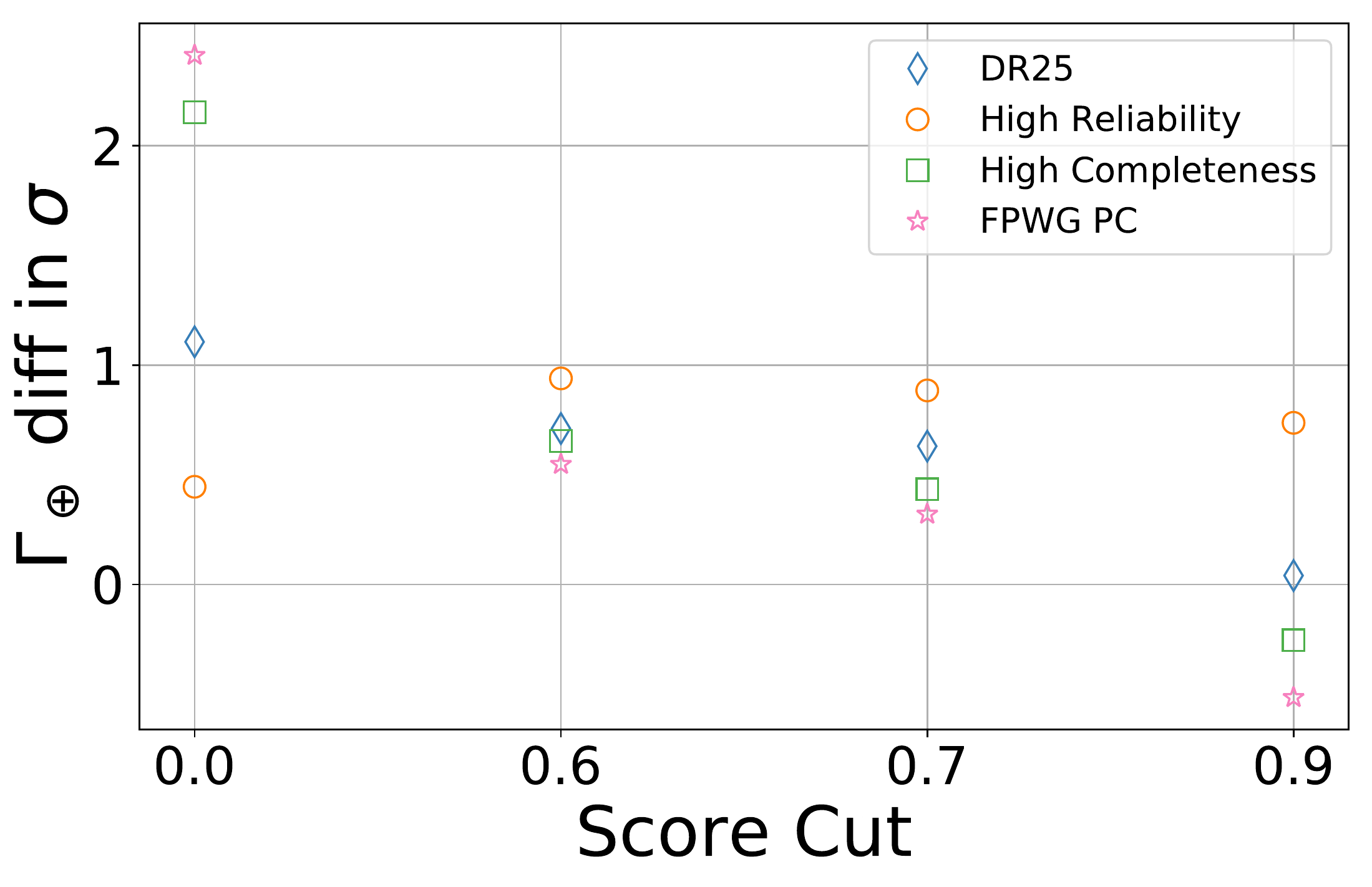} 
  \includegraphics[width=0.46\linewidth]{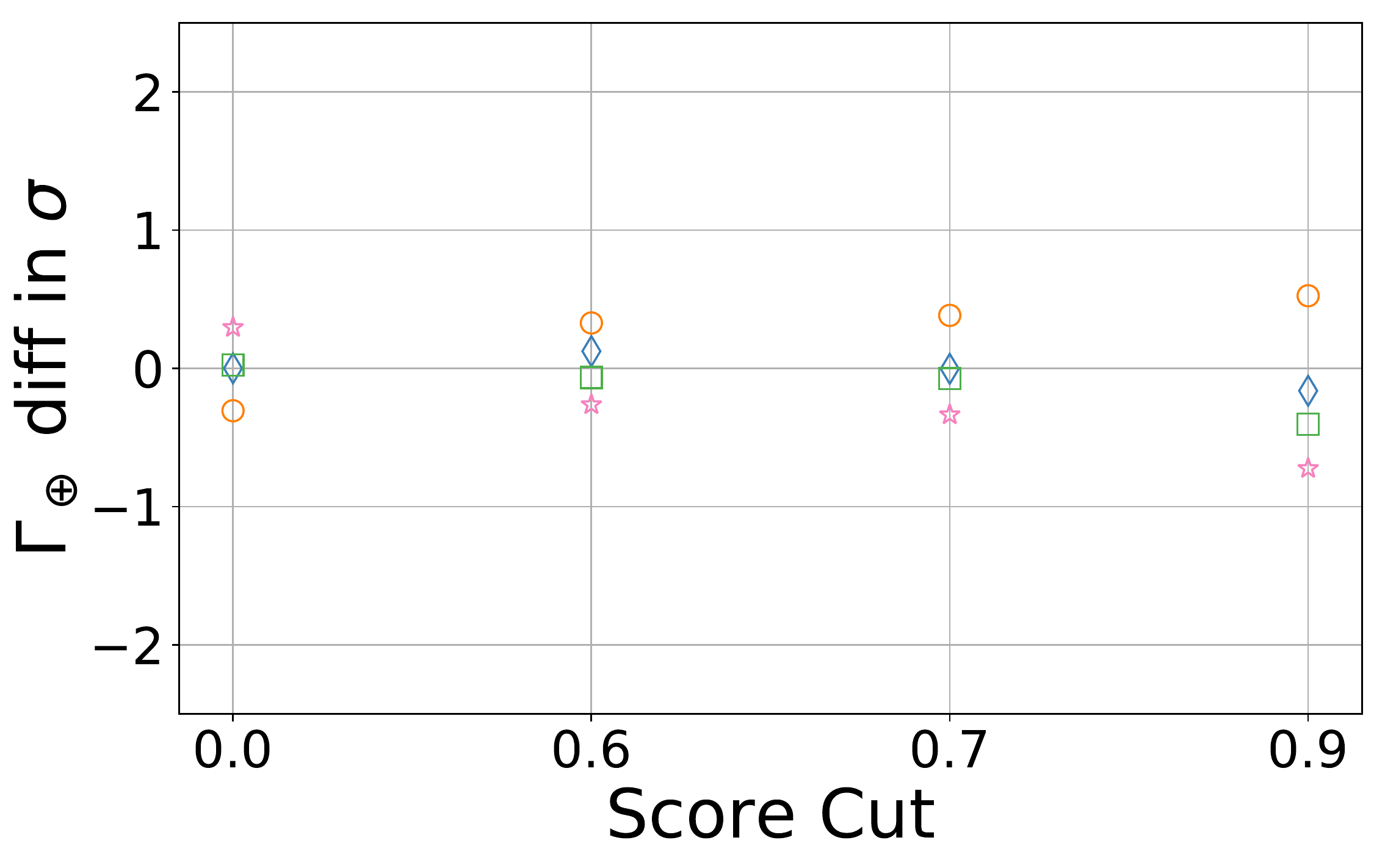} \\
  \includegraphics[width=0.46\linewidth]{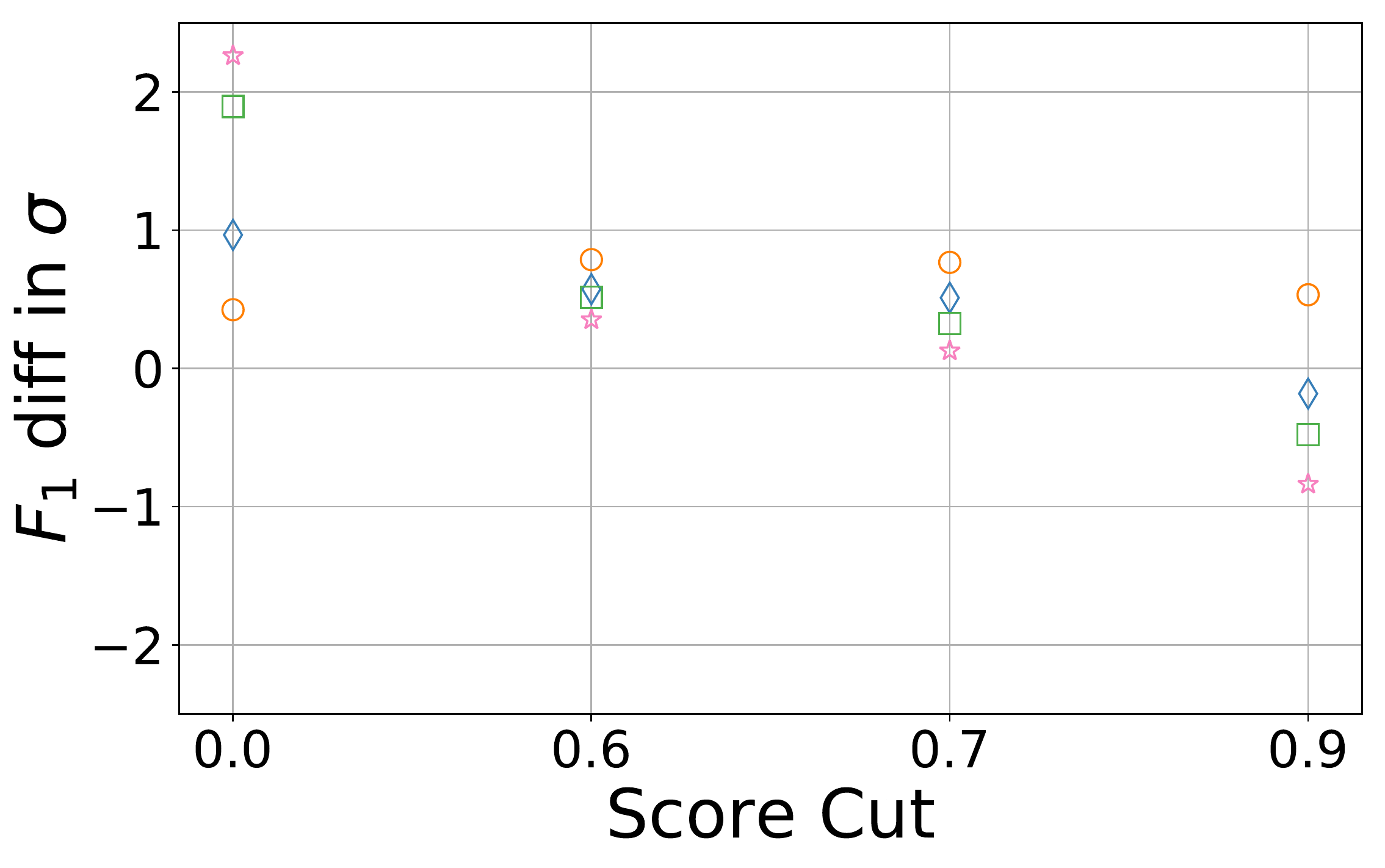}
  \includegraphics[width=0.46\linewidth]{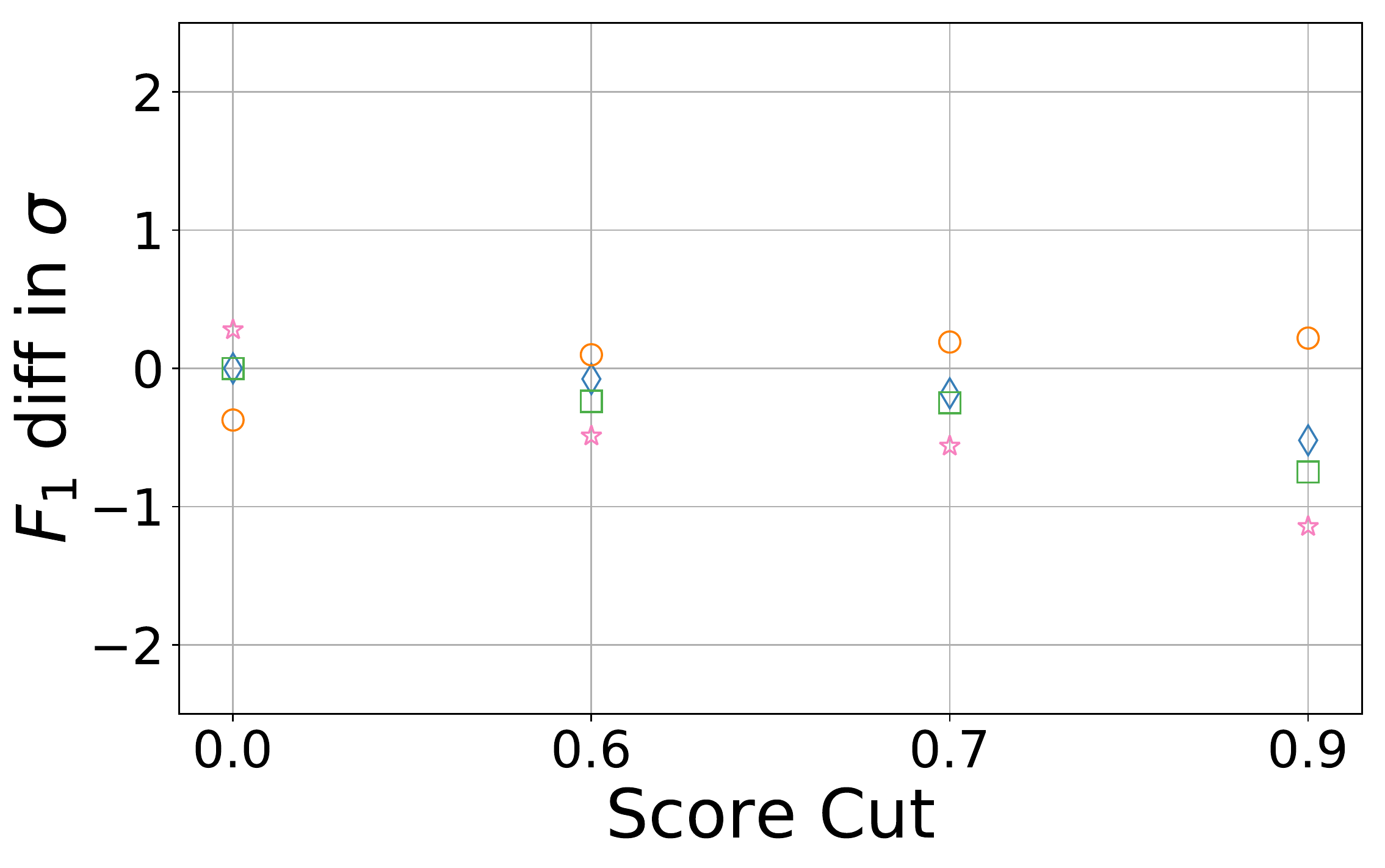} \\
  \includegraphics[width=0.46\linewidth]{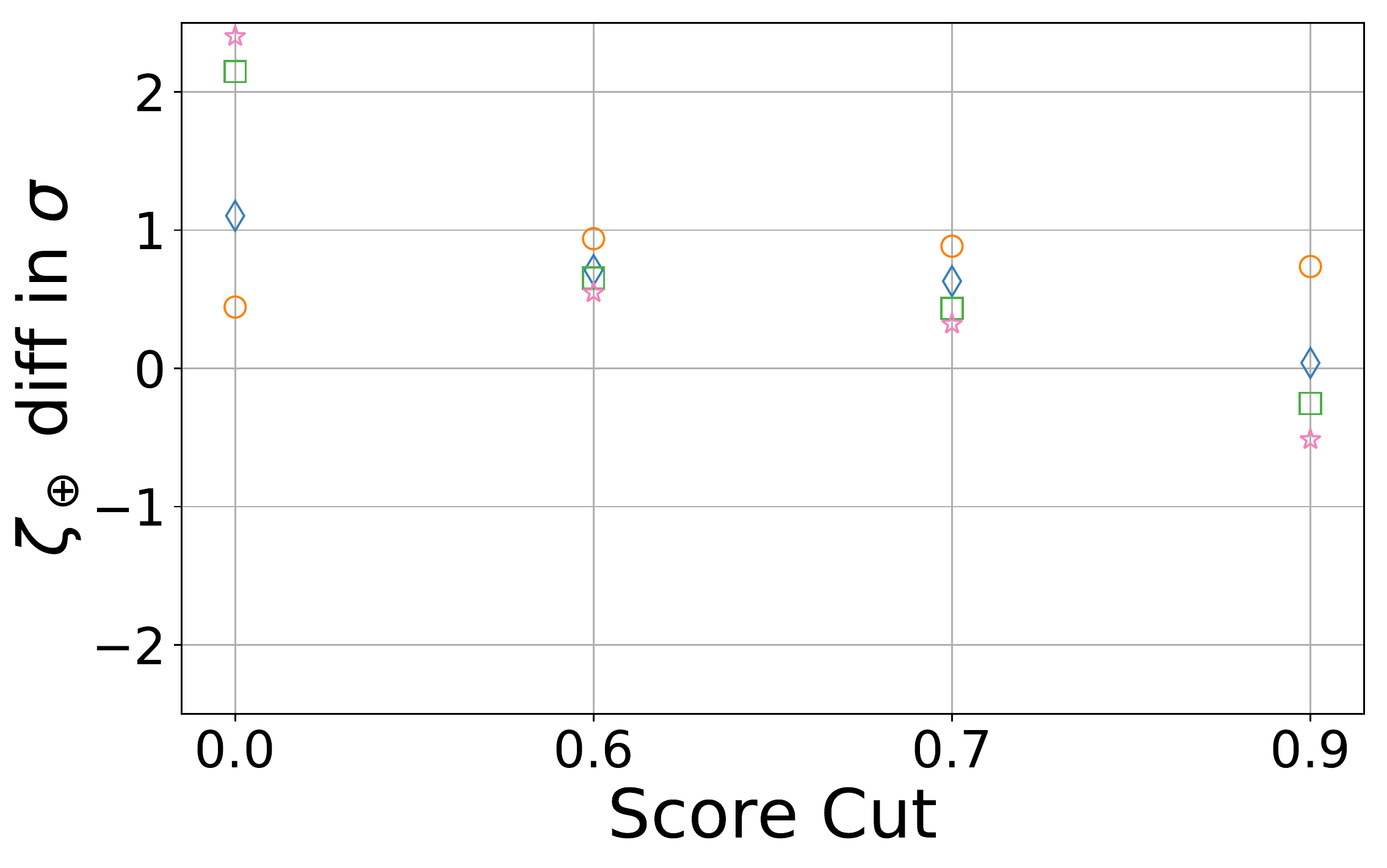}
  \includegraphics[width=0.46\linewidth]{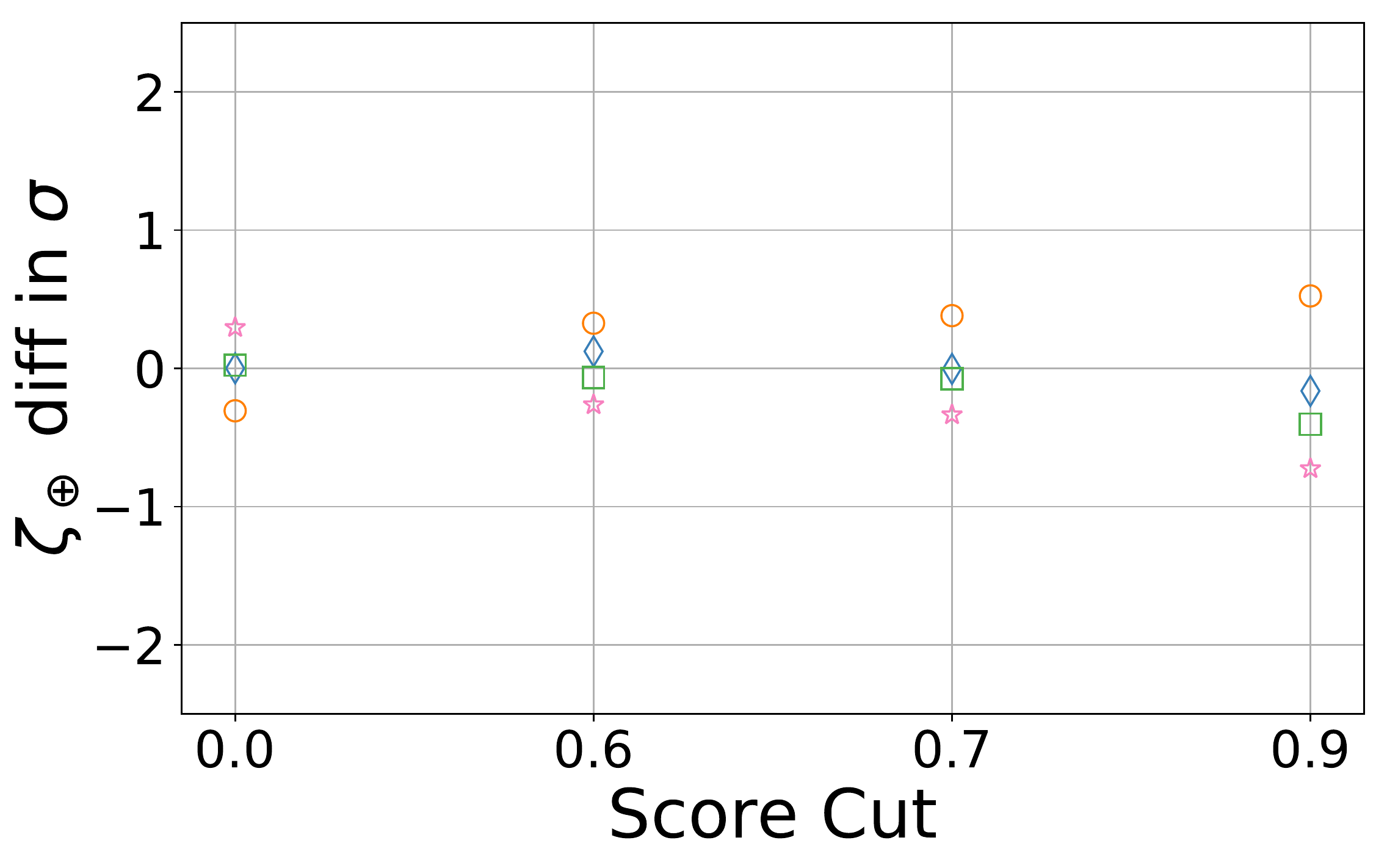} \\
  \includegraphics[width=0.46\linewidth]{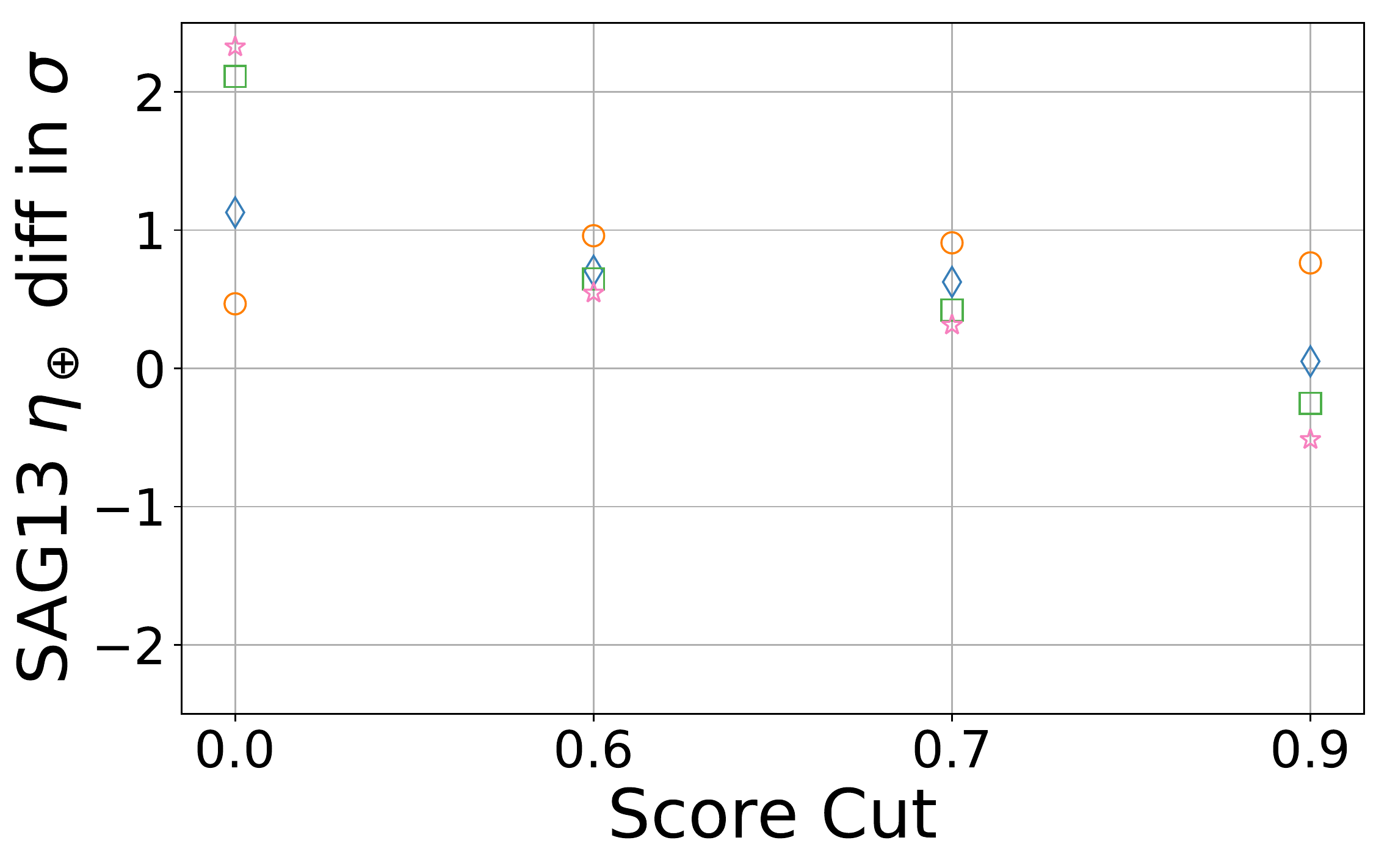}
  \includegraphics[width=0.46\linewidth]{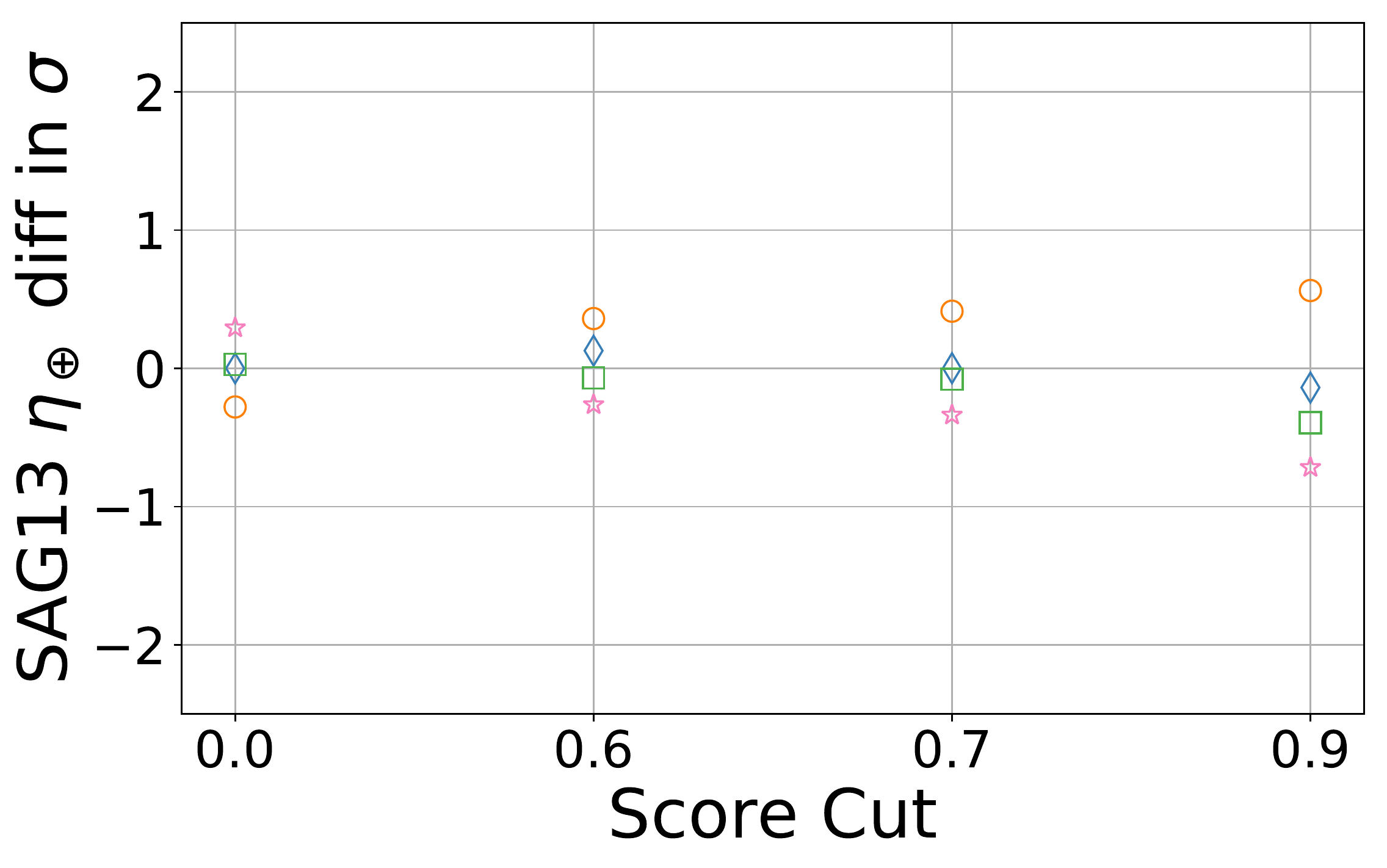}
   \caption{The difference in the medians of the catalogs from DR25 with no score cut, divided by the 1-sigma error in those differences for various occurrence rates and score cuts, computed with the Poisson method.  Left: without correcting for reliability.  Right: corrected for reliability.} \label{figure:scCompareSigma}
\end{figure*}

The results in Table~\ref{table:occurrenceResults} are shown graphically in Figure~\ref{figure:scCompare}.  In this figure we see that, though the medians are separated by less than the error bars, there is a consistent bias towards higher occurrence rates when not correcting for reliability compared to correcting for reliability except for a score cut of 0.9.  This is also seen in Figure~\ref{figure:scCompareSigma}, which shows the difference of each occurrence rate from the value for the DR25 catalog corrected for reliability without a score cut.  While this difference is less than $1 \sigma$, without reliability correction there is a consistent bias towards higher occurrence rates when using a score cut of 0.6 or 0.7.  When using a score cut of 0.9 without reliability correction, this bias disappears, but there is a larger spread of values across the catalogs, consistent with the maximum separation metric.  These biases disappear with reliability correction, implying that these biases are due to planets that have low reliability due to astrophysical false positive probability.

It is reasonable to conjecture that applying a score cut eliminates low-quality planet candidates, which would result in more accurate occurrence rates.  But as we have seen above, even with a high score cut there are low-reliability planet candidates due to astrophysical false positive probability, so an accurate occurrence rate requires reliability correction in any case.  For all score cuts considered above, score cut combined with reliability correction give results highly consistent with no score cut and reliability correction.  So there is no evidence that applying score cuts results in more accurate occurrence rates.


\renewcommand{\arraystretch}{1.25}
\begin{table*}[ht]
\centering
\caption{Occurrence rates using ABC}\label{table:occurrenceResultsABC}
\begin{tabular}{ r r r r r c}
\hline
\hline
         & \multicolumn{4}{c}{No Reliability Correction}  \\
Parameter & DR25 & High Reliability & High Completeness & FPWG PC & Max Separation $(\sigma)$ \\
\hline
$F_0$ & $0.596^{+0.092}_{-0.099}$ & $0.516^{+0.093}_{-0.080}$ & $0.818^{+0.144}_{-0.130}$ & $0.877^{+0.158}_{-0.138}$ & 2.18 \\
$\alpha$ & $0.440^{+0.524}_{-0.487}$ & $0.792^{+0.557}_{-0.571}$ & $-0.015^{+0.460}_{-0.438}$ & $-0.125^{+0.458}_{-0.453}$ & 1.25 \\
$\beta$ & $-0.562^{+0.154}_{-0.164}$ & $-0.663^{+0.167}_{-0.174}$ & $-0.278^{+0.166}_{-0.158}$ & $-0.252^{+0.177}_{-0.162}$ & 1.77 \\
$\Gamma_\oplus$ & $0.192^{+0.100}_{-0.070}$ & $0.132^{+0.074}_{-0.055}$ & $0.404^{+0.190}_{-0.136}$ & $0.463^{+0.219}_{-0.154}$ & 1.94 \\
$F_1$ & $0.185^{+0.031}_{-0.033}$ & $0.162^{+0.031}_{-0.028}$ & $0.229^{+0.038}_{-0.036}$ & $0.243^{+0.039}_{-0.035}$ & 1.71 \\
$\zeta_\oplus$ & $0.031^{+0.016}_{-0.011}$ & $0.021^{+0.012}_{-0.009}$ & $0.065^{+0.031}_{-0.022}$ & $0.074^{+0.036}_{-0.025}$ & 1.93 \\
SAG13 $\eta_\oplus$ & $0.272^{+0.155}_{-0.101}$ & $0.184^{+0.109}_{-0.075}$ & $0.635^{+0.348}_{-0.231}$ & $0.736^{+0.421}_{-0.263}$ & 1.94 \\
         & \multicolumn{4}{c}{With Reliability Correction}  \\
Parameter & DR25 & High Reliability & High Completeness & FPWG PC & Max Separation $(\sigma)$ \\
\hline
$F_0$ & $0.427^{+0.079}_{-0.065}$ & $0.415^{+0.076}_{-0.075}$ & $0.435^{+0.073}_{-0.070}$ & $0.464^{+0.076}_{-0.068}$ & 0.47 \\
$\alpha$ & $0.947^{+0.560}_{-0.610}$ & $1.149^{+0.592}_{-0.594}$ & $0.887^{+0.612}_{-0.498}$ & $0.777^{+0.536}_{-0.548}$ & 0.47 \\
$\beta$ & $-0.828^{+0.182}_{-0.161}$ & $-0.801^{+0.171}_{-0.181}$ & $-0.806^{+0.170}_{-0.173}$ & $-0.754^{+0.167}_{-0.177}$ & 0.29 \\
$\Gamma_\oplus$ & $0.086^{+0.056}_{-0.033}$ & $0.075^{+0.056}_{-0.031}$ & $0.090^{+0.053}_{-0.036}$ & $0.109^{+0.060}_{-0.040}$ & 0.49 \\
$F_1$ & $0.140^{+0.031}_{-0.025}$ & $0.131^{+0.030}_{-0.026}$ & $0.143^{+0.027}_{-0.027}$ & $0.152^{+0.030}_{-0.026}$ & 0.52 \\
$\zeta_\oplus$ & $0.014^{+0.009}_{-0.005}$ & $0.012^{+0.009}_{-0.005}$ & $0.015^{+0.008}_{-0.006}$ & $0.018^{+0.010}_{-0.006}$ & 0.49 \\
SAG13 $\eta_\oplus$ & $0.116^{+0.077}_{-0.045}$ & $0.103^{+0.076}_{-0.044}$ & $0.121^{+0.077}_{-0.047}$ & $0.148^{+0.087}_{-0.055}$ & 0.48 \\
\end{tabular}
\end{table*}

\begin{figure*}[ht]
  \centering
  \Large \hspace{0.2 in} No Reliability Correction \hspace{1 in} Corrected for Reliability\\
  \includegraphics[width=0.46\linewidth]{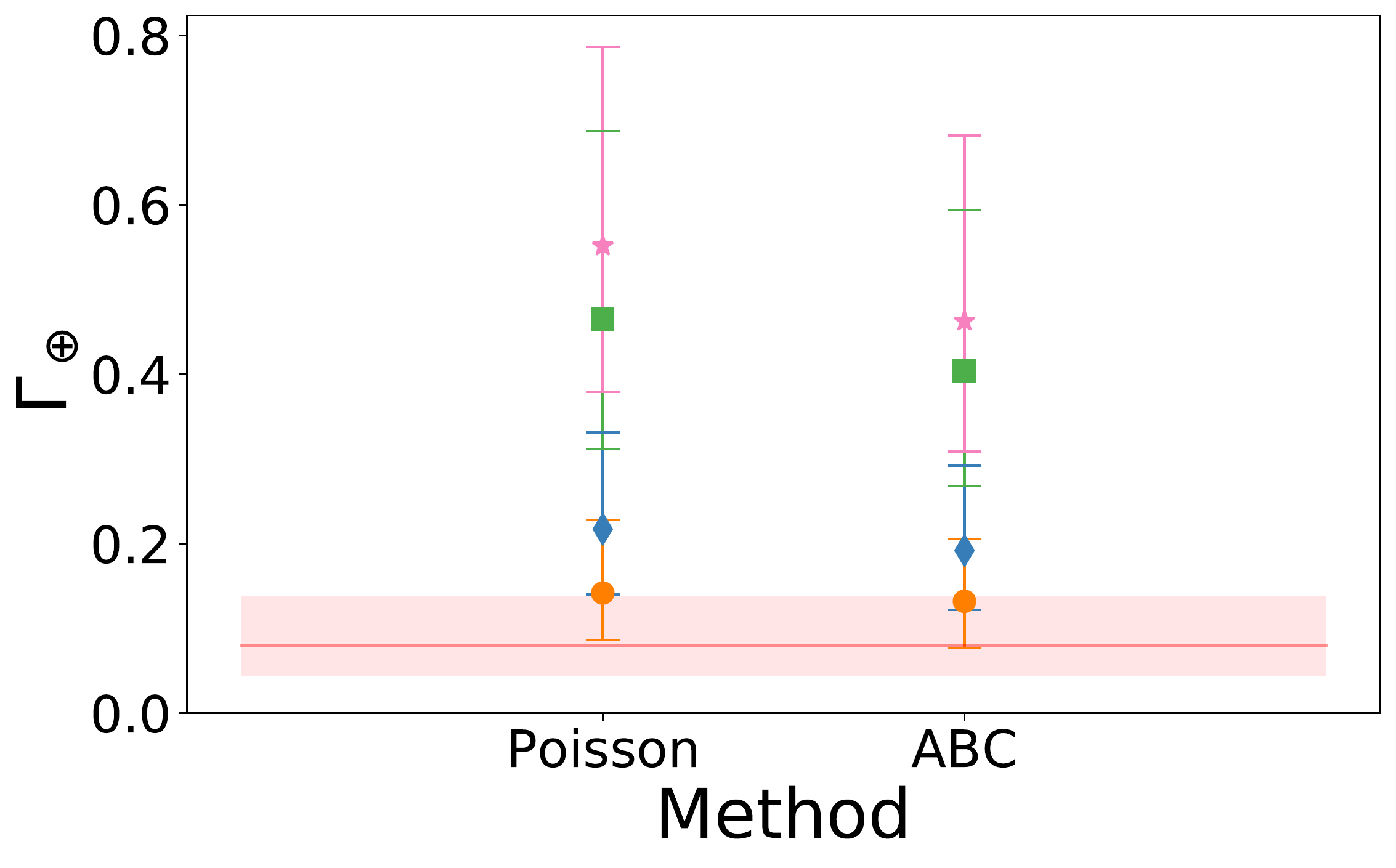} 
  \includegraphics[width=0.46\linewidth]{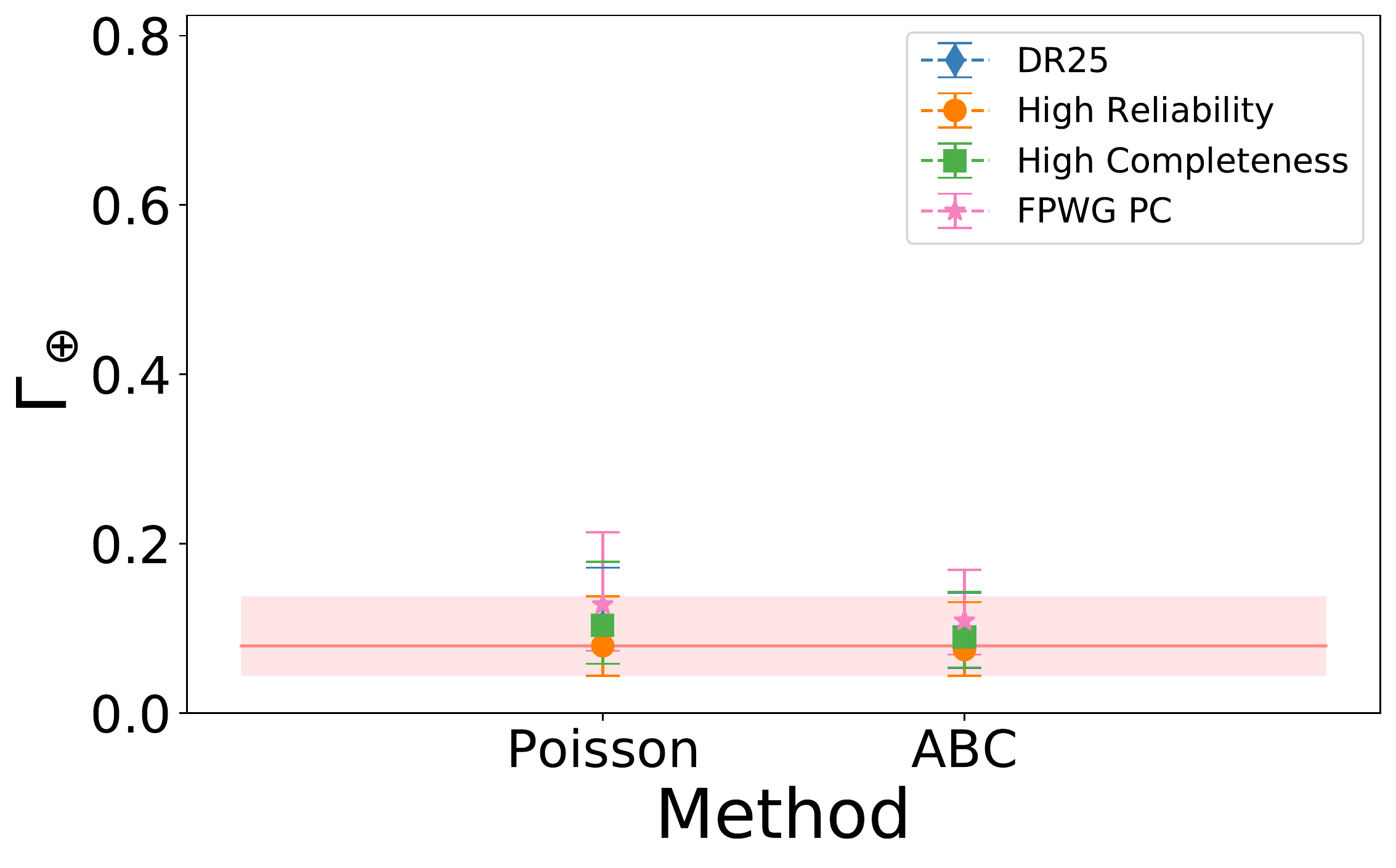} \\
  \includegraphics[width=0.46\linewidth]{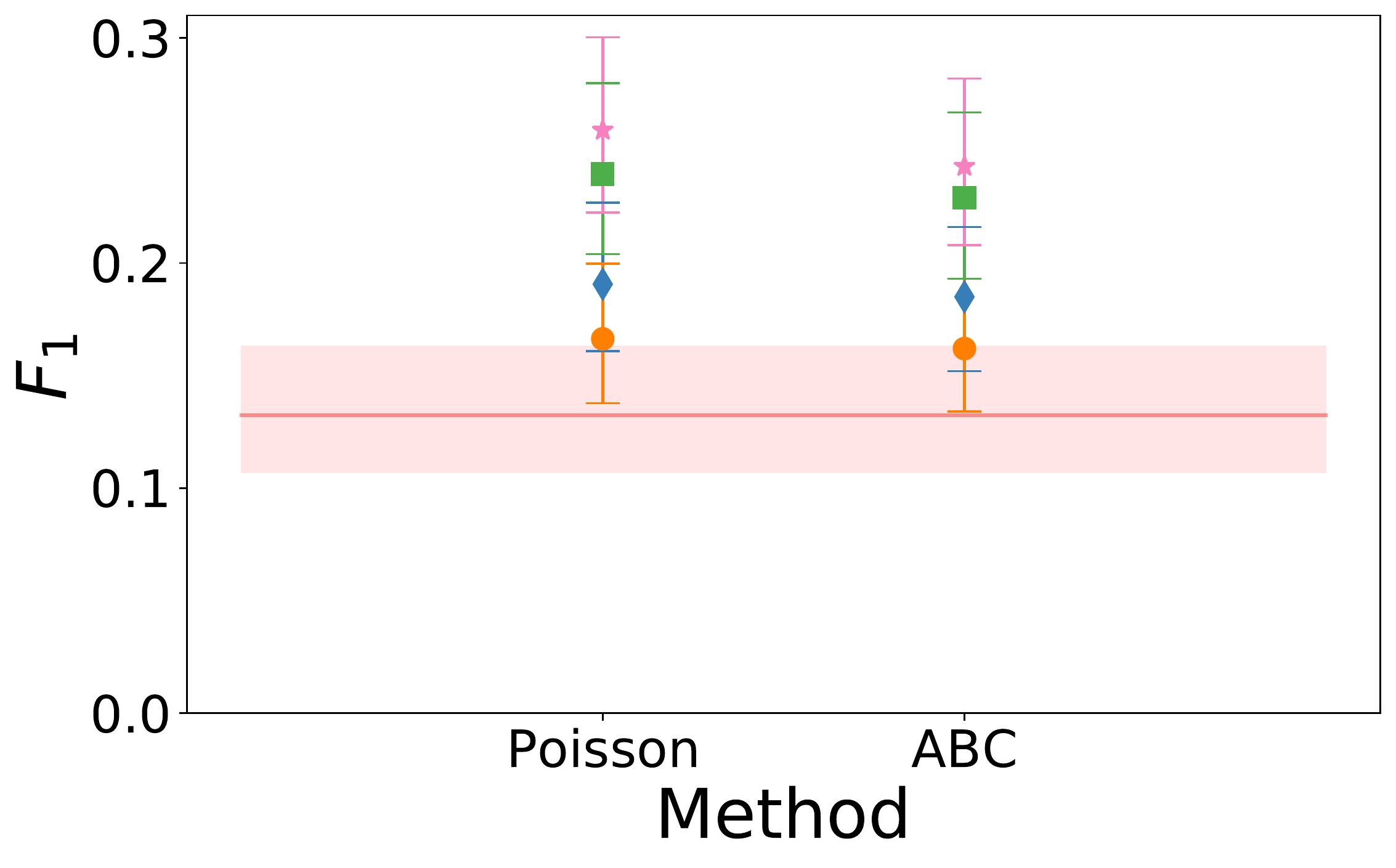}
  \includegraphics[width=0.46\linewidth]{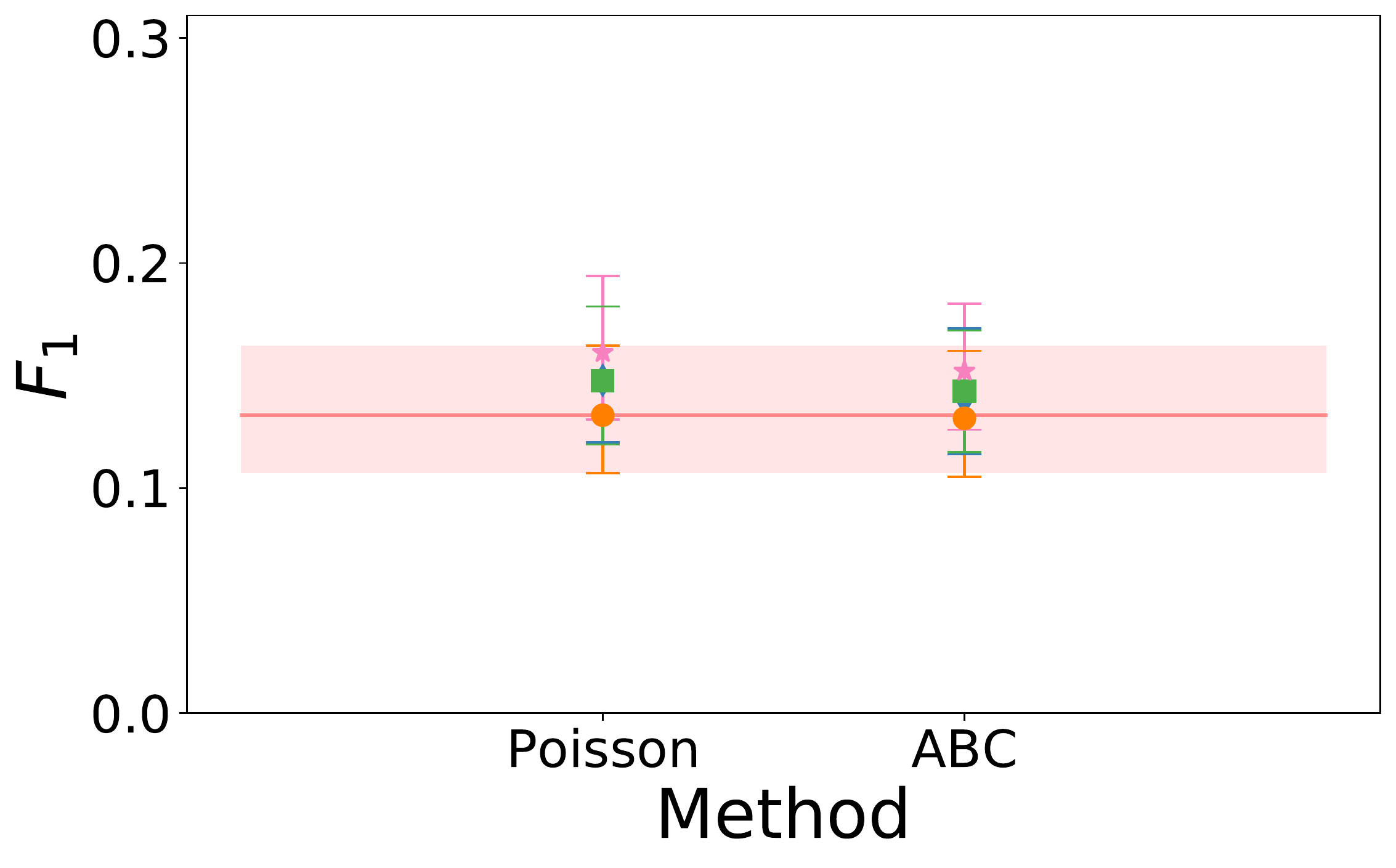} \\
  \includegraphics[width=0.46\linewidth]{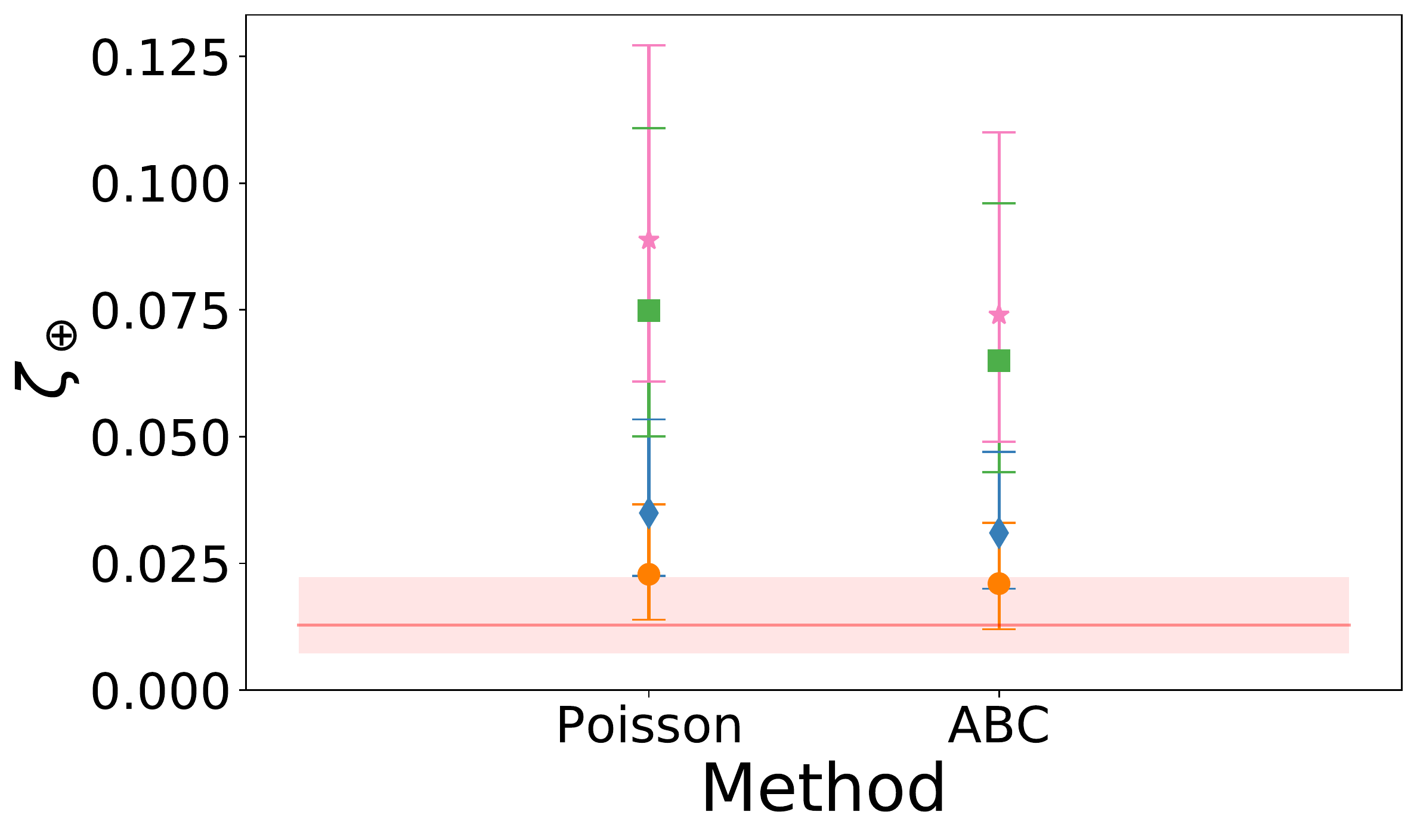}
  \includegraphics[width=0.46\linewidth]{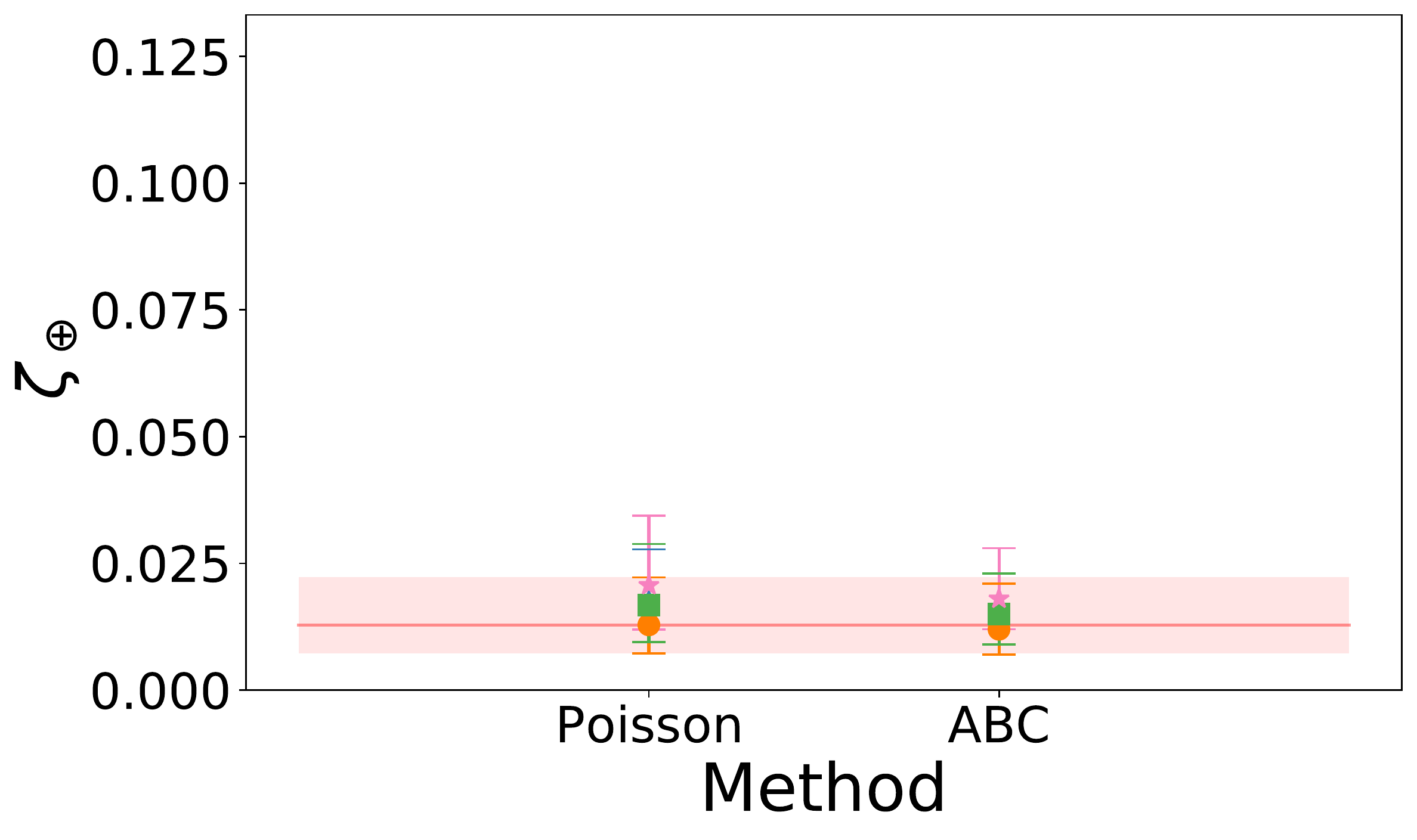} \\
  \includegraphics[width=0.46\linewidth]{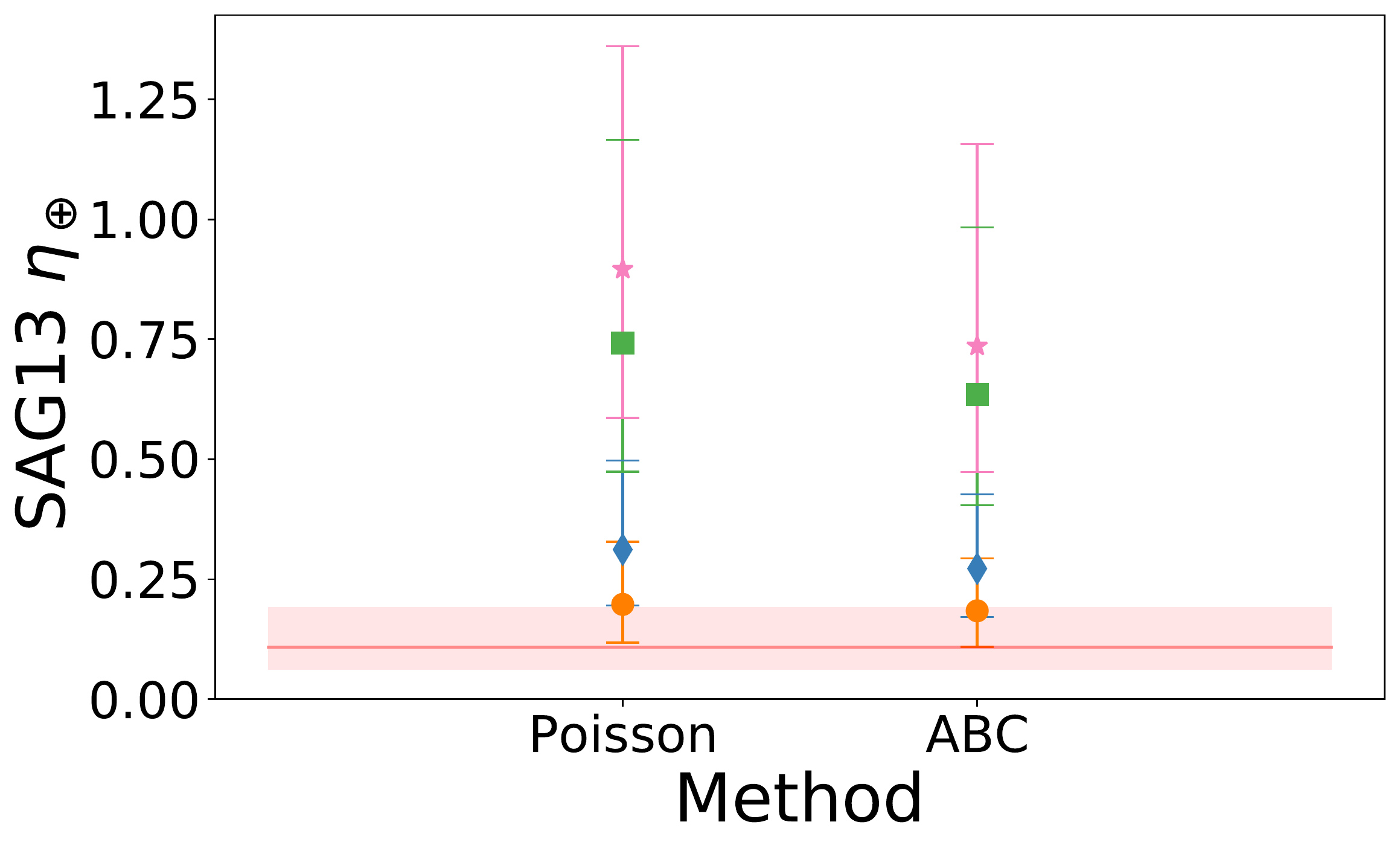}
  \includegraphics[width=0.46\linewidth]{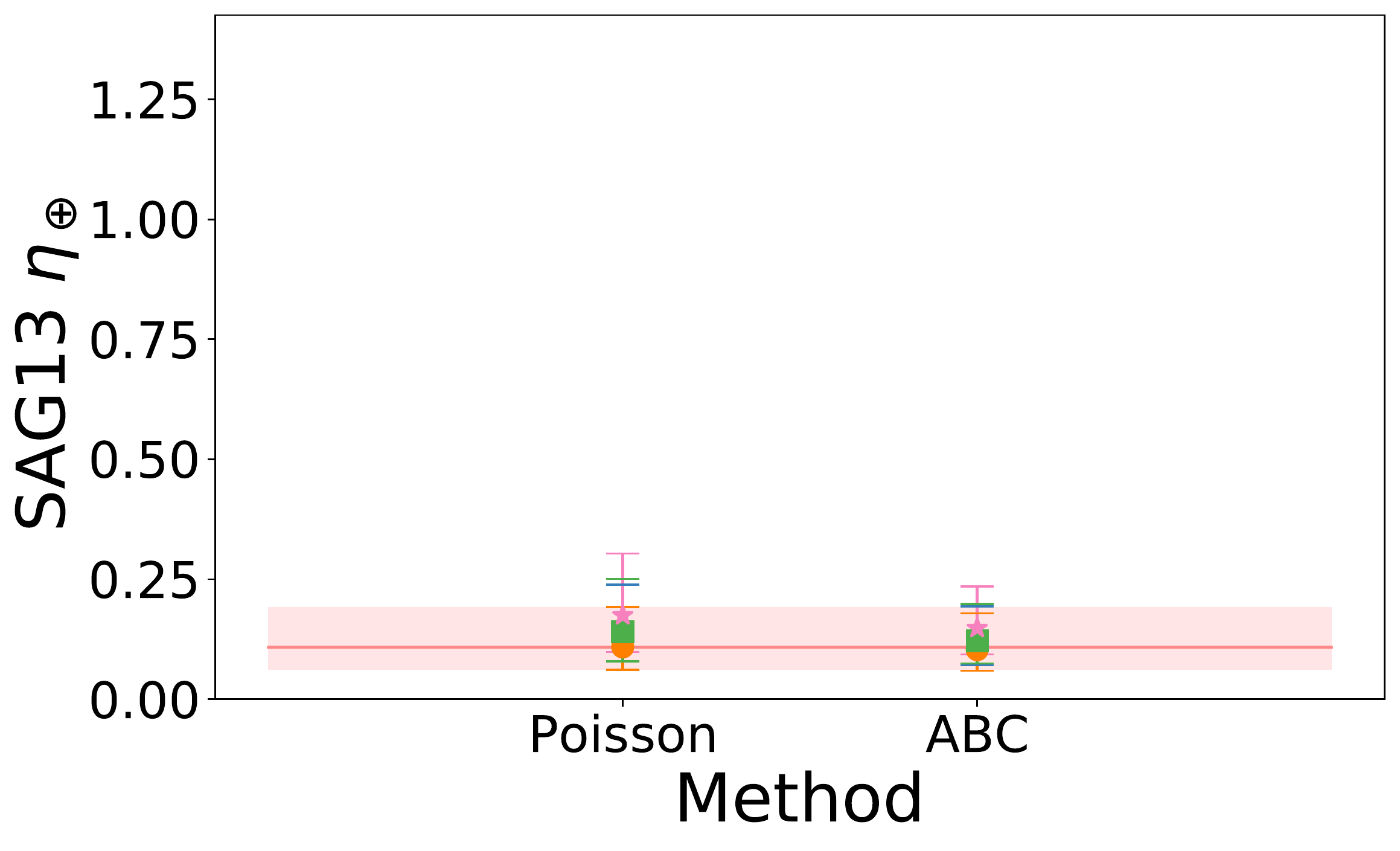}
   \caption{The median and 68\% confidence intervals for the four catalogs of various occurrence rates computed with the Poisson MCMC method and the ABC method.  Left: without correcting for reliability.  Right: corrected for reliability.  The red horizontal line and stripe are the median and 86\% confidence intervals for the DR25 catalog computed with the Poisson method.} \label{figure:scCompareABC}
\end{figure*}

\subsection{ABC Method}

We compute the rate function coefficients $\boldsymbol{\theta}$ using the ABC method described in \S\ref{section:occMethods} for the four catalogs and no score cut.  The results, including occurrence rates, are given in Table~\ref{table:occurrenceResultsABC}, and the occurrence rates are compared with those from the Poisson likelihood method in Figure~\ref{figure:scCompareABC}.  The results are very consistent with the Poisson likelihood method with no score cut, exhibiting the large variation of results from the different catalogs when not correcting for reliability and strong consistency when correcting for reliability. This is consistent with the results in \citet{kunimoto2020b}.  It is notable that the error bars using ABC are somewhat smaller than those from the Poisson likelihood.

\section{Discussion} \label{section:discussion}

In this paper we presented four planet candidate catalogs created from \Kepler\ DR25 detections and vetting metrics via the \Kepler\ Robovetter following \citet{Thompson2018}.  Each catalog uses different choices of vetting thresholds, chosen for differing balances between completeness and reliability.  The specific vetting thresholds for each catalog are equally reasonable and defensible, so each catalog can be considered as a legitimate, though imperfect, planet candidate catalog.  Therefore, occurrence rate measurements using these catalogs should provide consistent results.  By applying the Robovetter using these thresholds to observed, injected, inverted, and scrambled data we characterized the completeness and reliability of each catalog using the techniques of \citet{Bryson2020}.  We find that if we do not correct for reliability, occurrence rate estimates using these catalogs vary widely.  For example, comparing the range of distributions in the left panel of $\Gamma_\oplus$ in Figure~\ref{figure:allDist} with Figure 14 in \citet{Kunimoto2020a}, we see a range of values similar to that found in the literature.  When we correct for reliability, on the other hand, occurrence rate estimates for the various catalogs become very consistent, with a spread of medians well under $1 \sigma$.  {\bf This shows that correction for completeness and reliability is critical for robust occurrence rates and strongly suggests that the simulated false alarms in the inverted/scrambled data are a statistically accurate representation of the true false alarm population.}  

Using the criterion of consistent results with the four catalogs, we investigated the use of Robovetter disposition score cuts as a method of correcting for reliability.  We found that score cuts, which remove those planet candidates with score less than a threshold value, significantly improves the consistency of occurrence rates from the four catalogs without reliability correction, compared to not correcting for reliability with no score cut.  Using a score cut without reliability correction can produce results from the four catalogs consistent with correcting for reliability without a score cut.  However, using a score cut without reliability correction results in occurrence rates that are somewhat biased towards high occurrence rates relative to those with reliability correction.  This bias is removed when using a score cut and reliability correction, and implementing a score cut with reliability correction yields the same result as reliability correction without score cut.  Therefore we recommend always correcting for reliability when possible without a score cut, because score cuts remove data without providing any advantage.  If correcting for reliability is not possible, then a score cut is a reasonable alternative, but will give less accurate results.

We found that the above behavior of the occurrence rate calculation is the same for both the Poisson likelihood and Approximate Bayesian Computation methods, in spite of the dramatic difference in these method's treatment of completeness, reliability, and the statistics of the planet population model.

The occurrence rates presented in this paper are, like those in \citet{Bryson2020}, illustrative.  In particular, the occurrence rates $\zeta_\oplus$ and the SAG13 $\eta_\oplus$ involve significant extrapolation beyond where there is a significant amount of data.  We therefore treat these occurrence rates with some skepticism, though it is remarkable how robust these occurrence rates are against variations in Robovetter vetting thresholds, Robovetter disposition score cuts, and the Bayesian inference method. 

There are at least three aspects of the occurrence rate calculations presented in this paper that may compromise accuracy:
\begin{itemize}
    \item {\bf Incorrect population model.}  The product of independent power laws in period and radius in Equation~(\ref{eqn:powerLaw}) may not correctly describe the planet population.  There is ample evidence that exoplanet populations are significantly more complex \citep{Fulton2017, Mulders2019, Pascucci2019}, and may not be well-described by simple broken power laws.  Incorrect population models can lead to large inaccuracies when extrapolated as we do for $\zeta_\oplus$ and the SAG13 $\eta_\oplus$.
    \item {\bf Not accounting for planet multiplicity.}  \citet{zinkChristiansen2019} showed that the existence of short-period transiting planets can inhibit the detection of longer-period transiting planets on the same star. They estimate that longer-period occurrence rates may be as much as 16\% higher on individual stars that have short-period transits after correction for the impact of planet multiplicity on detection completeness.  However, it is difficult to correct for this effect in our methods, which rely on a uniform completeness characterization across the parent stellar population.
    \item {\bf Incomplete vetting metrics.}  We use the vetting metrics of \citet{Thompson2018}.  While these metrics are remarkably thorough, they do not fully exploit the \Kepler\ data.  For example, further vetting metrics based on pixel data can help distinguish astrophysical signals from instrumental artifacts.  Such metrics could potentially yield catalogs that are both more complete and more reliable, which may result in different, theoretically more accurate occurrence rates.
\end{itemize}
We expect that the robustness demonstrated in this paper would still apply with improved population models and vetting metrics.

Strictly speaking, the DR25 catalog is a catalog of objects that pass the Robovetter with a specific set of metrics.  We strongly believe that this catalog, when used with the associated measures of the completeness and reliability, provides a high-quality measurement of the true transiting planet population.  Using the same metrics but changing the metric thresholds as described in \S\ref{section:thresholdVariation} provide slightly different measurements of the same population, so we expect the resulting slightly different catalogs to be statistically consistent with each other. Adding different metrics, on the other hand, may potentially measure different populations of transiting planets and false alarms/positives, yielding a catalog that more closely matches the true underlying population.  Therefore new vetting metrics may yield statistically different occurrence rates which may be closer to the true value.


\section{Conclusions}
In this paper we explored the impact of using several alternative planet candidate catalogs derived from \Kepler\ data on exoplanet occurrence rates.  We find statistically consistent occurrence rates using these catalogs, so long as we correct for catalog completeness and reliability.  Ignoring reliability, however, results in statistically inconsistent occurrence rates between the catalogs.  This implies that a) completeness and reliability correction is necessary for accurate occurrence rates and b) the completeness and reliability of these catalogs are correctly statistically measured using injected, inverted, and scrambled data.  In particular, the false alarms in the injected and inverted data statistically represent the false alarms in the observed \Kepler\ data.  This result is independent of the computational method. 

We make the four planet catalogs we use, as well as the data required for their completeness and reliability characterization publicly available.  We recommend that other occurrence rate methods be tested using these catalogs to demonstrate that they yield statistically consistent results.  

This paper illustrates the importance of correcting for both completeness and reliability when performing demographic studies.  This lesson surely applies to any survey whose catalogs are incomplete and not fully reliable.  Our ability to characterize catalog completeness and reliability depends on being able to create our catalogs in a uniform and repeatable way, so the same catalog inclusion criteria can be applied to both observed and ground-truth data.  The ground-truth data must statistically represent both true and false detections.  

Large-scale transit surveys such as K2, TESS, and PLATO, as well as RV and microlensing surveys, provide wonderful opportunities for a deeper understanding of exoplanet demographics.  These surveys will require a similar ability to characterize their completeness and reliability in order to provide high-quality results.

\acknowledgments
We thank NASA, \Kepler\ management, and the Exoplanet Exploration Office for continued support of and encouragement for the analysis of \Kepler\ data. Gijs Mulders provided helpful comments.  We thank Bill Borucki and the \Kepler\ team for the excellent data that makes these studies possible.

\vspace{5mm}
\facilities{\Kepler}

\software{Python, Jupyter}

\clearpage

\appendix

\section{Robovetter vetting thresholds} \label{appendix}

\begin{figure*}[ht]
  \centering
  \includegraphics[width=0.48\linewidth]{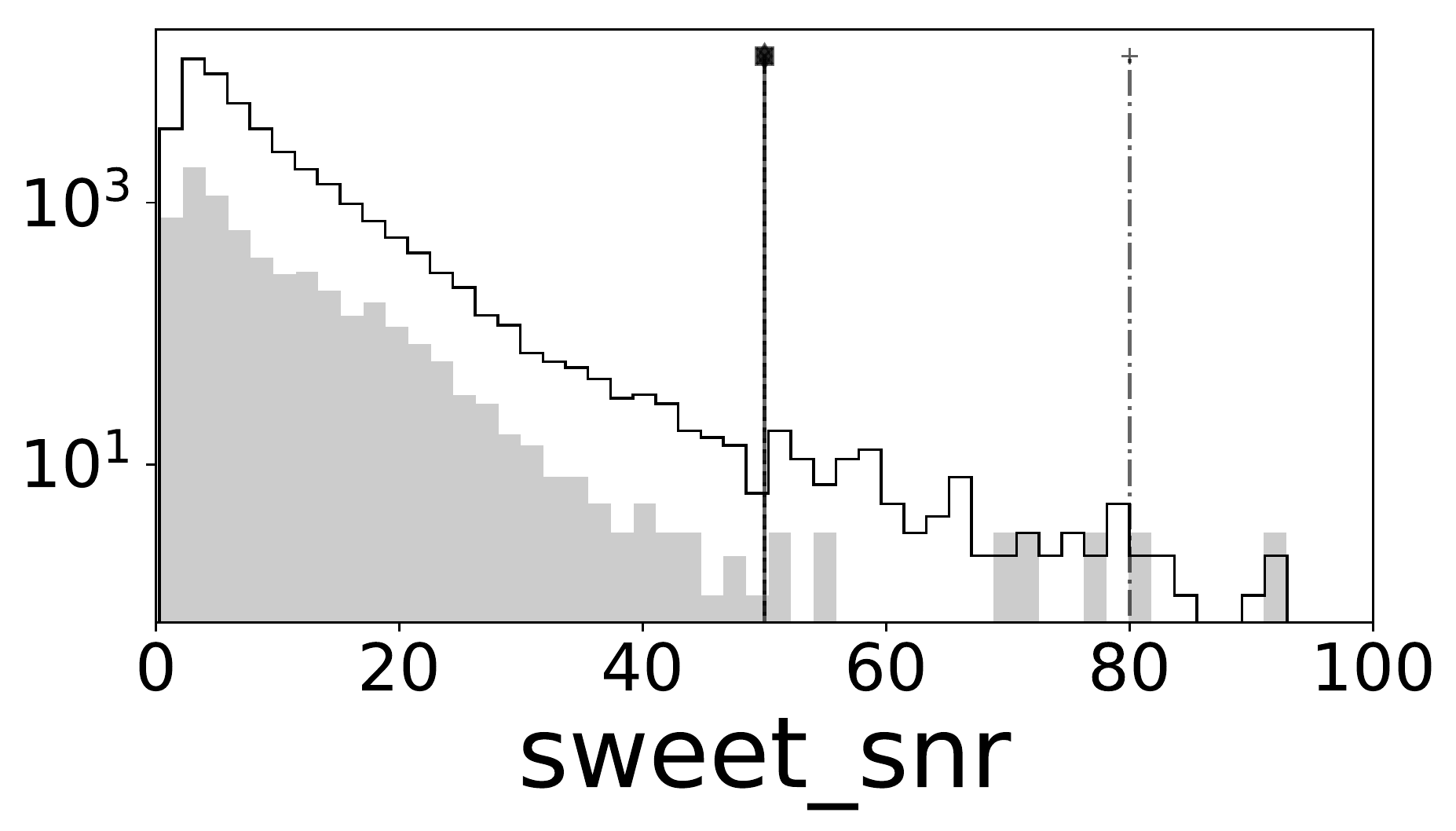} 
  \includegraphics[width=0.48\linewidth]{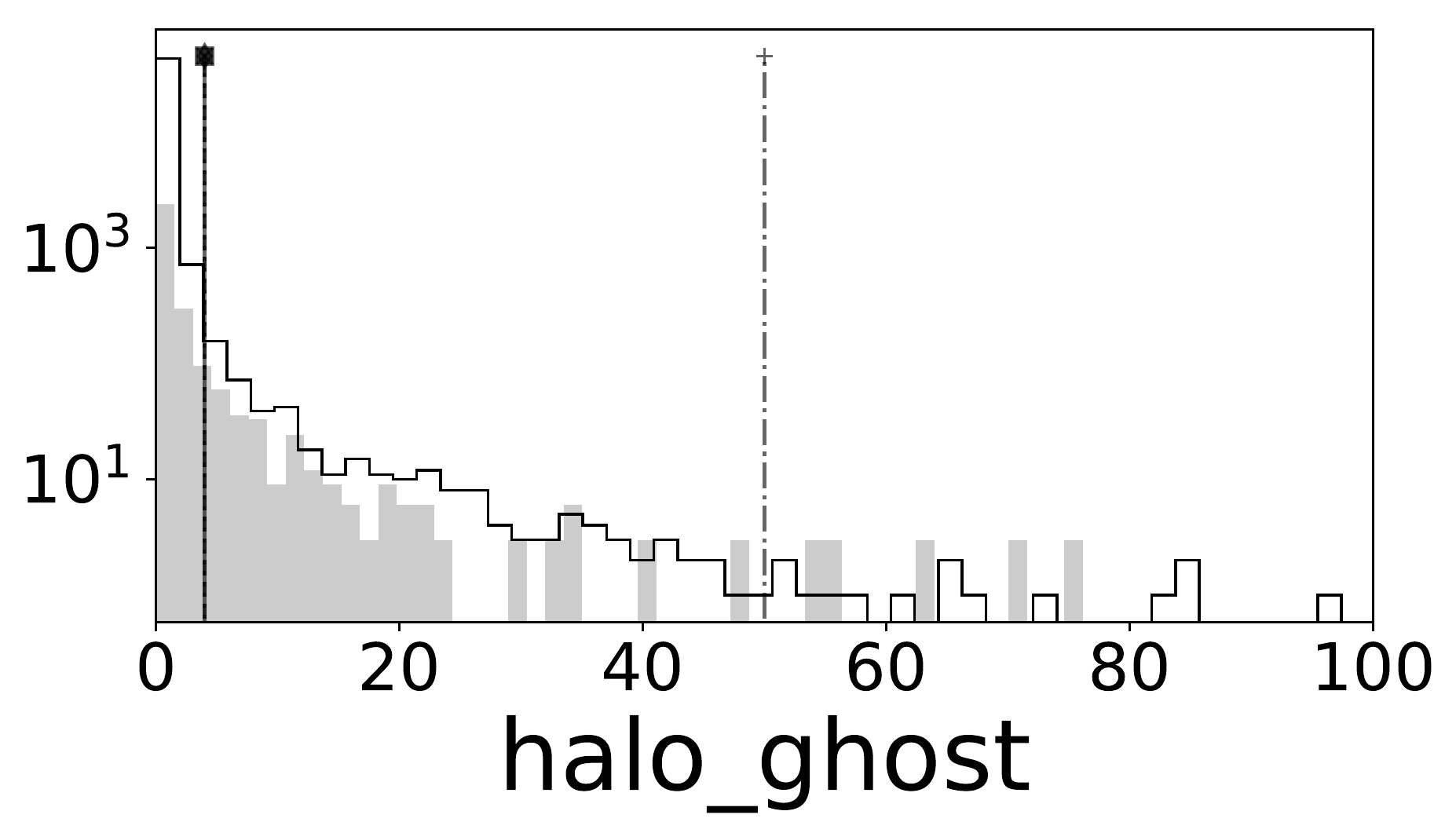} \\
  \includegraphics[width=0.48\linewidth]{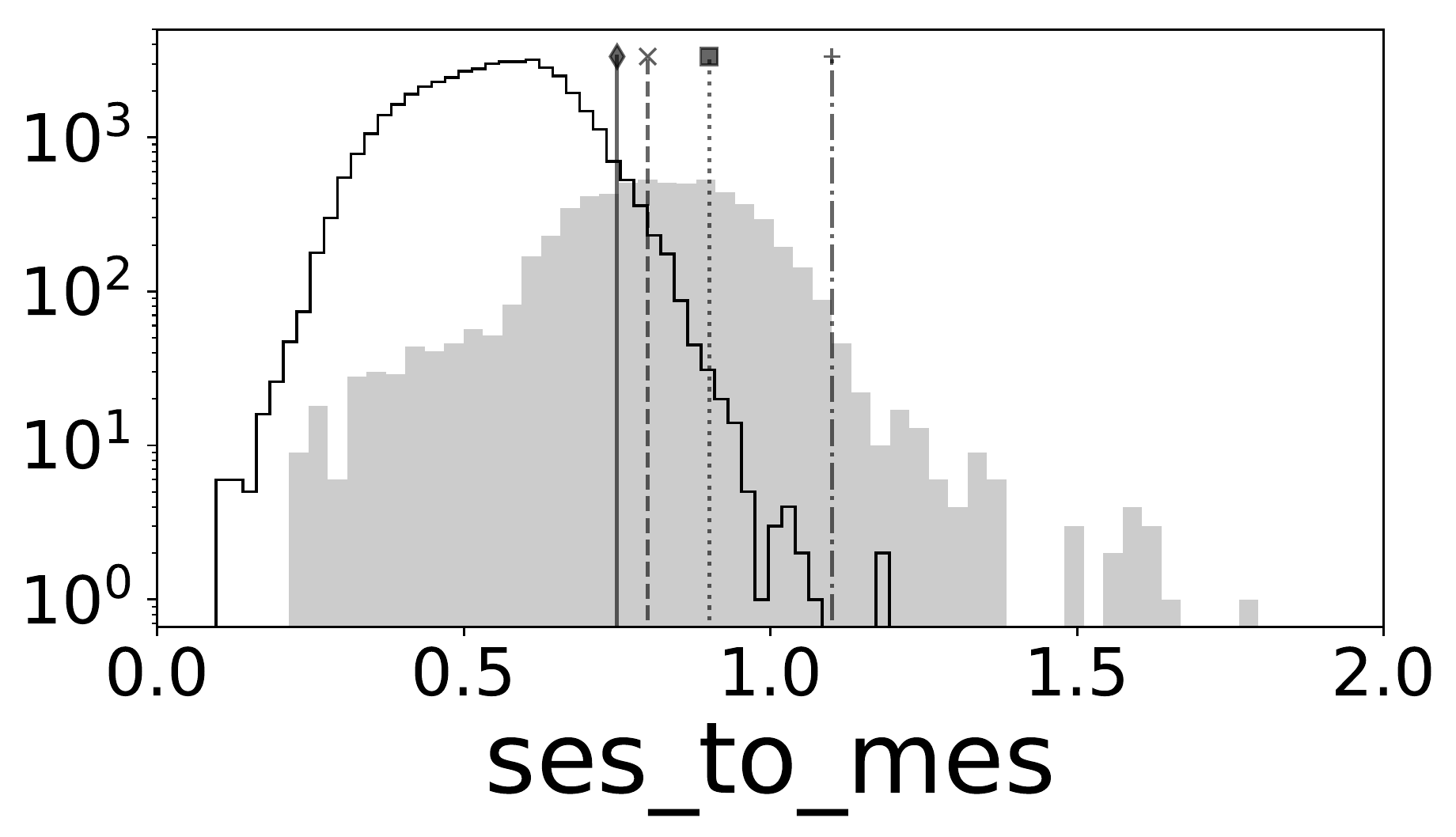} 
  \includegraphics[width=0.48\linewidth]{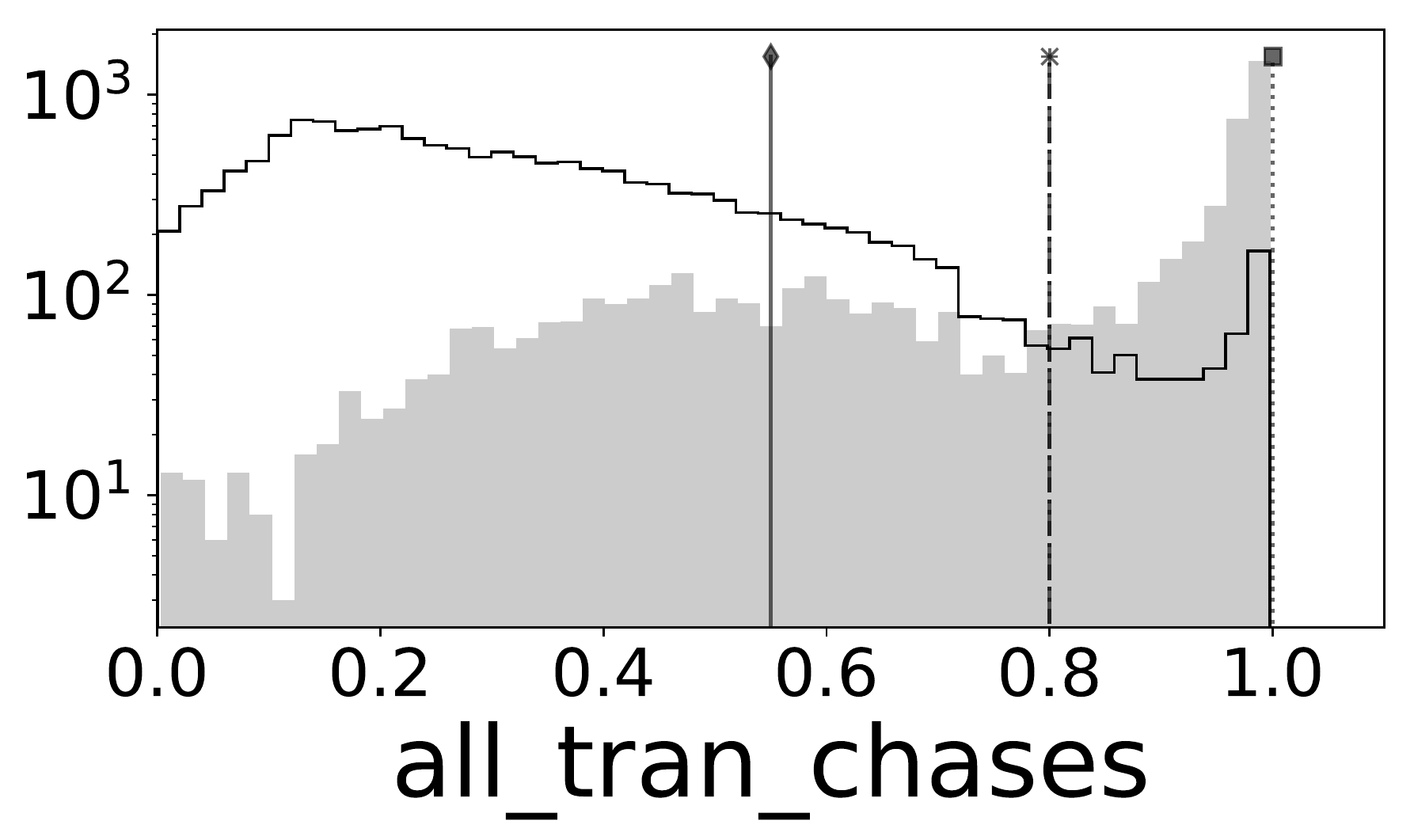} \\
  \includegraphics[width=0.48\linewidth]{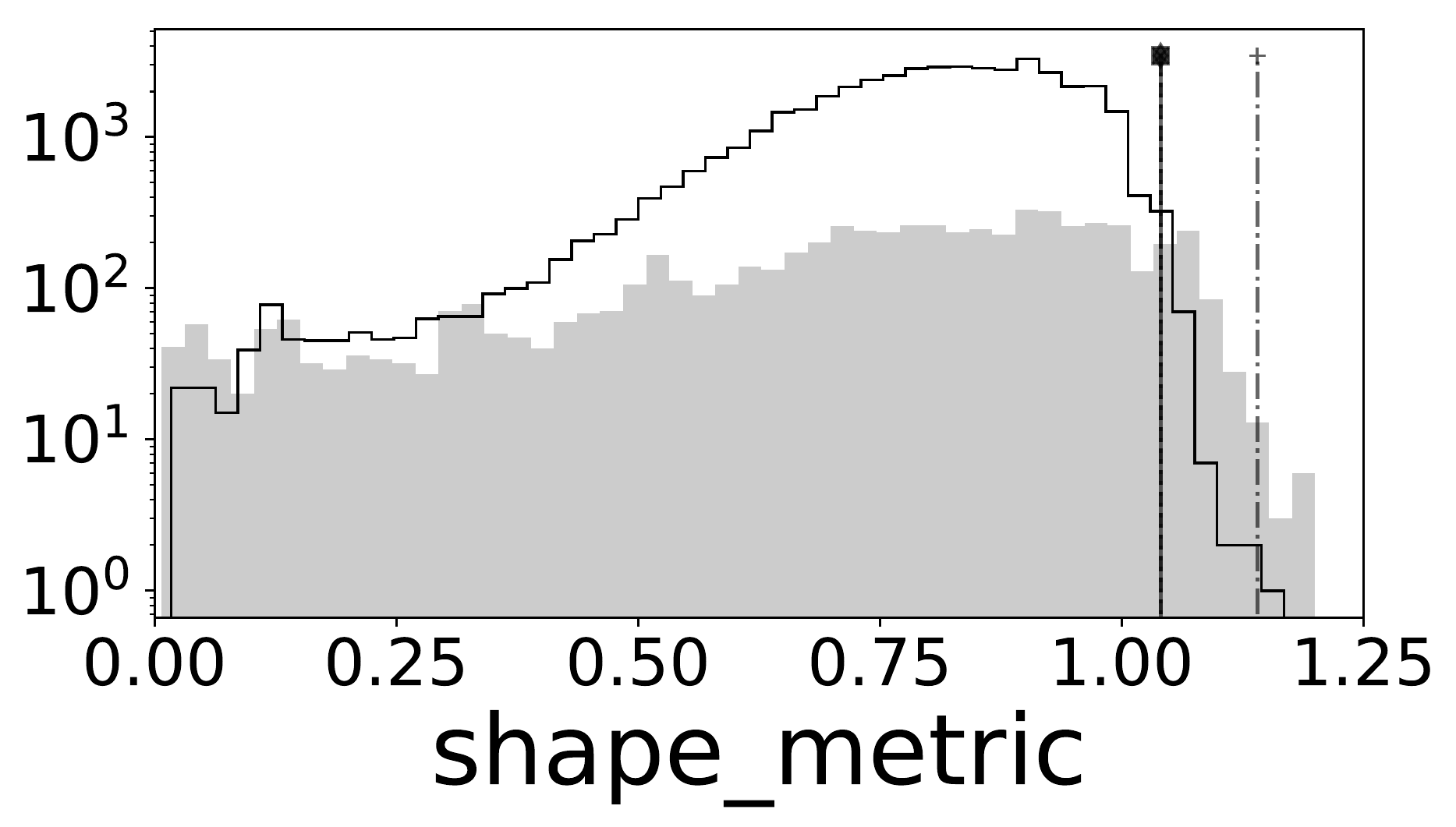} 
  \includegraphics[width=0.48\linewidth]{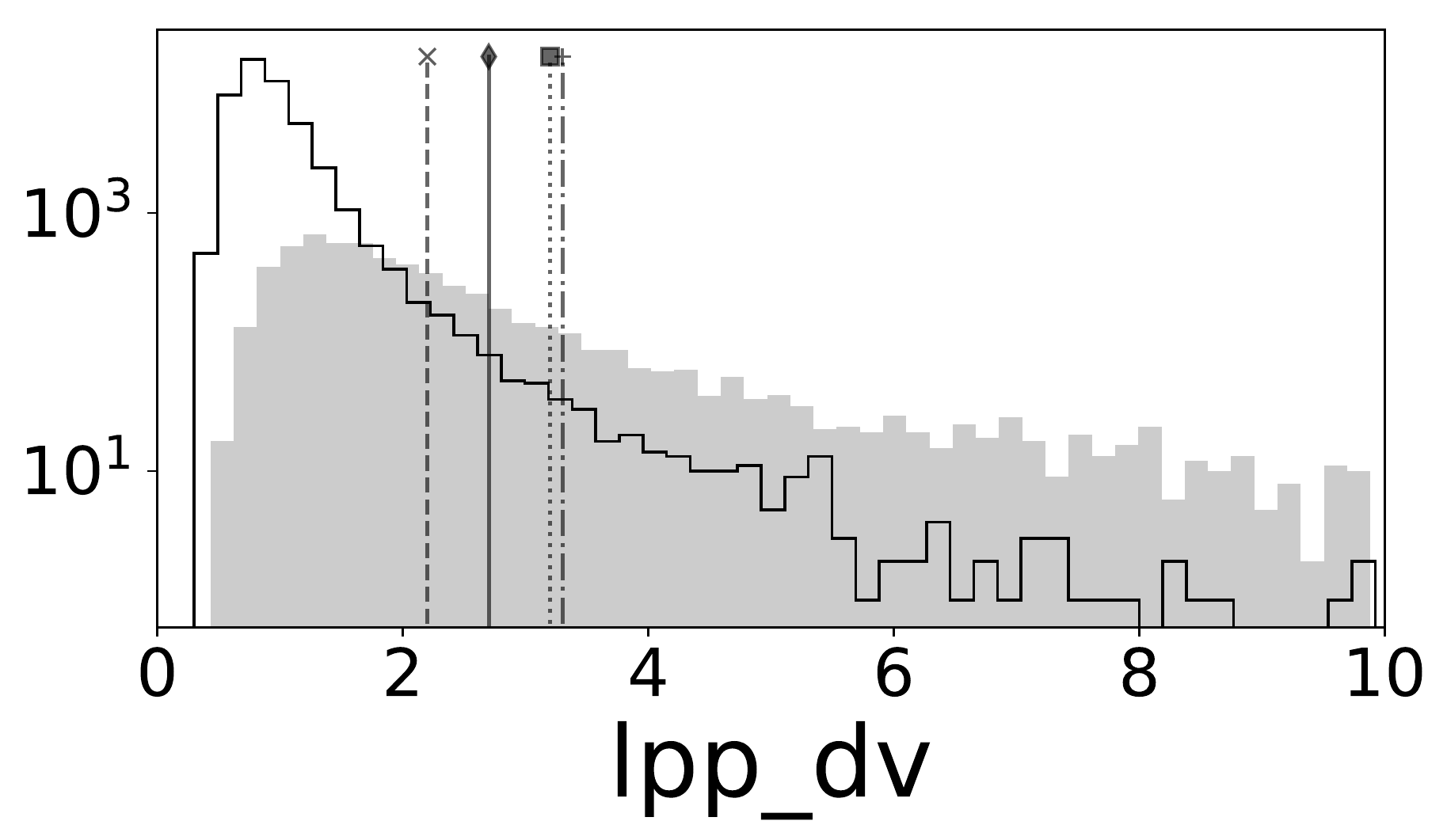} \\
  \includegraphics[width=0.48\linewidth]{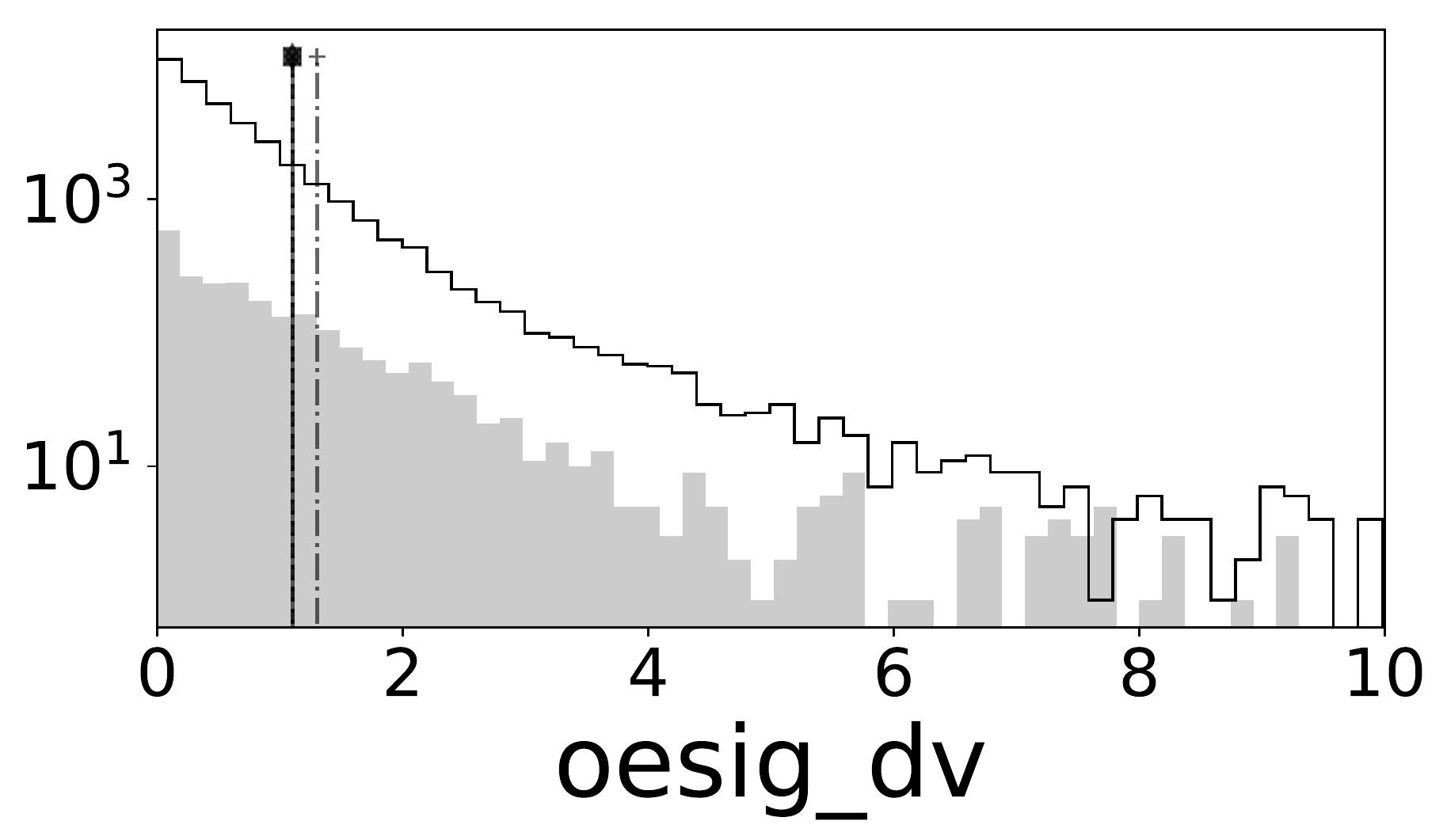} 
  \includegraphics[width=0.48\linewidth]{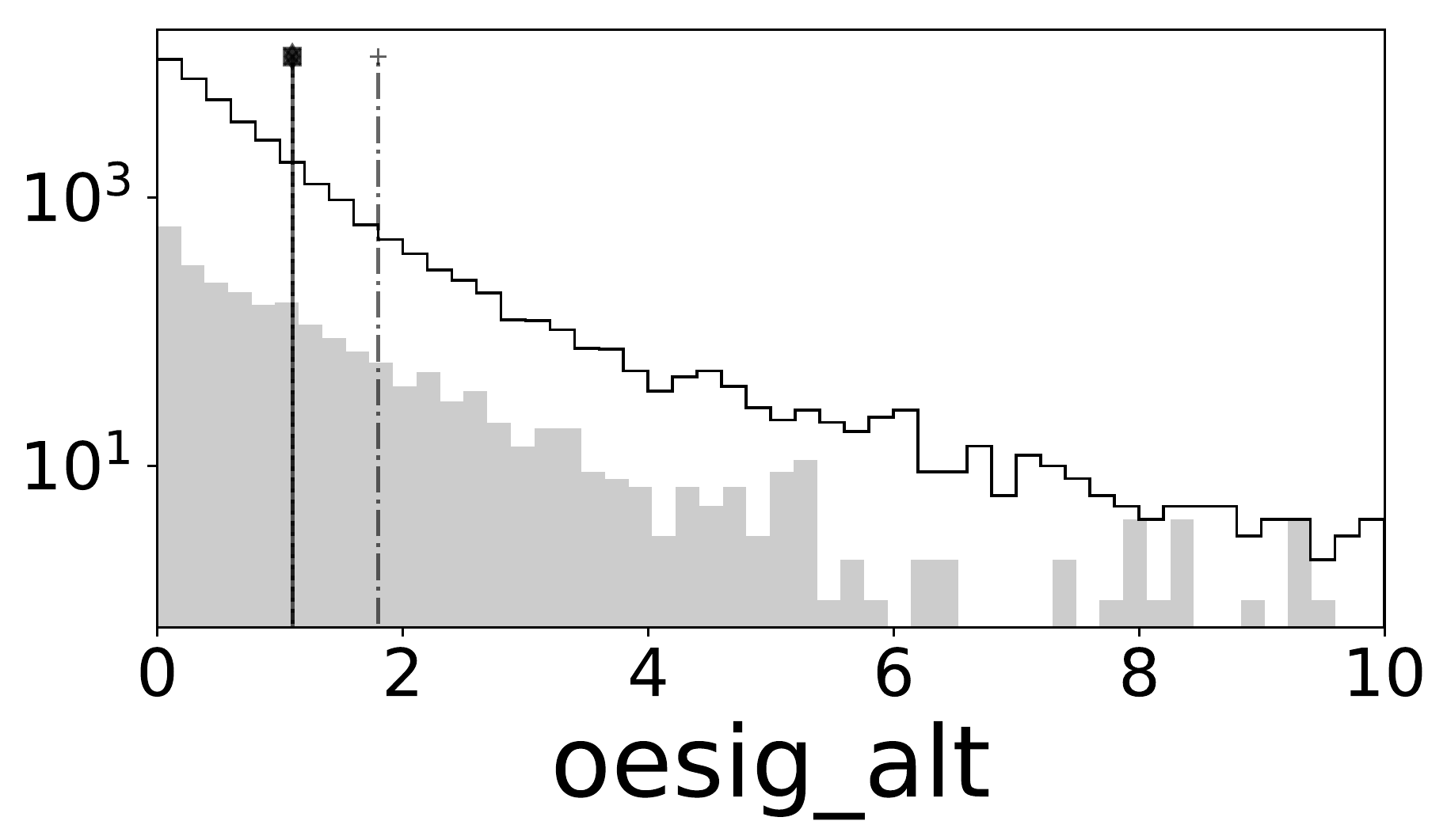} \\
   \caption{Robovetter metrics and thresholds.  Shaded histogram: metric distribution for false alarms from the inverted and scrambled data.  Line histogram: metric distribution for true transits from the injected data.  The thresholds given in Table~\ref{table:roboetterThresholds} are shown by the vertical lines: diamond solid line: high reliability; 'x' dashed line: DR25; square dotted line: high completeness; '+' dot-dashed line: FPWG PCs.} \label{figure:rvMetrics1}
\end{figure*}

\begin{figure*}[ht]
  \centering
  \includegraphics[width=0.48\linewidth]{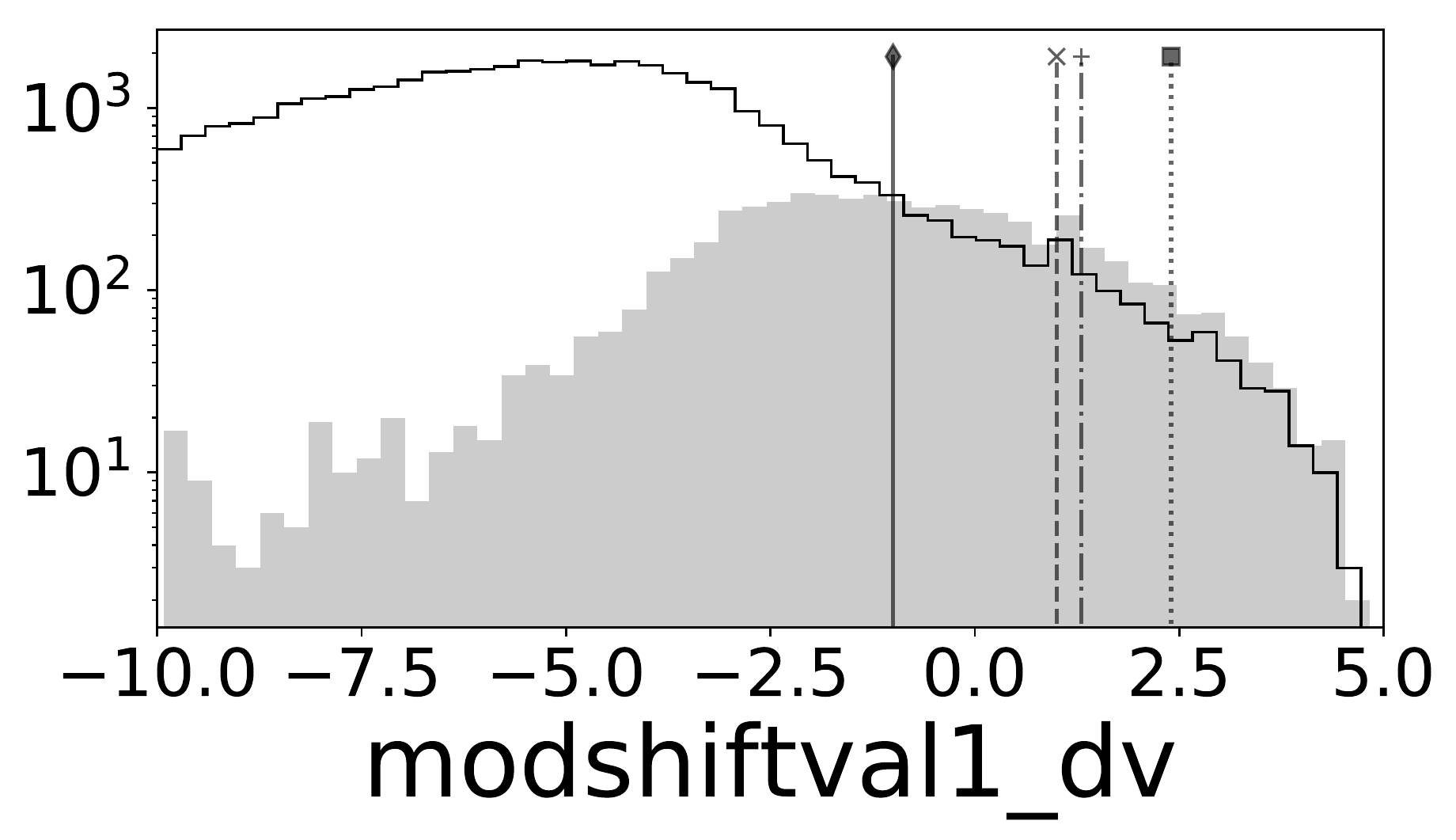} 
  \includegraphics[width=0.48\linewidth]{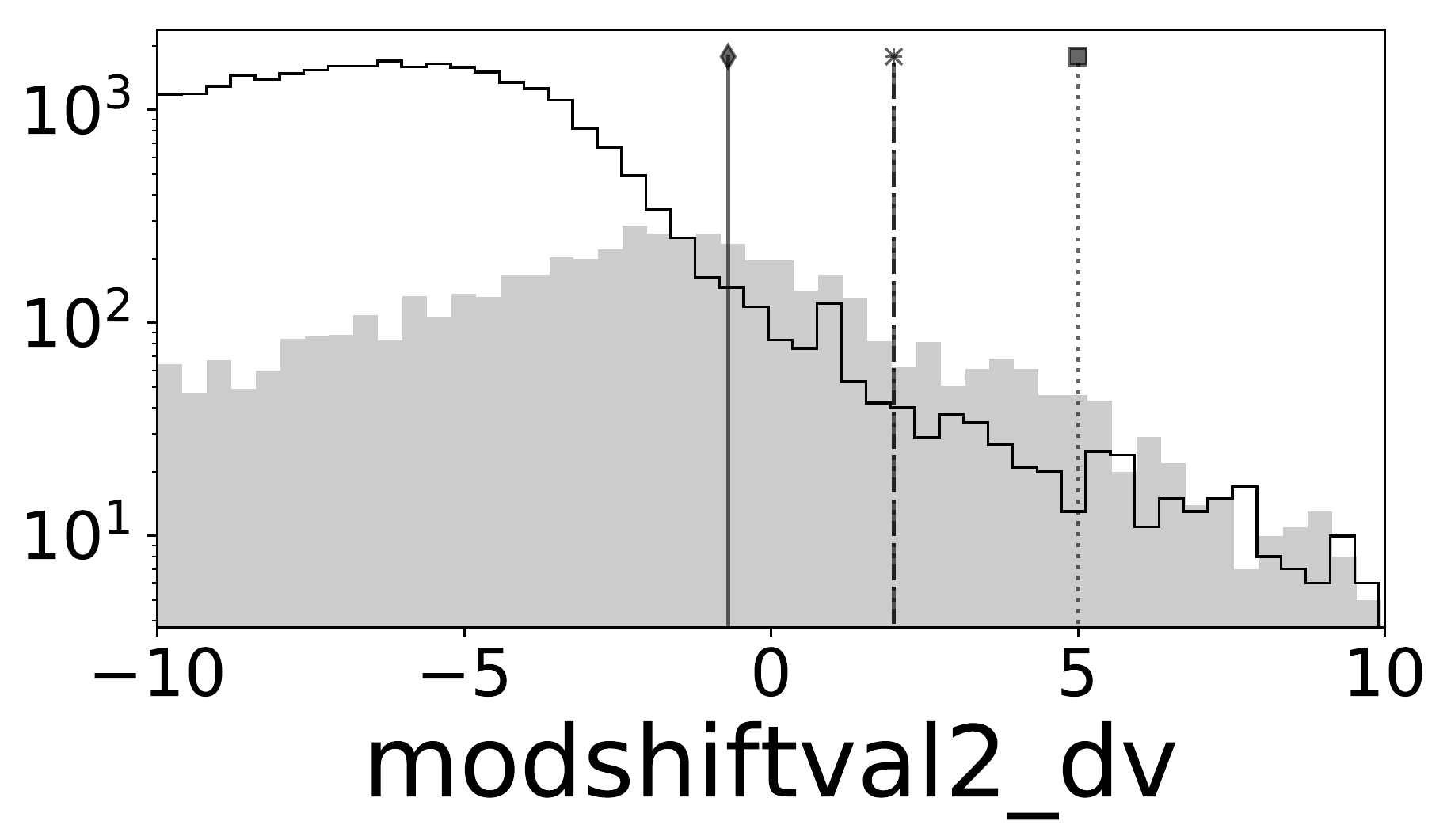} \\
  \includegraphics[width=0.48\linewidth]{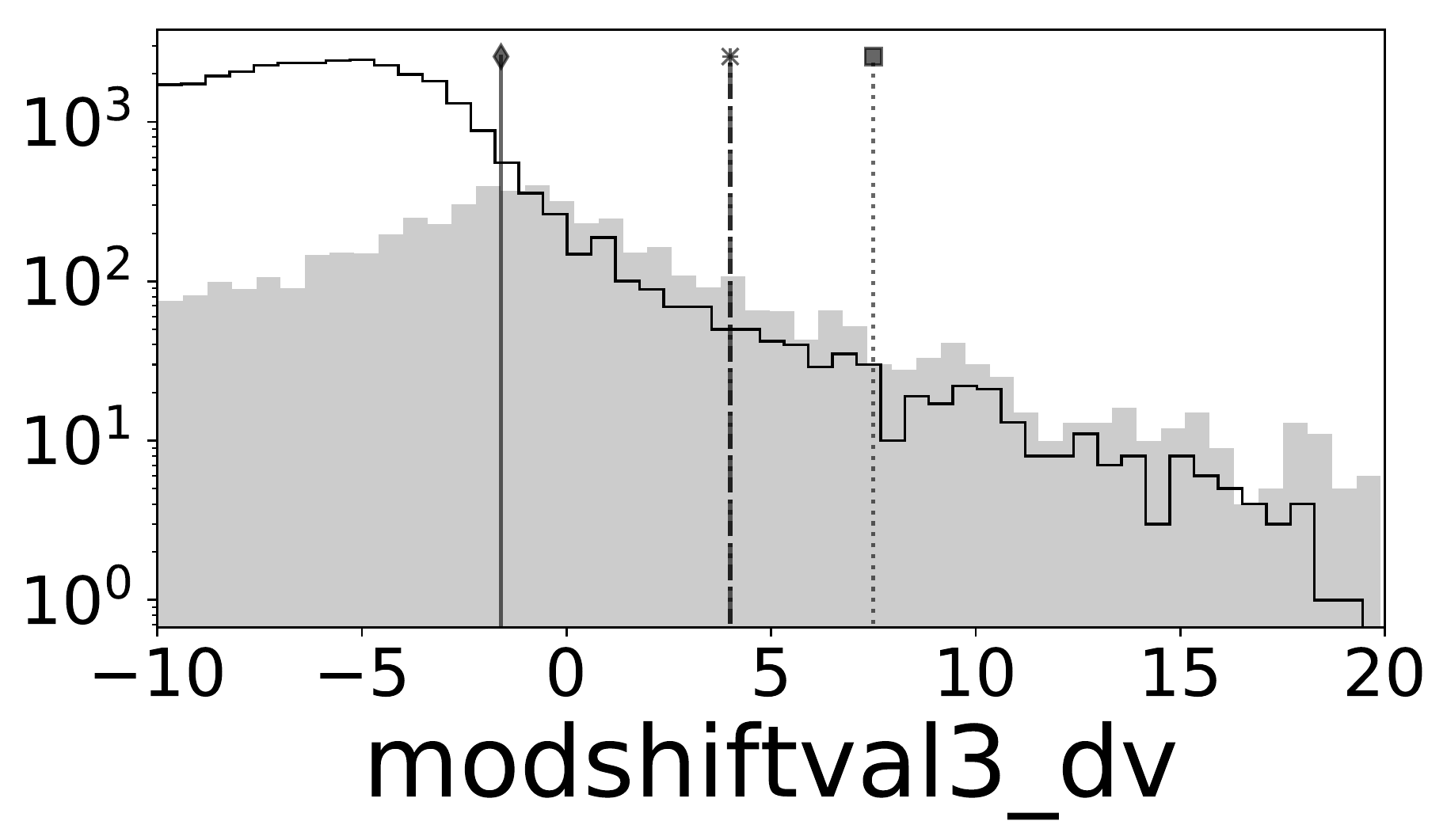} 
  \includegraphics[width=0.48\linewidth]{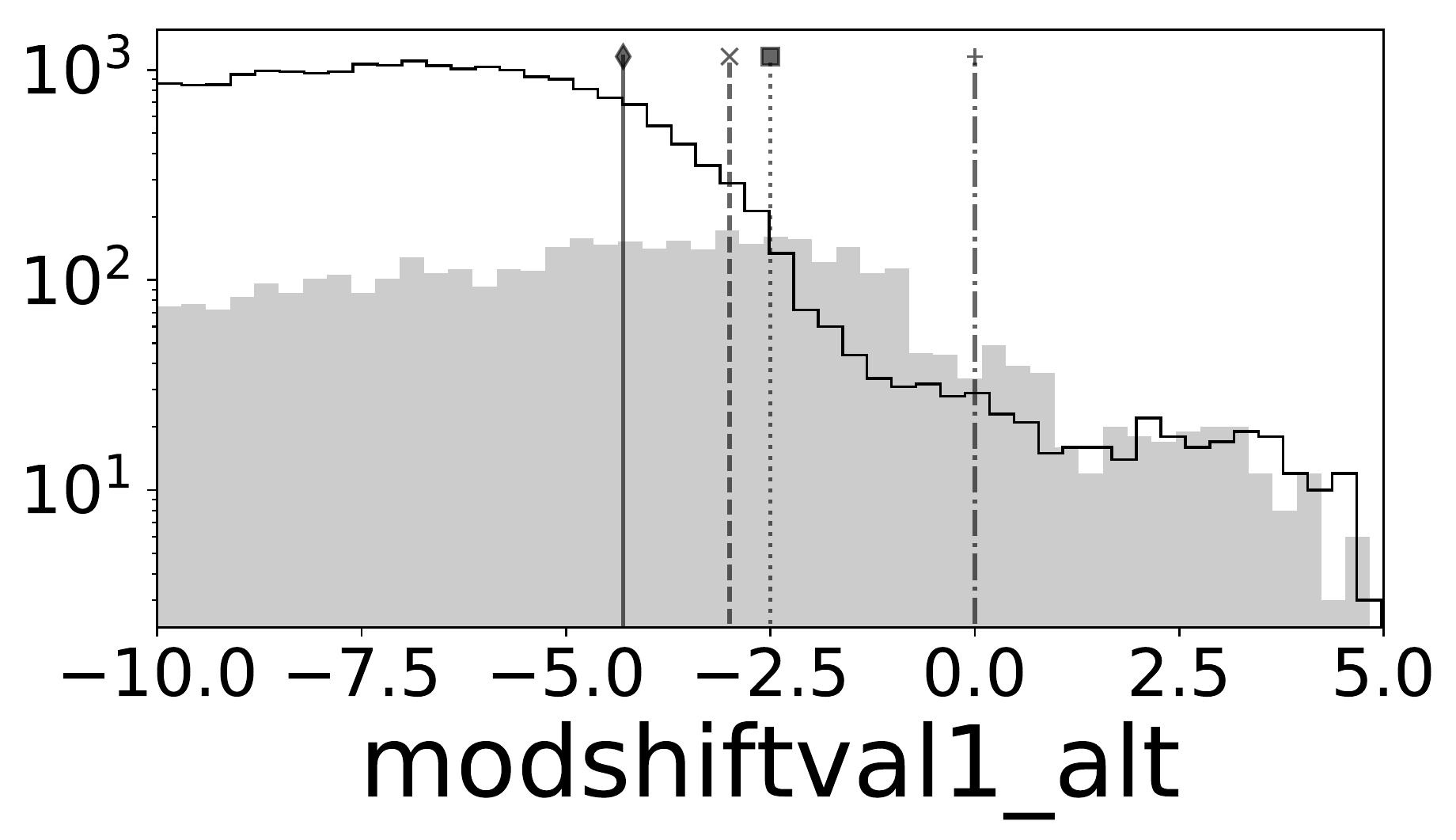} \\
  \includegraphics[width=0.48\linewidth]{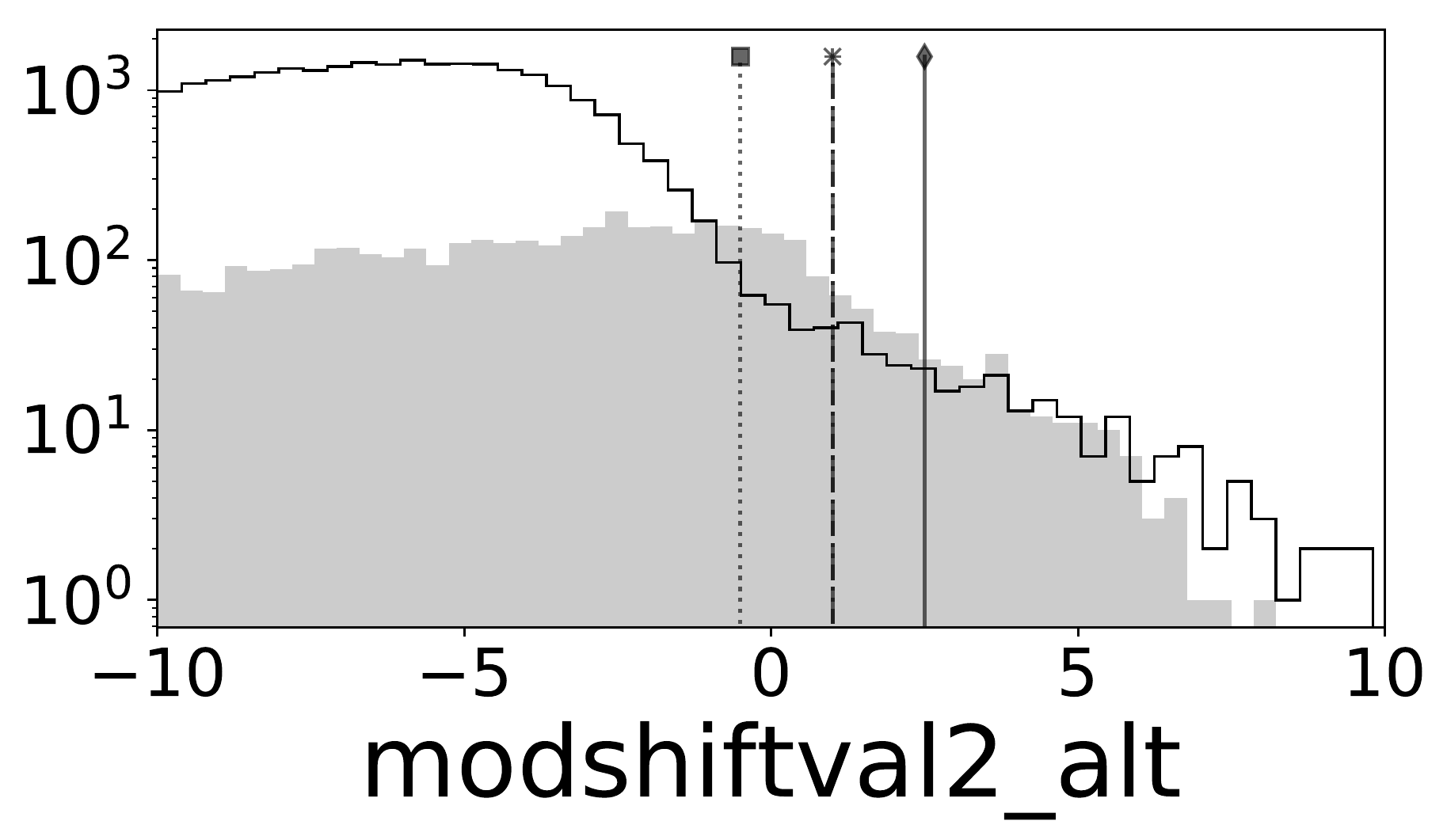} 
  \includegraphics[width=0.48\linewidth]{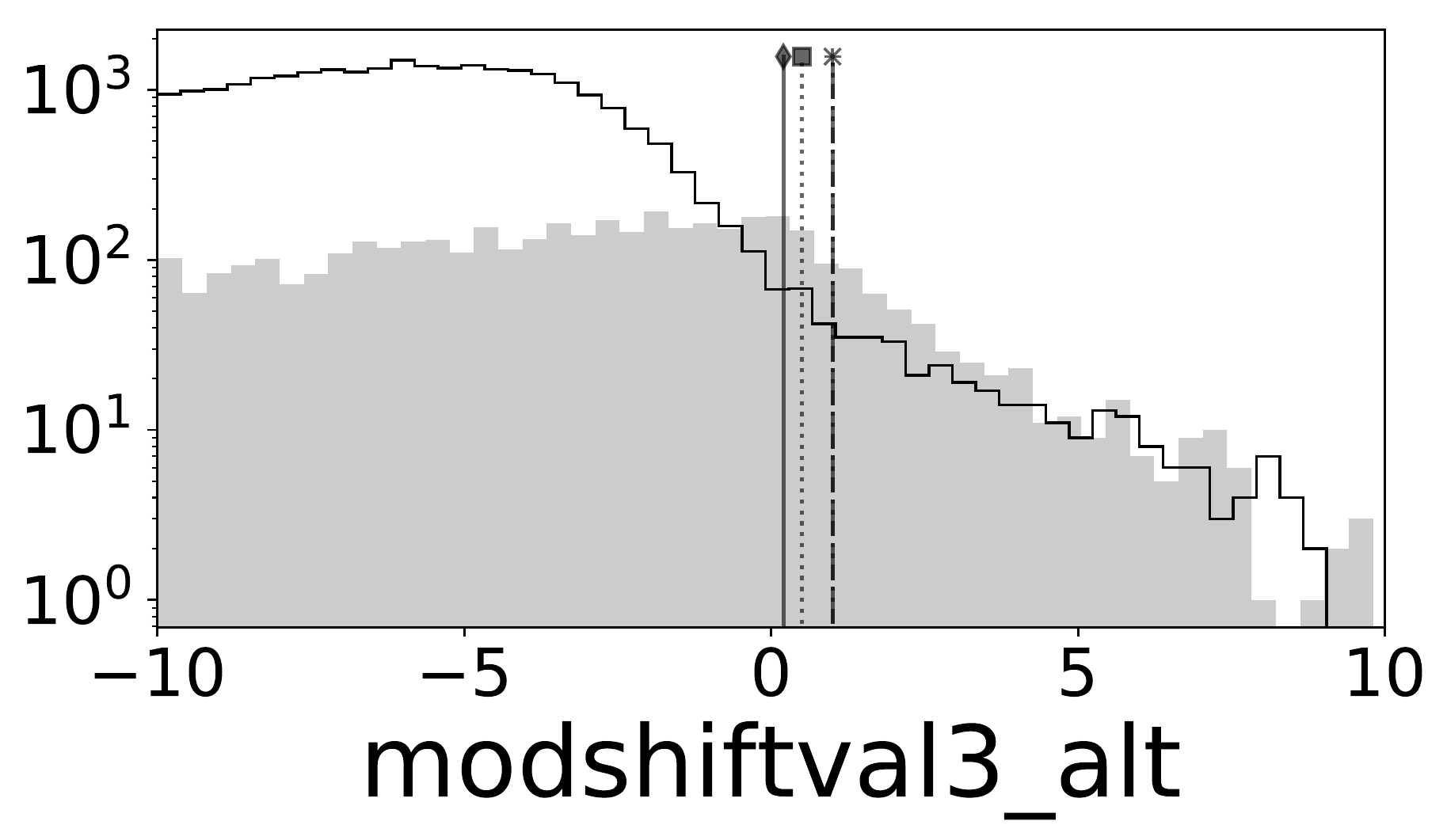} \\
  \includegraphics[width=0.48\linewidth]{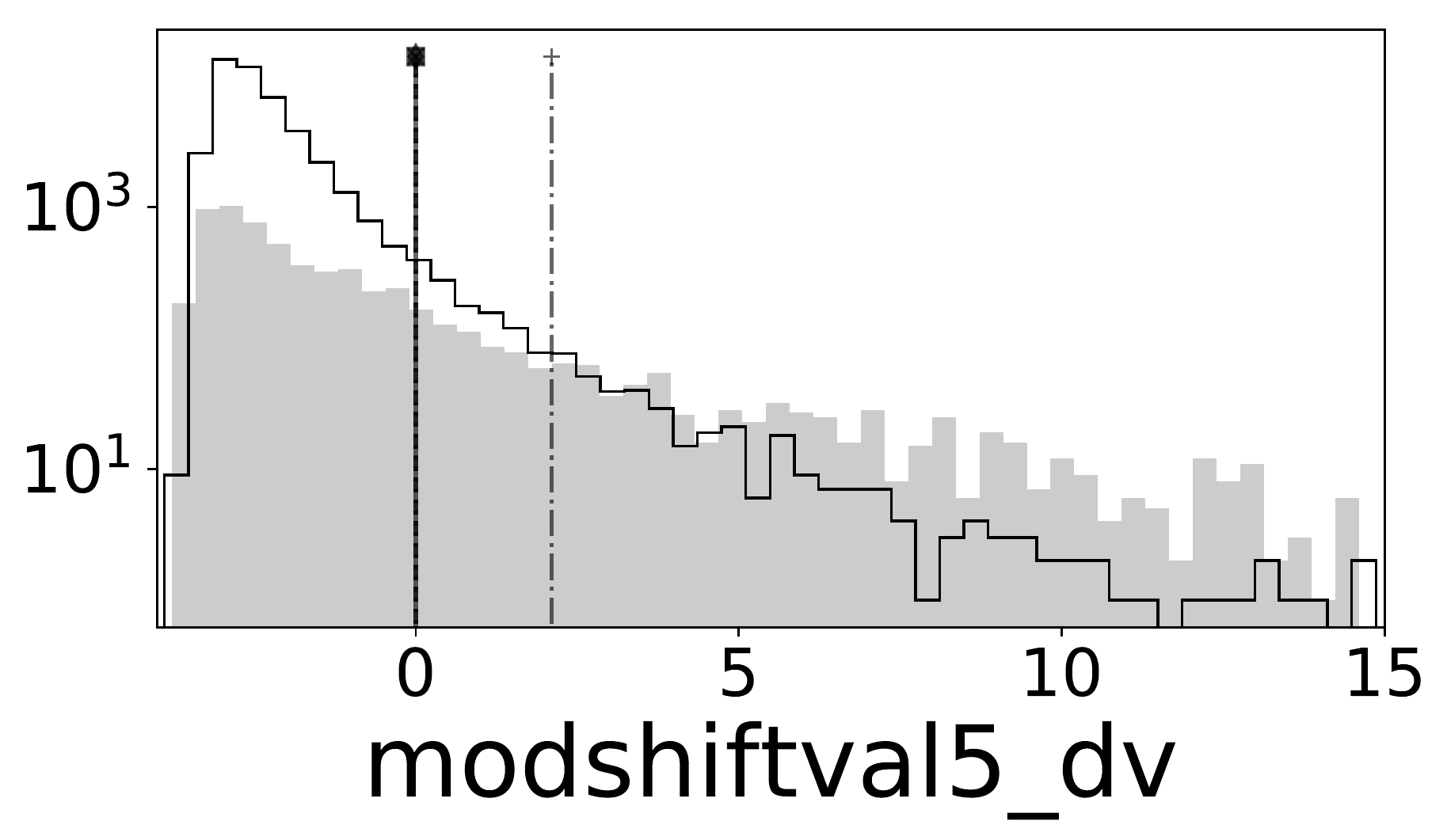}  \\
   \caption{Robovetter metrics and thresholds.  Shaded histogram: metric distribution for false alarms from the inverted and scrambled data.  Line histogram: metric distribution for true transits from the injected data.  The thresholds given in Table~\ref{table:roboetterThresholds} are shown by the vertical lines: diamond solid line: high reliability; 'x' dashed line: DR25; square dotted line: high completeness; '+' dot-dashed line: FPWG PCs.} \label{figure:rvMetrics2}
\end{figure*}

We created the catalogs described in \S\ref{section:thresholdVariation} by changing a subset of the DR25 Robovetter thresholds described in \citet{Thompson2018}.  Table~\ref{table:roboetterThresholds} shows those thresholds that were changed from the DR25 PC catalog -- all other thresholds were unchanged.  Figures~\ref{figure:rvMetrics1} and \ref{figure:rvMetrics2} show histograms of these metrics for the injected and inverted/scrambled data described in \S\ref{section:compRelProducts}.  Injected data provides true transits, while the inverted/scrambled data contains no true transits so any detected transit in the inverted/scrambled data is a false alarm.  Ideally, the thresholds would separate the injected (true transit) from the inverted/scrambled (no transit) populations.  We see, however, that for these metrics there is not a clean separation between data with true transits and data with no true transits, which makes a choice of threshold difficult.  \citet{Thompson2018} describes how these thresholds were chosen for a particular balance of completeness and reliability.  Our alternative thresholds provide different balances of completeness and reliability, and we see from Figures~\ref{figure:rvMetrics1} and \ref{figure:rvMetrics2} that there is considerable freedom in the choice of those thresholds. 

\begin{figure*}[ht]
  \centering
  \includegraphics[width=0.95\linewidth]{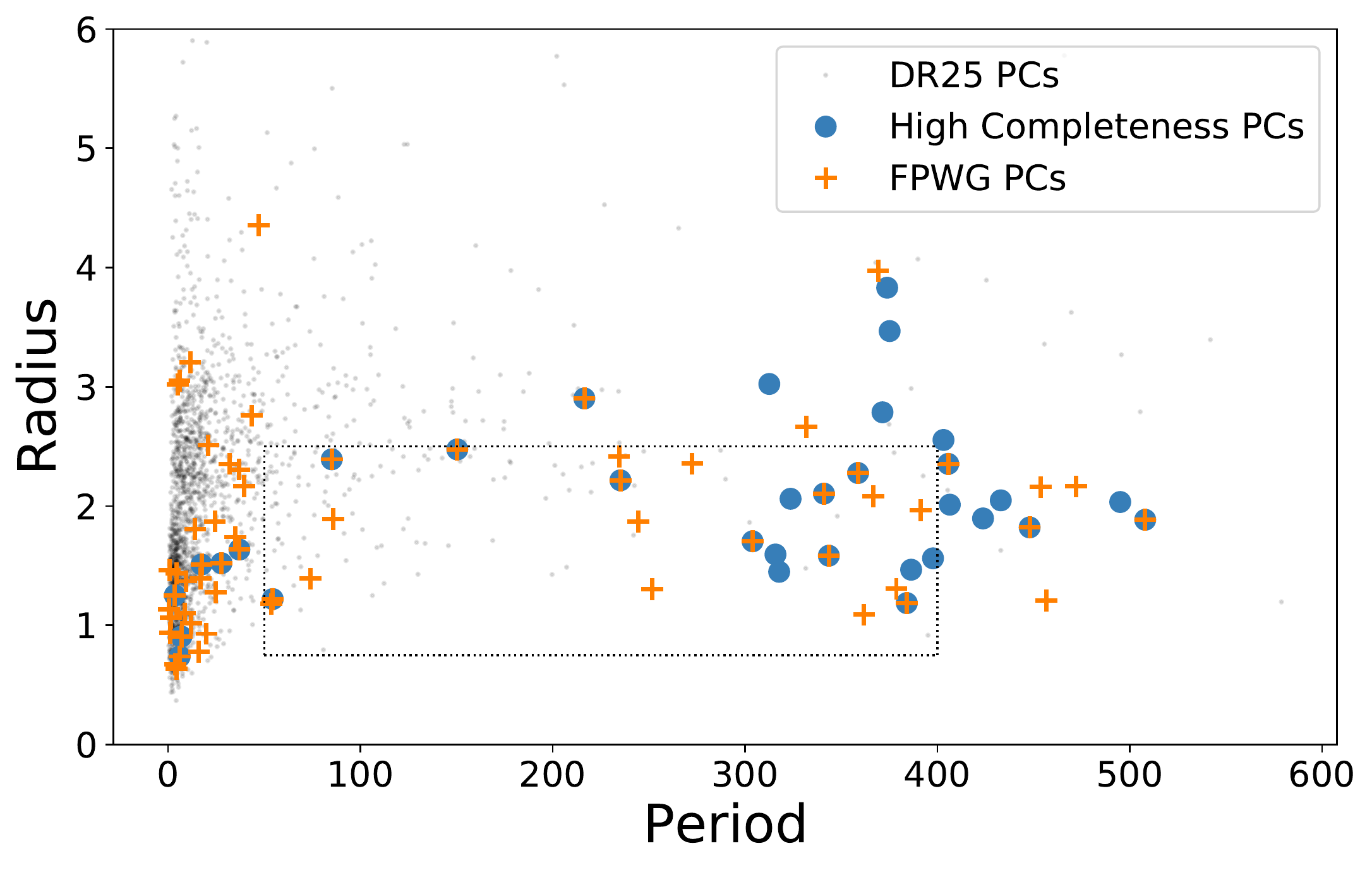} 
   \caption{Planet candidates resulting from the the high completeness and FPWG Robovetter thresholds that are not in the DR25 PC population.  The DR25 PC population is shown for comparison.  The dashed box is the period-radius range used when computing our population model parameters.} \label{figure:newPCs}
\end{figure*}

Table~\ref{table:planetNumbers} gives the total number of planets in each catalog, and how many remain after applying various score cuts.  Table~\ref{table:newPCs} lists the TCEs that were given planet candidate status in the high-completeness or FPWG PC catalogs that do not appear in the DR25 PC catalog (there were no new TCEs given PC status in the high-reliability catalog).  For each TCE in Table~\ref{table:newPCs} the false alarm reliability, computed as described in \S\ref{section:reliability} foe each catalog, is given for the catalog in which it appears.  A missing reliability value indicates that the TCE was not a PC in that catalog.   These new PCs are shown in Figure~\ref{figure:newPCs}.

\renewcommand{\arraystretch}{1}
\begin{table*}[ht]
\centering
\caption{Robovetter Thresholds}\label{table:roboetterThresholds}
\begin{tabular}{ r c c c c }
\hline
\hline
Robovetter Metric & DR25 & High Reliability & High Completeness & FPWG PC \\
\hline
SWEET\_THRESH & 50.0 & 50.0 & 50.0 & 80.0 \\
HALO\_GHOST\_THRESH & 4.0 & 4.0 & 4.0  & 50.0 \\
SES\_TO\_MES\_THRESH & 0.8 & 0.75 & 0.9 & 1.1 \\
ALL\_TRAN\_CHASES\_THRESH & 0.8 & 0.55 & 1.0 & 0.8 \\
SHAPE\_THRESH & 1.04 & 1.04 & 1.04 & 1.14 \\
MOD\_VAL1\_DV\_THRESH & 1.0 & -1.0 & 2.4 & 1.3 \\
MOD\_VAL2\_DV\_THRESH & 2.0 & -0.7 & 5.0 & 2.0 \\
MOD\_VAL3\_DV\_THRESH & 4.0 & -1.6 & 7.5 & 4.0 \\
MOD\_VAL1\_ALT\_THRESH & -3.0 & -4.3 & -2.5 & 0.0 \\
MOD\_VAL2\_ALT\_THRESH & 1.0 & 2.5 & -0.5 & 1.0 \\
MOD\_VAL3\_ALT\_THRESH & 1.0 & 0.2 & 0.5 & 1.0 \\
LPP\_DV\_THRESH & 2.2 & 2.7 & 2.8 & 3.3 \\
LPP\_ALT\_THRESH & 3.2 & 3.2 & 3.2 & 3.2 \\
RV\_OE\_DV\_THRESH & 1.1 & 1.1 & 1.1 & 1.3 \\
RV\_OE\_ALT\_THRESH & 1.1 & 1.1 & 1.1 & 1.8 \\
MOD\_VAL5\_DV\_THRESH & 0.0 & 0.0 & 0.0 & 2.1 \\
\end{tabular}
\end{table*}

\renewcommand{\arraystretch}{1.25}
\begin{table*}[ht]
\centering
\caption{Number of planet candidates in each scenario and score cut}\label{table:planetNumbers}
\begin{tabular}{ r c c c c}
\hline
\hline
Score Cut & DR25 & High Reliability & High Completeness & FPWG PC  \\
\hline
0.0  & 1894  & 1849  & 1928  & 1976 \\
0.6  & 1837  & 1809  & 1837  & 1837 \\
0.7  & 1820  & 1796  & 1820  & 1820 \\
0.9  & 1705  & 1693  & 1705  & 1705 \\
\end{tabular}
\end{table*}

\renewcommand{\arraystretch}{1}
\startlongtable
\begin{deluxetable*}{ r c c c c }
\centering
\tablecaption{New PCs in the High Completeness and FPWG PC catalogs \label{table:newPCs}}
\startdata
\tablehead{TCE ID & Period & Radius & HC Reliability & FPWG PC Reliability\\ & (Days) & ($R_{\oplus}$) &  & }
012403968-01 & 0.59 & 1.14 & - & 0.73 \\
011904151-02 & 0.84 & 1.46 & - & 1.0 \\
007838675-01 & 1.01 & 0.94 & - & 1.0 \\
005009688-01 & 1.38 & 1.07 & - & 0.93 \\
011601357-01 & 3.55 & 1.25 & 0.97 & 0.94 \\
007935997-01 & 3.88 & 0.67 & - & 0.3 \\
005376067-01 & 3.95 & 47.62 & - & 0.0 \\
005688683-02 & 4.45 & 0.64 & - & 0.33 \\
009415108-01 & 4.51 & 1.44 & - & 1.0 \\
009842890-01 & 4.99 & 3.02 & - & 0.41 \\
005872150-02 & 5.92 & 3.05 & - & 1.0 \\
012021943-01 & 6.1 & 0.74 & 0.96 & 0.92 \\
005177859-01 & 6.98 & 0.91 & 1.0 & 1.0 \\
005449777-01 & 7.22 & 52.99 & - & 0.0 \\
006768616-02 & 8.82 & 1.1 & - & 0.97 \\
011599038-02 & 9.28 & 1.37 & - & 0.96 \\
011702948-01 & 9.77 & 12.74 & - & 0.21 \\
010149023-01 & 9.96 & 28.98 & - & 0.0 \\
009119458-01 & 11.53 & 3.21 & - & 0.96 \\
010019399-01 & 11.81 & 10.8 & - & 0.04 \\
008750503-01 & 11.93 & 1.02 & - & 0.86 \\
006599919-01 & 13.61 & 19.47 & - & 0.05 \\
012061969-01 & 14.09 & 1.81 & - & 1.0 \\
008326342-01 & 14.41 & 32.2 & - & 0.0 \\
009763612-01 & 16.05 & 0.78 & - & 0.97 \\
007811537-02 & 16.93 & 1.4 & - & 0.88 \\
011599038-03 & 17.42 & 1.51 & 0.98 & 0.93 \\
010811496-01 & 19.9 & 20.79 & - & 0.07 \\
009347066-01 & 20.0 & 0.93 & - & 0.8 \\
009729691-02 & 21.0 & 2.51 & - & 1.0 \\
003219643-01 & 24.34 & 1.87 & - & 0.26 \\
006436505-02 & 24.72 & 1.28 & - & 0.97 \\
006025124-01 & 26.84 & 16.26 & - & 0.08 \\
005981058-01 & 27.77 & 1.52 & 0.0 & 0.0 \\
004263293-03 & 32.13 & 2.36 & - & 1.0 \\
010019763-01 & 32.5 & 11.37 & - & 0.27 \\
008008913-01 & 35.0 & 1.74 & - & 0.65 \\
011854636-01 & 37.02 & 2.31 & - & 0.54 \\
008938937-03 & 37.11 & 1.64 & 0.95 & 0.92 \\
006381846-03 & 39.6 & 2.17 & - & 1.0 \\
011045383-01 & 41.17 & 23.82 & - & 0.04 \\
005640085-02 & 43.59 & 2.76 & - & 1.0 \\
003355104-01 & 47.06 & 4.36 & - & 0.13 \\
009932970-01 & 52.97 & 29.4 & - & 0.01 \\
011774991-02 & 53.58 & 1.18 & - & 1.0 \\
005871116-01 & 54.43 & 1.22 & 0.1 & 0.09 \\
004371172-01 & 73.99 & 1.39 & - & 0.28 \\
006690171-01 & 85.06 & 21.33 & - & 0.09 \\
003218844-01 & 85.11 & 2.39 & 0.85 & 0.78 \\
006182508-01 & 85.98 & 1.89 & - & 0.82 \\
010666242-01 & 87.24 & 16.29 & - & 0.13 \\
006471021-01 & 125.63 & 8.27 & - & 0.98 \\
007813039-01 & 141.73 & 10.64 & - & 0.01 \\
005706595-03 & 150.38 & 2.48 & 0.81 & 0.78 \\
005015459-01 & 158.32 & 31.09 & - & 0.0 \\
011909686-01 & 185.95 & 64.4 & - & 0.0 \\
004902202-01 & 216.46 & 2.9 & 0.88 & 0.92 \\
012644020-01 & 234.52 & 2.42 & - & 0.36 \\
006032318-01 & 235.21 & 2.22 & 0.39 & 0.4 \\
009209808-01 & 244.55 & 1.87 & - & 0.26 \\
010387742-02 & 251.75 & 1.31 & - & 0.38 \\
012117215-01 & 272.54 & 2.36 & - & 0.32 \\
008223655-01 & 280.16 & 21.74 & - & 0.03 \\
003854101-01 & 293.51 & 46.27 & - & 0.0 \\
007900114-01 & 303.91 & 1.71 & 0.68 & 0.74 \\
006600492-01 & 312.61 & 3.02 & 0.01 & - \\
007762886-02 & 315.81 & 1.59 & 0.3 & - \\
002010152-01 & 317.75 & 1.45 & 0.18 & - \\
008832676-02 & 323.67 & 2.06 & 0.13 & - \\
008742735-01 & 331.93 & 2.67 & - & 0.47 \\
007664272-01 & 341.01 & 2.11 & 0.07 & 0.09 \\
005638699-01 & 343.56 & 1.58 & 0.23 & 0.3 \\
010010452-01 & 358.74 & 2.28 & 0.0 & 0.0 \\
004557341-01 & 361.9 & 1.09 & - & 0.11 \\
006681618-01 & 364.42 & 34.2 & 0.0 & - \\
010338529-01 & 366.75 & 2.08 & - & 0.15 \\
007757698-01 & 369.18 & 3.98 & - & 0.28 \\
007751294-01 & 371.46 & 2.79 & 0.21 & - \\
010205598-03 & 373.89 & 3.83 & 0.05 & - \\
005872139-01 & 375.18 & 3.47 & 0.28 & - \\
010585887-01 & 378.65 & 1.31 & - & 0.19 \\
004763020-01 & 384.1 & 1.19 & 0.18 & 0.2 \\
009710611-02 & 386.35 & 1.47 & 0.16 & - \\
011348086-01 & 391.42 & 1.96 & - & 0.46 \\
011760931-01 & 397.73 & 1.56 & 0.29 & - \\
009771576-01 & 398.72 & 21.45 & - & 0.0 \\
009026007-01 & 403.16 & 2.55 & 0.56 & - \\
004276445-01 & 405.65 & 2.35 & 0.26 & 0.3 \\
006113752-01 & 406.46 & 2.01 & 0.59 & - \\
011124353-02 & 423.73 & 1.9 & 0.37 & - \\
005775090-01 & 432.97 & 2.05 & 0.07 & - \\
008808064-01 & 447.97 & 1.82 & 0.38 & 0.42 \\
010014875-01 & 453.65 & 2.16 & - & 0.3 \\
009239670-01 & 456.55 & 1.21 & - & 0.14 \\
004472143-01 & 472.07 & 2.17 & - & 0.04 \\
012121118-01 & 495.07 & 2.03 & 0.47 & - \\
004645492-01 & 508.04 & 1.89 & 0.42 & 0.39 \\
\enddata
\end{deluxetable*}

\clearpage

\bibliography{refs}

\end{document}